\documentclass[trackchanges, twocolumn]{aastex701}

\usepackage[T1]{fontenc}
\usepackage{textgreek, gensymb, needspace, array, amsmath, afterpage}
\usepackage[table]{xcolor}

\newcommand{\badrow}{\rowcolor{red!12}}

\newcolumntype{L}{>{\hspace{6pt}}l<{\hspace{6pt}}}
\newcolumntype{M}{>{\hspace{6pt}$}r<{$\hspace{6pt}}}
\newcolumntype{C}{>{$}r<{$}}

\newcommand{\tightcutinhead}[1]{%
  \noalign{\vskip 2pt}%
  \tableline
  \multicolumn{5}{c}{#1}\\
  \tableline
  \noalign{\vskip 2pt}%
}

\makeatletter
\newcommand{\maketablepageslandscape}{%
  \let\LS@rot\@empty
  \Gin@pdftexfalse

  \@tempdima=\paperwidth
  \paperwidth=\paperheight
  \paperheight=\@tempdima
  \pdfpagewidth=\paperwidth
  \pdfpageheight=\paperheight
  \pdfpageattr{/Rotate 0}

  \textwidth=\hsize
  \hoffset=0pt
  \voffset=0pt

  \oddsidemargin=\dimexpr
    (\paperwidth-\textwidth)/2-1in\relax
  \evensidemargin=\oddsidemargin
  \topmargin=\dimexpr
    (\paperheight-\textheight)/2-1in
    -\headheight-\headsep\relax
}
\makeatother

\received{November 26, 2026}
\submitjournal{ApJ}

\shorttitle{Mid-to-Late M-Dwarf Metallicities}
\shortauthors{Pass et al.}
\graphicspath{{./}{figures/}}
\begin{document}

\title{Calibrating a Near-IR Metallicity Relation for Mid-to-Late M Dwarfs}

\author[0000-0002-1533-9029]{Emily K. Pass}
\affiliation{Kavli Institute for Astrophysics and Space Research, Massachusetts Institute of Technology, Cambridge, MA 02139, USA}
\affiliation{Center for Astrophysics $\vert$ Harvard \& Smithsonian, 60 Garden Street, Cambridge, MA 02138, USA}
\email{epass@mit.edu}

\author[0000-0001-9911-7388]{David W. Latham}
\affiliation{Center for Astrophysics $\vert$ Harvard \& Smithsonian, 60 Garden Street, Cambridge, MA 02138, USA}
\email{}

\author[0000-0002-9003-484X]{David Charbonneau}
\affiliation{Center for Astrophysics $\vert$ Harvard \& Smithsonian, 60 Garden Street, Cambridge, MA 02138, USA}
\email{}

\author[0000-0001-7246-5438]{Andrew Vanderburg}
\affiliation{Center for Astrophysics $\vert$ Harvard \& Smithsonian, 60 Garden Street, Cambridge, MA 02138, USA}
\email{}

\author[0000-0001-6031-9513]{Jennifer G. Winters}
\affiliation{Department of Physics, Photonics and Optical Engineering, Bridgewater State University, 131 Summer Street, \\ Bridgewater, MA 02325, USA}
\affiliation{Center for Astrophysics $\vert$ Harvard \& Smithsonian, 60 Garden Street, Cambridge, MA 02138, USA}
\email{}

\author[0000-0002-1617-8917]{Phillip A. Cargile}
\affiliation{Center for Astrophysics $\vert$ Harvard \& Smithsonian, 60 Garden Street, Cambridge, MA 02138, USA}
\email{}

\author[0000-0001-6637-5401]{Allyson Bieryla}
\affiliation{Center for Astrophysics $\vert$ Harvard \& Smithsonian, 60 Garden Street, Cambridge, MA 02138, USA}
\email{}

\author{Perry Berlind}
\affiliation{Center for Astrophysics $\vert$ Harvard \& Smithsonian, 60 Garden Street, Cambridge, MA 02138, USA}
\email{}

\author[0000-0002-2830-5661]{Michael L. Calkins}
\affiliation{Center for Astrophysics $\vert$ Harvard \& Smithsonian, 60 Garden Street, Cambridge, MA 02138, USA}
\email{}

\author[0000-0002-9789-5474]{Gilbert A. Esquerdo}
\affiliation{Center for Astrophysics $\vert$ Harvard \& Smithsonian, 60 Garden Street, Cambridge, MA 02138, USA}
\email{}

\author[0000-0003-3594-1823]{Jessica Mink}
\affiliation{Center for Astrophysics $\vert$ Harvard \& Smithsonian, 60 Garden Street, Cambridge, MA 02138, USA}
\email{}



\begin{abstract}

\noindent Measuring M-dwarf metallicities is challenging due to these stars' ubiquitous molecular bands that completely disguise the continuum, particularly given the limitations of theoretical models. In this work, we develop a new calibration that uses absolute $K$--band magnitude and the central line of the near-infrared Ca II triplet at 8542\AA\ to estimate [Fe/H] for inactive, mid-to-late M dwarfs. We show that existing relations are discrepant in the low-mass M-dwarf regime, highlighting the need for mid-to-late M dwarfs with external constraints on metallicity to calibrate this indicator. To this end, we use the high-resolution TRES and MAROON-X spectrographs to calibrate our relation using low-mass M dwarfs in wide binaries with FGK stars for which we homogeneously measure metallicities using the TRES--\texttt{uberMS} code. We observe 69 wide binaries and downselect to a curated sample of 36 pairs, avoiding primaries with unreliable properties, active M dwarfs, and unresolved multiples. Our relation produces median uncertainties of $\pm0.07$~dex in [Fe/H] (or $\pm0.05$~dex in [\textalpha/H]). We apply this relation to 118 single, inactive stars from a volume-complete sample of mid-to-late M dwarfs within 15pc. There are 19 known planet hosts in this sample, and we find tentative evidence (94\% confidence) that these planet hosts are preferentially metal rich. Our FGK--M observing campaign also identifies two new M-dwarf spectroscopic binaries that are members of higher-order multiples, HD~90089~BD and 2MASS~J1705+3101~AB, and we provide full orbital characterization of the inner systems.

\end{abstract}

%\keywords{stars: abundances --- stars: low-mass --- techniques: spectroscopic}


\section{Introduction}
\label{sec:intro}

There are many reasons why stellar metallicity is an important property for exoplanets. One of the most fundamental is that precise stellar parameters are needed to obtain precise planetary parameters \citep[e.g.,][]{Muirhead2012, Tayar2022}, but also, stellar metallicity correlates with planet occurrence \citep[at least for giant planets;][]{Fischer2005} and a diverse array of planetary properties \citep[e.g.,][]{Spalding2022, An2023, Zhu2023, Hua2025}. As nearby M dwarfs represent the most favorable hosts for the detection and characterization of Earth-like planets with present and near-future instrumentation \citep{Charbonneau2007, Morley2017}, there is a particular need to determine the metallicities of these small stars. However, such metallicity measurements are challenging. Due to the cool temperatures of these stars and the omnipresent molecular bands that obscure the continuum, techniques that work well for Sun-like stars \citep[e.g.,][]{Valenti2005} fail in the low-mass regime. 

Despite the challenges posed by the complex spectra of M dwarfs, there have been many attempts to extend the use of theoretical models for metallicity determination to lower stellar masses \citep[e.g.,][]{Woolf2005, Onehag2012, Veyette2017, Rajpurohit2018, Passegger2018, Souto2020}. However, the robustness of such methods is still a topic of debate: M-dwarf theoretical models are known to be inaccurate \citep[e.g.,][]{Parsons2018, Dieterich2021, Olander2021, Glidden2026}, and current model-based methods produce estimates that do not correlate well with other works \citep[e.g.,][]{Fouque2018}.

Empirical calibrations circumvent the need for accurate theoretical models, and numerous such calibrations have also been developed \citep[e.g.,][]{RojasAyala2012, Neves2012, Terrien2012, Mann2013, Mann2014, Newton2014}. These calibrations link equivalent widths (EWs) of various spectral features in M dwarfs to metallicity using wide-binary pairs where the primary is an F-, G-, or K-type star and the secondary is the M dwarf of interest. As each pair formed from the same molecular cloud, one can assume they share a common metallicity \citep{Bonfils2005}. This assumption has been validated for wide binaries where both components are FGK stars \citep{Desidera2004, Desidera2006}. Such methods are capable of measuring M-dwarf metallicities to better than $\pm$0.1 dex, although they still cannot match the roughly $\pm$0.03 dex uncertainties achievable for FGK stars.

\begin{figure}
    \includegraphics[width=0.5\textwidth]{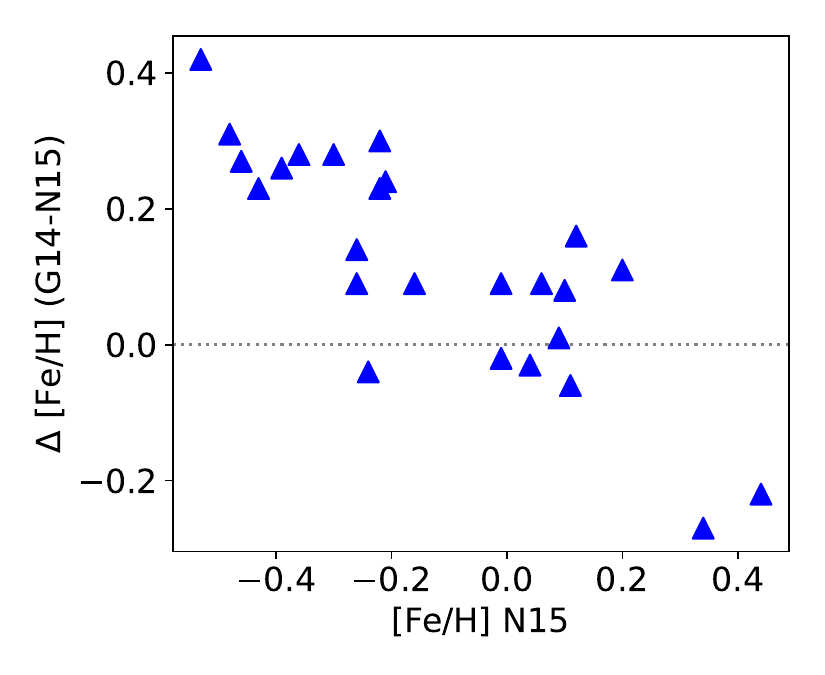}\vspace{-0.4cm}
    \caption{[Fe/H] estimates for 25 low-mass ($<0.3$M$_\odot$) M dwarfs that were studied in both \citet{Newton2015}, which uses the \citet{Mann2013, Mann2014} infrared relations, and \citet{Gaidos2014_conch}, which uses the \citet{Mann2013} optical relation. There are strong systematic differences between the two relations: when the infrared relation gives [Fe/H]=$-$0.4\,dex, the optical relation deviates by 0.3\,dex (i.e., giving values of $-$0.1\,dex). For ease of interpretation, unity is marked by the dotted line.}
    \label{fig:g14}
\end{figure}

Arguably the most widely used empirical relation is from \citet[][hereafter M+]{Mann2013, Mann2014}, who calibrate different relations for optical, $J$--, $H$--, or $K$--band spectra. These relations produce [Fe/H] estimates with errors of $\pm$0.08~dex in the infrared and 0.10--0.13~dex in the optical \citep{Mann2015}. \citet{Gaidos2014_conch} and \citet{Newton2015} estimate metallicities for a large number of M dwarfs based on the M+ relations, with \citet{Gaidos2014_conch} using the optical and \citet{Newton2015} using the $K$ band. However, the application of these relations to large samples of M dwarfs reveals a problem: the optical and infrared relations are discrepant for fully convective M dwarfs. In Figure~\ref{fig:g14}, we highlight the 25 low-mass M dwarfs ($<0.3$M$_\odot$\footnote{To ensure consistency with the rest of this work, we calculate these masses using the \citet{Benedict2016} $K$--band mass--luminosity relation, 2MASS magnitudes, and Gaia DR3 parallaxes, rather than using each work's tabulated mass estimates. This relation has an RMS error of $\pm$0.014M$_\odot$ and is less sensitive to metallicity effects than optical-based relations \citep{Delfosse2000}.}) that were studied in both works. The agreement is poor at both low and high metallicities; for example, when the infrared relation yields [Fe/H]=$-$0.4\,dex, the optical relation is more metal rich by 0.3\,dex. To explain this discrepancy, we note that the visible, $J$--, and $H$--band relations are only calibrated down to a spectral type of M6, with a separate relation available that uses $K$--band spectra and that is applicable for M4.5–M9.5 dwarfs. The discrepancy between the mid-M dwarfs in the \citet{Gaidos2014_conch} and \citet{Newton2015} samples may suggest that the optical relation breaks down even earlier than M6, and it therefore should not be used to determine [Fe/H] for low-mass M dwarfs.

This discussion brings us to the present work: we aim to measure metallicities from a volume-complete survey of nearby mid-to-late M-dwarfs \citep{Winters2021, Winters2026}, which provides high-resolution optical spectra ($\sim$4000--9000\AA) for all 0.1--0.3M$_\odot$ M dwarfs within 15\,pc. As discussed above, the M+ relations are not calibrated for optical spectra of the lowest-mass stars. The MCAL method \citep{Neves2014} uses optical spectra, but is similarly limited to M0--M5 dwarfs. No existing calibration is applicable to these data. This incompatibility motivates the new calibration we present here, which links the equivalent width of the central 8542\AA\ line of the near-IR Ca II triplet (CaT) to metallicity.

In this work, we use a sample of FGK--M wide binaries to calibrate a novel spectroscopic metallicity relation for inactive, mid-to-late M dwarfs. In Section~\ref{sec:fgk-m}, we describe the observation, reduction, and analysis of the binary sample. In Section~\ref{sec:calib}, we use these stars to calibrate our CaT-based metallicity relation. In Section~\ref{sec:met}, we apply our calibration to the 15pc M-dwarf survey and discuss population-level trends. We conclude in Section~\ref{sec:conclusion}.

\section{FGK--M Sample}
\label{sec:fgk-m}

\subsection{Observations}
Our FGK--M sample is divided between two instruments. The first subsample comprises 56 nearby ($\leq50$~pc) mid-to-late M dwarfs (0.10--0.35M$_\odot$) with widely separated ($\geq4"$) FGK primaries that are observable with the Tillinghast Reflector Echelle Spectrograph (TRES; 3900–9100\AA; \citealt{Szentgyorgyi2007}) on the 1.5 m telescope at the Fred Lawrence Whipple Observatory in Arizona. Most of these pairs are well studied in the literature, although we also supplement our target list with a search for common proper motion pairs using Gaia \citep{Gaia2016}; that method is discussed below. From 2020--2025, we gathered $R$=44,000 TRES spectra of each primary and secondary, obtaining multiple observations of the M-dwarf components with the goal of attaining a coadded per-pixel signal-to-noise ratio (SNR) of 30 in TRES order 49, near the calcium infrared triplet.

Even with coaddition, TRES is limited to M dwarfs brighter than $G$=14.5~mag, resulting in few M-dwarf secondaries in the lower end of our mass regime. To address this imbalance, we pursued a complementary observing campaign with MAROON-X at the \hbox{8.1 m} Gemini-North telescope, an $R$=85,000 spectrograph spanning the wavelength range 5000--9200\AA\ \citep{Seifahrt2016, Seifahrt2018, Seifahrt2020, Seifahrt2022}. These MAROON-X observations were observed under programs GN-2022B-Q-323 and GN-2023B-Q-325 (PI: Pass). We also collected TRES observations of their FGK primaries.

To select our MAROON-X sample, we searched for wide ($\geq4"$) common proper motion pairs in the Gaia DR3 catalog \citep{Gaia2023} using the proper motion and position angle cuts of \citet[][their Equations 1 and 2]{Montes2018}, requiring parallax agreement within 0.5 mas. We cross-matched with the 2MASS catalog \citep{Cutri2003} to obtain $K$--band magnitudes and used the \citet{Benedict2016} $K$--band relation to estimate masses for the M dwarfs, selecting all the pairs within 50pc that contain an M dwarf with mass $<0.3$M$_\odot$ and an FGK primary. We took care to mitigate contamination by close binaries and young, active stars: we rejected stars with a known unresolved companion, stars with a large renormalised unit weight error (RUWE; i.e., an astrometric perturbation likely due to an unresolved companion) in Gaia DR3, pairs that are members of young moving groups, pairs for which gyrochronology of the primary suggests youth, M dwarfs with measured H\textalpha\ emission or large $v$sin$i$ in the literature, M dwarfs for which we observe rapid rotation in Transiting Exoplanet Survey Satellite \citep[TESS;][]{Ricker2015} light curves, and overluminous M dwarfs. Many of our MAROON-X pairs were not identified as binaries before Gaia and thus represent novel targets that have not been used to calibrate other metallicity relations in the literature. We verify that these new wide binaries have proper motion differences consistent with a bound Keplerian orbit using the criterion of \citet[][their Equation 3]{el-Badry2021}.

Despite only partial program completion, we were able to observe 26 secondaries with MAROON-X: 15 novel targets and 11 targets that overlap with the TRES subsample. The overlapping targets allow us to evaluate whether there are systematic offsets between measurements from the two different instruments (Section~\ref{sec:ew_calib}). After combining the TRES and MAROON-X data sets, our sample comprises 71 pairs. All of these binaries are listed in Table~\ref{tab:stellar-pairs}, along with their astrometric properties from Gaia DR3 \citep{Gaia2023}.

Unfortunately, in two cases (TIC~29923115 and TIC~441592532) the primary star was mistakenly observed with MAROON-X instead of the M dwarf. We therefore remove these pairs from our sample, reducing the number of systems to 69. We nonetheless report our measured properties from the TRES spectra of their primaries (HD~10226 and HD~59201) in case these data are useful to the reader.

\afterpage{%
  \onecolumngrid 
\begin{longrotatetable}
\maketablepageslandscape
\begin{deluxetable*}{llllrrrrrrrrrcc}
\tabletypesize{\fontsize{10}{11}\selectfont}
\setlength{\tabcolsep}{1.5pt}
\renewcommand{\arraystretch}{1}
\setlength{\extrarowheight}{0pt}
\tablewidth{0pt}
\tablecolumns{15}
\tablecaption{Gaia DR3 astrometric data (separations, parallaxes, proper motions, and radial velocities) for 71 FGK--M pairs\label{tab:stellar-pairs}}
\tablehead{
\multicolumn{2}{c}{Name} & \multicolumn{2}{c}{Gaia DR3 Source ID} &
\multicolumn{1}{c}{Sep.} & \multicolumn{2}{c}{$\pi$} &
\multicolumn{2}{c}{$\mu_{\alpha *}$} & \multicolumn{2}{c}{$\mu_\delta$} &
\multicolumn{2}{c}{RV} & Used?\hspace{0.2cm} & Note \\
Primary & M Dwarf & Primary & M Dwarf & ($^{\prime\prime}$) &
\multicolumn{2}{c}{(mas)} & \multicolumn{2}{c}{(mas\,yr$^{-1}$)} &
\multicolumn{2}{c}{(mas\,yr$^{-1}$)} & \multicolumn{2}{c}{(km\,s$^{-1}$)} & & \\
 & & & & & A & B & A & B & A & B & A & B & & }
\startdata
HD~6660 & GJ~53.1~B & 2791782794564103808 & 2791782794564103680 & $9$ & $48.7$ & $48.7$ & $103$ & $97$ & $-490$ & $-495$ & $6.6$ & $9.3$ & N & CP \\
HIP~6339 & BD+30206~B & 310361075612875392 & 310361140039987584 & $10$ & $34.4$ & $34.7$ & $560$ & $541$ & $-110$ & $-122$ & $25.1$ & $22.1$ & Y &  \\
HD~8828 & LP~587-63 & 2534009921293834240 & 2534008512544580480 & $180$ & $32.2$ & $32.2$ & $299$ & $299$ & $-357$ & $-357$ & $13.6$ & $12.8$ & Y &  \\
HD~9826 & GJ~61~B & 348020482735930112 & 348020242217448576 & $56$ & $74.2$ & $74.2$ & $-172$ & $-172$ & $-382$ & $-384$ & $-28.8$ & $-28.3$ & Y &  \\
HD~10008 & G~271-110 & 2477815222028038272 & 2478001486169801216 & $612$ & $41.6$ & $41.7$ & $172$ & $174$ & $-98$ & $-100$ & $11.5$ & $\cdots$ & N & AM \\
HD~10226 & TIC~29923115 & 2463995460378587904 & 2463995460377935616 & $8$ & $20.9$ & $20.7$ & $239$ & $237$ & $90$ & $96$ & $13.5$ & $\cdots$ & N & W \\
HD~16160 & GJ~105~B & 19316224572460416 & 18565464288396416 & $164$ & $138.3$ & $138.4$ & $1779$ & $1802$ & $1477$ & $1450$ & $25.2$ & $26.1$ & N & CP \\
HD~16417 & TIC~122555697 & 5062172841616745856 & 5062172669818053760 & $45$ & $39.3$ & $39.3$ & $-18$ & $-17$ & $-259$ & $-265$ & $11.2$ & $\cdots$ & Y &  \\
HD~16714 & TIC~122574948 & 5062313166791762176 & 5062312410877517056 & $32$ & $24.7$ & $24.7$ & $77$ & $77$ & $-394$ & $-394$ & $24.0$ & $28.8$ & Y &  \\
HD~18143 & GJ~118.2~C & 116037204451525376 & 116037376250217984 & $44$ & $44.4$ & $44.5$ & $265$ & $277$ & $-193$ & $-176$ & $31.9$ & $32.5$ & Y &  \\
HD~18404 & 2MA0258+20 & 108421608959951488 & 108421402801188864 & $15$ & $30.6$ & $30.6$ & $234$ & $238$ & $-32$ & $-24$ & $27.9$ & $\cdots$ & Y &  \\
HD~21962 & TIC~435293563 & 42768532716776704 & 42768945033709568 & $467$ & $20.7$ & $20.5$ & $-24$ & $-24$ & $-96$ & $-96$ & $-6.2$ & $-0.8$ & N & AM \\
HIP~17346 & TIC~89148224 & 5085376076015097600 & 5085376179094311680 & $35$ & $38.4$ & $38.4$ & $24$ & $29$ & $-374$ & $-373$ & $16.0$ & $17.8$ & N & CP \\
HD~24421 & LEP0355+52 & 251738448519542144 & 251726010294299136 & $236$ & $25.3$ & $25.2$ & $-120$ & $-121$ & $109$ & $109$ & $-33.4$ & $-33.1$ & Y &  \\
HD~24421 & HD~24421~B & 251738448519542144 & 251738620318225792 & $52$ & $25.3$ & $25.2$ & $-120$ & $-117$ & $109$ & $108$ & $-33.4$ & $-33.4$ & Y &  \\
HD~24409 & HD~24409~C & 473347456640406016 & 473347456640408448 & $10$ & $46.0$ & $45.8$ & $-276$ & $-292$ & $152$ & $176$ & $-15.7$ & $-20.1$ & Y &  \\
HD~26965 & GJ~166~C & 3195919528989223040 & 3195919254111314816 & $78$ & $199.6$ & $199.5$ & $-2240$ & $-2247$ & $-3422$ & $-3410$ & $-42.5$ & $-44.1$ & N & AM \\
HIP~20240 & TIC~167455127 & 3192532758297814656 & 3192532758297814784 & $18$ & $29.4$ & $29.5$ & $-58$ & $-65$ & $-13$ & $-10$ & $-7.2$ & $\cdots$ & N & AM \\
HD~35681 & 2MA0528+33 & 182461725546788480 & 182461725544178816 & $10$ & $29.2$ & $29.2$ & $20$ & $24$ & $-210$ & $-204$ & $12.1$ & $\cdots$ & N & AM \\
HD~245409 & LEP0536+11 & 3339921875389105152 & 3339914973377653760 & $157$ & $87.5$ & $87.3$ & $-3$ & $-3$ & $-56$ & $-61$ & $21.4$ & $21.6$ & N & AM \\
HD~40397 & TIC~39543069 & 3022685626580347392 & 3022685905755527296 & $89$ & $41.5$ & $41.5$ & $75$ & $79$ & $-200$ & $-213$ & $143.4$ & $146.0$ & Y &  \\
HD~43147 & 2MA0616+44 & 962379266249460736 & 962379270546799872 & $18$ & $23.8$ & $23.9$ & $-256$ & $-255$ & $-333$ & $-336$ & $-34.6$ & $-31.5$ & Y &  \\
HD~44985 & TIC~206680308 & 3330686669209149824 & 3330686463046060288 & $9$ & $30.1$ & $30.0$ & $-123$ & $-132$ & $-57$ & $-52$ & $32.4$ & $33.2$ & Y &  \\
HD~51067 & LP~16-395 & 1115378031539193984 & 1115378443855487104 & $61$ & $25.8$ & $25.4$ & $-91$ & $-89$ & $-256$ & $-250$ & $12.6$ & $14.0$ & N & AM \\
HD~48974 & LP~4-248 & 1149610570154926464 & 1149611738386037248 & $297$ & $19.8$ & $19.8$ & $26$ & $27$ & $-221$ & $-223$ & $-26.2$ & $-25.5$ & N & BM \\
HD~57625 & TIC~444981081 & 989016898234332416 & 989016790859521280 & $9$ & $22.6$ & $22.5$ & $-55$ & $-62$ & $-59$ & $-59$ & $4.7$ & $\cdots$ & N & AM \\
HD~59201 & TIC~441592532 & 1111224007889044736 & 1111224007889044992 & $7$ & $27.1$ & $27.1$ & $120$ & $119$ & $333$ & $341$ & $-28.5$ & $-25.6$ & N & W \\
HD~60912 & TIC~453409628 & 982266480954408704 & 982263903974100352 & $106$ & $22.6$ & $22.5$ & $-79$ & $-80$ & $-62$ & $-61$ & $7.8$ & $\cdots$ & N & AM \\
HD~67458 & TIC~154234106 & 5596797039059063680 & 5596797073418770560 & $9$ & $39.0$ & $39.0$ & $340$ & $341$ & $-355$ & $-353$ & $-15.7$ & $-15.4$ & Y &  \\
HD~68638 & G~194-18 & 1035000055055287680 & 1034999982042706048 & $60$ & $30.8$ & $30.7$ & $-315$ & $-314$ & $-223$ & $-219$ & $17.5$ & $18.0$ & N & BP \\
HD~75302 & LEP0849+03~W & 581100382135253760 & 581103581886377728 & $159$ & $33.0$ & $33.0$ & $-148$ & $-150$ & $60$ & $58$ & $10.1$ & $\cdots$ & N & AM \\
HD~75732 & GJ~324~B & 704967037090946688 & 704966762213039488 & $85$ & $79.4$ & $79.7$ & $-486$ & $-481$ & $-234$ & $-245$ & $27.2$ & $27.8$ & Y &  \\
HD~77599 & NLTT~20915 & 1036004149689903872 & 1036004046610688000 & $56$ & $21.5$ & $21.4$ & $-214$ & $-214$ & $2$ & $2$ & $26.3$ & $25.8$ & Y &  \\
HD~90089 & HD~90089~B & 1146337324038514176 & 1146337324038513536 & $13$ & $43.4$ & $44.2$ & $-86$ & $-105$ & $20$ & $37$ & $5.8$ & $59.0$ & N & BM \\
HD~93380 & HD~93380~B & 5470514142863169280 & 5470514211582645248 & $50$ & $47.0$ & $46.8$ & $-142$ & $-138$ & $-110$ & $-112$ & $-11.9$ & $-12.6$ & Y & CP \\
HD~93811 & LP~128-32 & 837178843564106112 & 837177915851182080 & $178$ & $37.4$ & $37.9$ & $-191$ & $-191$ & $45$ & $3$ & $-13.3$ & $-0.2$ & N & AM \\
HD~96937 & 2MA1109+02 & 3811594088617869056 & 3811594122977607680 & $10$ & $34.0$ & $34.1$ & $-277$ & $-285$ & $39$ & $35$ & $10.1$ & $9.9$ & Y &  \\
HD~101650 & 2MA1141$-$23 & 3540033580054935808 & 3540033305177028480 & $77$ & $23.7$ & $23.7$ & $-171$ & $-174$ & $-37$ & $-37$ & $-2.2$ & $-1.7$ & N & CP \\
HD~102574 & 2MA1148$-$10 & 3587572064793827328 & 3587571858635396352 & $18$ & $22.4$ & $22.5$ & $-105$ & $-108$ & $-109$ & $-104$ & $14.6$ & $14.8$ & Y &  \\
HIP~58817 & LP~794-16 & 3567910941462154880 & 3567818445046215168 & $324$ & $26.8$ & $26.8$ & $390$ & $391$ & $-444$ & $-446$ & $27.0$ & $28.6$ & N & CP \\
HD~110869 & LP~95-334 & 1578083104672582144 & 1578083104672582784 & $35$ & $22.1$ & $22.1$ & $-213$ & $-213$ & $27$ & $27$ & $-9.1$ & $-8.8$ & Y &  \\
HD~111398 & 2MA1248+12 & 3928040444539772800 & 3928040341460587904 & $76$ & $27.6$ & $27.5$ & $233$ & $232$ & $-140$ & $-141$ & $3.3$ & $6.2$ & Y &  \\
LP~676-69 & LP~676-70 & 3678639905636817152 & 3678639905636817408 & $17$ & $21.7$ & $21.7$ & $26$ & $26$ & $-233$ & $-237$ & $5.9$ & $5.0$ & N & CP \\
HD~125455 & GJ~544~B & 3642413746280598784 & 3642410791343099008 & $15$ & $48.9$ & $48.8$ & $-633$ & $-633$ & $-121$ & $-115$ & $-10.0$ & $-10.6$ & Y &  \\
HD~127871 & G~66-6 & 1176167143218779776 & 1176167040139596032 & $74$ & $31.0$ & $31.0$ & $158$ & $159$ & $-513$ & $-512$ & $30.5$ & $31.4$ & Y &  \\
HD~129502 & 2MA1443$-$05 & 6337717036911729024 & 6337717243069148032 & $42$ & $52.2$ & $53.6$ & $110$ & $97$ & $-316$ & $-317$ & $1.6$ & $\cdots$ & N & AM \\
HD~129715 & LP~858-23 & 6278792868749734912 & 6278792971828981888 & $69$ & $33.0$ & $32.9$ & $-106$ & $-109$ & $-339$ & $-339$ & $8.6$ & $8.3$ & N & CP \\
HIP~76042 & G~201-50 & 1401760747118886016 & 1401772395070195840 & $160$ & $30.8$ & $30.8$ & $155$ & $156$ & $-299$ & $-300$ & $13.2$ & $11.1$ & N & CP \\
HD~139477 & LP~99-392 & 1603267143629157120 & 1603272950424941440 & $43$ & $51.2$ & $51.3$ & $172$ & $171$ & $-163$ & $-164$ & $-9.0$ & $-9.4$ & N & AM \\
HD~141272 & HD~141272~B & 4422809716282811776 & 4422809711989181824 & $18$ & $46.2$ & $46.1$ & $-176$ & $-179$ & $-166$ & $-172$ & $-26.6$ & $-25.3$ & Y &  \\
HD~144579 & GJ~611~B & 1379500756257031808 & 1379500928055726848 & $70$ & $69.6$ & $69.6$ & $-571$ & $-563$ & $53$ & $42$ & $-59.4$ & $-58.2$ & Y &  \\
HIP~82042 & 2MA1645+33 & 1314452651925646592 & 1314452651925645952 & $8$ & $27.8$ & $27.7$ & $164$ & $174$ & $-155$ & $-154$ & $29.0$ & $28.8$ & Y & CP \\
HIP~83630 & LEP1705+31~W & 1310147789025317888 & 1310148544939563776 & $24$ & $20.8$ & $20.8$ & $18$ & $17$ & $185$ & $184$ & $-39.1$ & $-80.6$ & N & BM \\
HD~156985 & LP~138-36 & 1416124285987232256 & 1416123117756120960 & $183$ & $54.0$ & $53.9$ & $17$ & $18$ & $-195$ & $-195$ & $-4.8$ & $\cdots$ & Y & CP \\
HD~165401 & 2MA1805+04 & 4471087175915086080 & 4471087175915093632 & $15$ & $41.5$ & $41.4$ & $-14$ & $-18$ & $-319$ & $-316$ & $-118.0$ & $-116.6$ & Y &  \\
HD~171314 & GJ~718~B & 4535144998232287232 & 4535145719786807552 & $49$ & $42.7$ & $42.7$ & $-176$ & $-170$ & $-473$ & $-475$ & $38.3$ & $38.4$ & N & CP \\
HD~175225 & LP~141-7 & 2146204847257954944 & 2146209387041088512 & $207$ & $37.7$ & $37.9$ & $-16$ & $-45$ & $248$ & $274$ & $1.9$ & $1.8$ & Y &  \\
HD~183870 & HD~183870~B & 4187837450005193088 & 4187836934609063552 & $209$ & $56.6$ & $56.6$ & $235$ & $236$ & $18$ & $23$ & $-49.0$ & $-49.0$ & N & AM \\
HD~185414 & 2MA1936+57 & 2142179058450057856 & 2142179058447029760 & $10$ & $41.8$ & $41.8$ & $5$ & $-9$ & $-204$ & $-180$ & $-16.7$ & $-15.9$ & Y &  \\
HD~189733 & 2MA2000+22 & 1827242816201846144 & 1827242816176111360 & $11$ & $50.6$ & $50.6$ & $-3$ & $-12$ & $-250$ & $-254$ & $-2.5$ & $1.5$ & Y &  \\
HD~190360 & GJ~777~B & 2029433521248546304 & 2029432043779954432 & $178$ & $62.5$ & $62.5$ & $683$ & $681$ & $-526$ & $-526$ & $-45.3$ & $-44.0$ & Y &  \\
HD~191785 & GJ~783.2~B & 1809360187275432832 & 1809359465717142656 & $104$ & $48.9$ & $48.9$ & $-415$ & $-417$ & $398$ & $400$ & $-49.4$ & $-50.1$ & Y &  \\
HD~191902 & TIC~14307740 & 6879590383224917120 & 6879589661670401536 & $114$ & $24.6$ & $24.5$ & $168$ & $167$ & $-222$ & $-221$ & $-41.0$ & $-40.8$ & Y & CP \\
HD~192031 & HD~192031~B & 6874580149255288832 & 6874580527212411776 & $10$ & $28.2$ & $28.2$ & $-322$ & $-319$ & $-267$ & $-265$ & $24.3$ & $24.2$ & Y &  \\
HIP~102582 & TIC~383384600 & 1751680734976457728 & 1751680833758839936 & $15$ & $31.5$ & $31.5$ & $90$ & $88$ & $-592$ & $-590$ & $58.6$ & $\cdots$ & N & CP \\
HD~206963 & TIC~190927603 & 1977653223715346688 & 1977653258075087488 & $18$ & $22.0$ & $22.1$ & $-13$ & $-14$ & $24$ & $24$ & $5.3$ & $5.7$ & N & HP \\
HD~207491 & GJ~838.1~B & 2700084315081609344 & 2697080449314615168 & $435$ & $44.3$ & $44.2$ & $542$ & $542$ & $-27$ & $-27$ & $-11.0$ & $-9.1$ & Y & CP \\
HD~211472 & G~232-62 & 2005798350576886144 & 2005804153064109056 & $77$ & $45.6$ & $45.6$ & $213$ & $211$ & $70$ & $67$ & $-7.8$ & $-8.5$ & N & AM \\
HD~217142 & 2MA2256+76 & 2281114756485256704 & 2281128259862423680 & $242$ & $40.7$ & $40.8$ & $34$ & $35$ & $-36$ & $-37$ & $-9.4$ & $-8.9$ & N & AM \\
HD~216803 & LP~876-10~C & 6604147121141267712 & 6623351805412369024 & $26738$ & $131.6$ & $130.3$ & $330$ & $332$ & $-159$ & $-184$ & $6.8$ & $5.8$ & N & AM \\
HD~220140 & LP~12-90 & 2282846074981010688 & 2282790893241302144 & $963$ & $52.8$ & $52.8$ & $203$ & $207$ & $72$ & $65$ & $-17.3$ & $\cdots$ & N & AM \\
\enddata
\tablecomments{The "Used?" column flags the stars that are ultimately included in this work's metallicity calibration. The "Note" column uses the following abbreviations: CP = Cool Primary; HP = Hot Primary; AM = Active M; BM = Binary M; BP = Binary Primary; W = Wrong star observed.}
\end{deluxetable*}
\end{longrotatetable}  \twocolumngrid
}

\subsection{Data reduction}
We reduce our TRES data using a modified version of its standard pipeline \citep[][including flat-fielding, cosmic-ray rejection,
echelle order extraction, and wavelength calibration with ThAr
spectra]{Buchhave2010}, with three changes. For one, we also reduce and subtract the sky fiber, as there are some prominent skylines in our wavelength range of interest. Two, we swap the order of the scattered light and pixel variance corrections to mitigate residual fringing, which is significant in the reddest TRES orders. And three, we propagate uncertainties throughout the reduction process.

Once reduced, we apply a correction to the nominal blaze function based on nightly monitoring of Sirius and Vega; this method is described in \citet{Pass2025} and corrects for changes in the blaze over time, fitting the change in shape relative to an arbitrary reference date. In Section~\ref{sec:ew_calib}, we discuss an overlap sample of M dwarfs that also have high-SNR, telluric-corrected template spectra from CARMENES; to better remove the instrumental response function, we estimate an additional correction to the blaze by fitting a smoothed function to the averaged residuals of this comparison.

Subsequent processing differs between the primaries and secondaries; the method for the primaries follows \citet{Pass2025} and is summarized in the next section. For the M dwarfs, we calculate the radial velocity (RV) of each observation by cross-correlating with a template of Barnard's Star in TRES order 41 (7065--7165\AA), which is dominated by TiO bandhead features. We then shift the data to the star's rest frame and coadd our observations, fitting for the normalization and telluric strength of each individual spectrum to maximize agreement. For computational efficiency, we use \texttt{TelFit} \citep{Gullikson2014} to build a grid of telluric models over our wavelength range of interest (TRES order 49; 8480--8640\AA) instead of applying \texttt{TelFit} directly to each spectrum. We find that this grid can be parameterized by a single `strength' parameter, as the resolution, temperature, and pressure do not change greatly over our data set, and increased airmass and humidity cause similar morphological changes to the telluric spectrum.

The MAROON-X data are reduced using the standard MAROON-X reduction pipeline \citep{Seifahrt2020}. We combine the optimal extraction fluxes from the three object fibers. The sky fiber is inoperational for MAROON-X, preventing us from applying the same sky subtraction method as above; instead, we mask wavelength regions associated with skylines. We follow the same \texttt{TelFit}-based telluric correction procedure but use a grid appropriate to MAROON-X's higher resolution and order boundaries.

\begin{deluxetable*}{@{}LMMMMM@{}}
\renewcommand{\arraystretch}{0.88}
\tablecaption{TRES-\texttt{uberMS} results for primary stars \label{tab:stellar_parameters}}
\tabletypesize{\scriptsize}
\setlength{\tabcolsep}{0pt}

\tablehead{
\colhead{\begin{tabular}{c}Name\\[-0.4ex]\phantom{(K)}\end{tabular}} &
\colhead{\begin{tabular}{c}$T_{\rm eff}$\\[-0.4ex](K)\end{tabular}} &
\colhead{\begin{tabular}{c}$\log g$\\[-0.4ex](dex)\end{tabular}} &
\colhead{\begin{tabular}{c}{[Fe/H]}\\[-0.4ex](dex)\end{tabular}} &
\colhead{\begin{tabular}{c}{[$\alpha$/Fe]}\\[-0.4ex](dex)\end{tabular}} &
\colhead{\begin{tabular}{c}$v\sin i$\\[-0.4ex](kms$^{-1}$)\end{tabular}}
}

\startdata
\badrow HD~129502$^*$ & 6698 & 3.932 & -0.165 & 0.069 & 46.3 \\
\badrow HD~206963 & 6471 & 4.013 & -0.141 & 0.016 & 41.3 \\
\badrow HD~90089$^{*\dag}$ & 6460 & 4.153 & -0.354 & -0.018 & 55.7 \\
HD~18404 & 6429 & 3.985 & -0.008 & 0.011 & 27.0 \\
HD~35681$^*$ & 6370 & 4.298 & 0.082 & 0.007 & 8.6 \\
HD~60912$^*$ & 6314 & 4.063 & -0.112 & 0.051 & 11.6 \\
HD~21962$^*$ & 6213 & 4.056 & 0.034 & 0.034 & 15.2 \\
HD~24421 & 6182 & 4.190 & -0.282 & 0.087 & 4.4 \\
HD~51067$^*$ & 6148 & 4.413 & 0.147 & 0.017 & 6.9 \\
HD~9826 & 6146 & 4.083 & 0.072 & 0.042 & 9.9 \\
HD~44985 & 6077 & 4.421 & -0.013 & 0.054 & 4.2 \\
HD~57625$^*$ & 6038 & 4.433 & 0.073 & 0.044 & 4.2 \\
HD~102574 & 6033 & 3.932 & 0.166 & 0.049 & 8.5 \\
HD~10226 & 5999 & 4.437 & 0.163 & 0.050 & 5.3 \\
HD~77599 & 5995 & 4.473 & 0.060 & 0.042 & 4.8 \\
HD~67458 & 5957 & 4.483 & -0.100 & 0.081 & 4.4 \\
HD~185414 & 5891 & 4.493 & -0.044 & 0.058 & 4.0 \\
HD~16417 & 5866 & 4.156 & 0.155 & 0.065 & 4.4 \\
HD~165401 & 5836 & 4.467 & -0.343 & 0.245 & 4.3 \\
HD~110869 & 5823 & 4.466 & 0.159 & 0.051 & 4.0 \\
HD~111398 & 5720 & 4.235 & 0.097 & 0.093 & 4.0 \\
HD~24409 & 5705 & 4.393 & -0.037 & 0.041 & 3.8 \\
HD~75302$^*$ & 5699 & 4.526 & 0.111 & 0.047 & 4.1 \\
HD~48974$^\dag$ & 5547 & 4.500 & -0.083 & 0.083 & 3.5 \\
HD~16714 & 5543 & 4.481 & -0.122 & 0.104 & 3.5 \\
HD~190360 & 5525 & 4.296 & 0.225 & 0.068 & 3.8 \\
HD~40397 & 5510 & 4.386 & -0.066 & 0.160 & 3.5 \\
HD~96937 & 5388 & 4.469 & 0.134 & 0.050 & 3.5 \\
HD~8828 & 5361 & 4.474 & -0.118 & 0.101 & 3.1 \\
HIP~83630$^{*\dag}$ & 5354 &4.528 &-0.136 &0.084 & 3.1 \\
HD~175225 & 5285 & 3.809 & 0.181 & 0.045 & 4.0 \\
HD~10008$^*$ & 5249 & 4.522 & -0.027 & 0.042 & 3.4 \\
HD~141272 & 5235 & 4.506 & -0.038 & 0.043 & 4.3 \\
HD~211472$^*$ & 5217 & 4.527 & -0.007 & 0.035 & 4.7 \\
HD~75732 & 5205 & 4.403 & 0.332 & 0.024 & 3.1 \\
HD~192031 & 5196 & 4.439 & -0.817 & 0.379 & 2.7 \\
HD~59201 & 5185 & 4.438 & -0.061 & 0.144 & 3.2 \\
HD~144579 & 5161 & 4.386 & -0.658 & 0.352 & 2.2 \\
HD~43147 & 5133 & 4.459 & -0.375 & 0.189 & 2.9 \\
HD~191785 & 5072 & 4.357 & -0.098 & 0.211 & 3.3 \\
HD~125455 & 5034 & 4.452 & -0.151 & 0.105 & 2.9 \\
HD~18143 & 5033 & 4.423 & 0.214 & 0.038 & 3.0 \\
HD~26965$^*$ & 5027 & 4.332 & -0.279 & 0.276 & 2.8 \\
HD~220140$^*$ & 4990 & 4.270 & -0.093 & 0.016 & 14.7 \\
HD~93811$^*$ & 4979 & 4.512 & -0.032 & 0.016 & 4.6 \\
HD~189733 & 4955 & 4.478 & -0.015 & 0.044 & 4.1 \\
HIP~6339 & 4942 & 4.489 & 0.036 & 0.063 & 3.0 \\
HD~183870$^*$ & 4930 & 4.513 & -0.038 & 0.037 & 3.7 \\
HD~217142$^*$ & 4911 & 4.475 & -0.057 & 0.043 & 3.5 \\
HD~127871 & 4885 & 4.442 & -0.047 & 0.066 & 2.9 \\
\badrow HIP~82042 & 4767 & 4.441 & -0.322 & 0.117 & 2.8 \\
\badrow HD~156985 & 4727 & 4.413 & -0.162 & 0.079 & 2.6 \\
\badrow HD~16160 & 4717 & 4.341 & -0.141 & 0.179 & 2.5 \\
\badrow HD~191902 & 4662 & 4.439 & -0.117 & 0.020 & 2.6 \\
\badrow HD~139477$^*$ & 4641 & 4.446 & -0.101 & 0.008 & 3.7 \\
\badrow HD~207491 & 4626 & 4.374 & -0.269 & 0.088 & 2.7 \\
\badrow HD~216803$^*$ & 4588 & 4.420 & -0.118 & 0.015 & 3.7 \\
\badrow HD~6660 & 4578 & 4.355 & 0.056 & 0.002 & 2.3 \\
\badrow HD~129715 & 4543 & 4.303 & 0.090 & 0.025 & 2.2 \\
\badrow HIP~20240$^*$ & 4537 & 4.369 & -0.136 & 0.007 & 3.7 \\
\badrow HIP~102582 & 4529 & 4.365 & -0.524 & 0.330 & 2.1 \\
\badrow HIP~17346 & 4506 & 4.302 & -0.115 & 0.053 & 2.8 \\
\badrow HD~171314 & 4503 & 4.309 & -0.115 & 0.051 & 2.5 \\
\badrow HD~93380 & 4477 & 4.555 & -0.647 & 0.190 & 1.9 \\
\badrow HD~101650 & 4471 & 4.396 & -0.384 & 0.234 & 1.5 \\
\badrow LP~676-69 & 4463 & 4.480 & -0.289 & 0.087 & 1.3 \\
\badrow HIP~76042 & 4434 & 4.514 & -0.408 & 0.147 & 2.6 \\
\badrow HIP~58817 & 4425 & 4.544 & -0.547 & 0.193 & 1.1 \\
\badrow HD~245409$^*$ & 4113 & 4.795 & -0.756 & 0.137 & 2.9 \\
\enddata
\vspace{-0.5em}
\tablecomments{Red rows indicate parameters outside the range where TRES-\texttt{uberMS} is reliable. Primaries with active companions are marked with *\ and binary companions with \dag.}
\end{deluxetable*}

\begin{figure*}\hspace{-0.4cm} %cfa/metallicity_plot.py
    \includegraphics[width=1.0\textwidth]{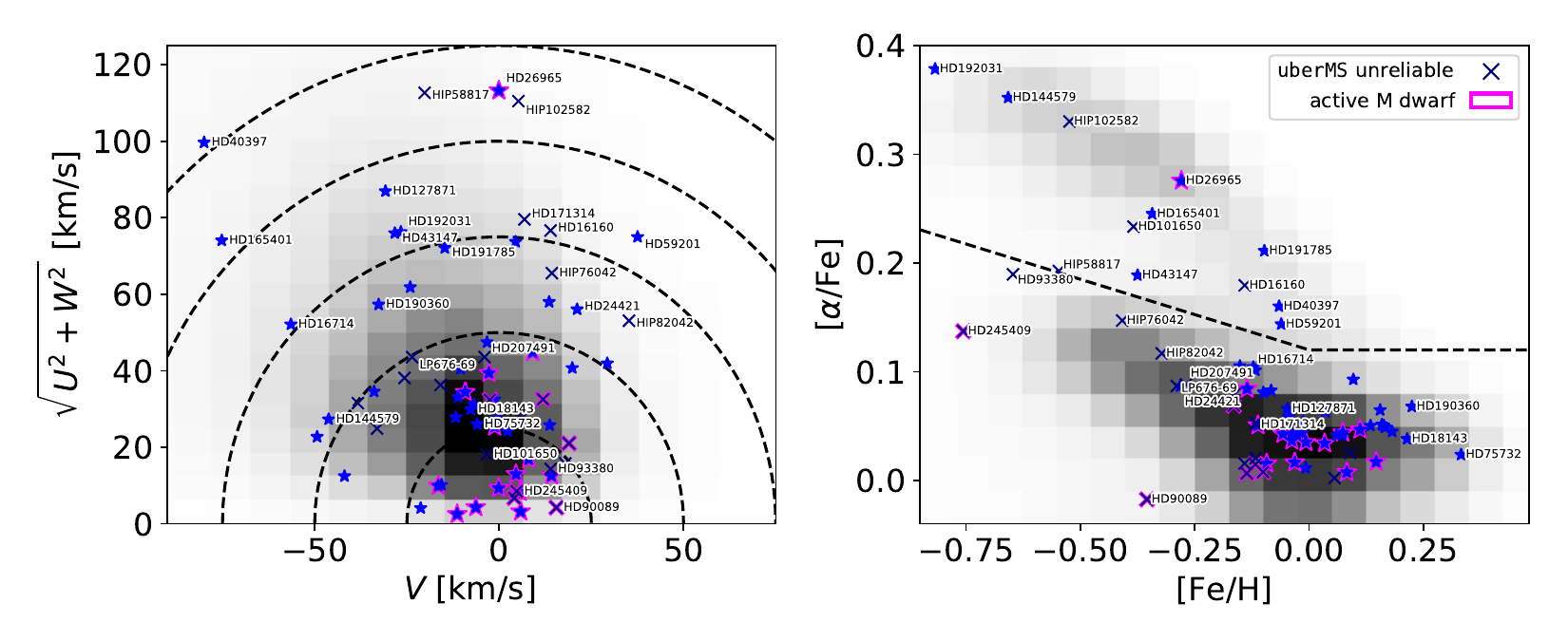}
    \vspace{-0.5cm}
    \caption{The galactic kinematics and chemistries of our FGK--M pairs, with points labeled by their primary. In the left panel, we show a Toomre diagram, with the dashed lines indicating contours of constant velocity. UVW space motions are calculated following \citet{Johnson1987} using the Gaia data from Table~\ref{tab:stellar-pairs} and are relative to the local standard of rest \citep{Schonrich2010}. The right panel shows a Tinsley-Wallerstein diagram, including the dividing line between high- and low-[\textalpha/Fe] from \citet{Weinberg2019}. Our sample is in blue, with Xs denoting stars with potentially erroneous \texttt{uberMS} stellar parameters (highlighted in red in Table~\ref{tab:stellar_parameters}). We use a pink halo to mark stars whose M-dwarf secondary shows H\textalpha\ activity; these stars are generally younger, and so these points typically fall near the origin of the left panel and in the low-[\textalpha/Fe] mode of the right panel. The background heatmaps show the APOGEE sample \citep{Majewski2017} for reference.}
    \label{fig:sample}
    \includegraphics[width=1.\textwidth]{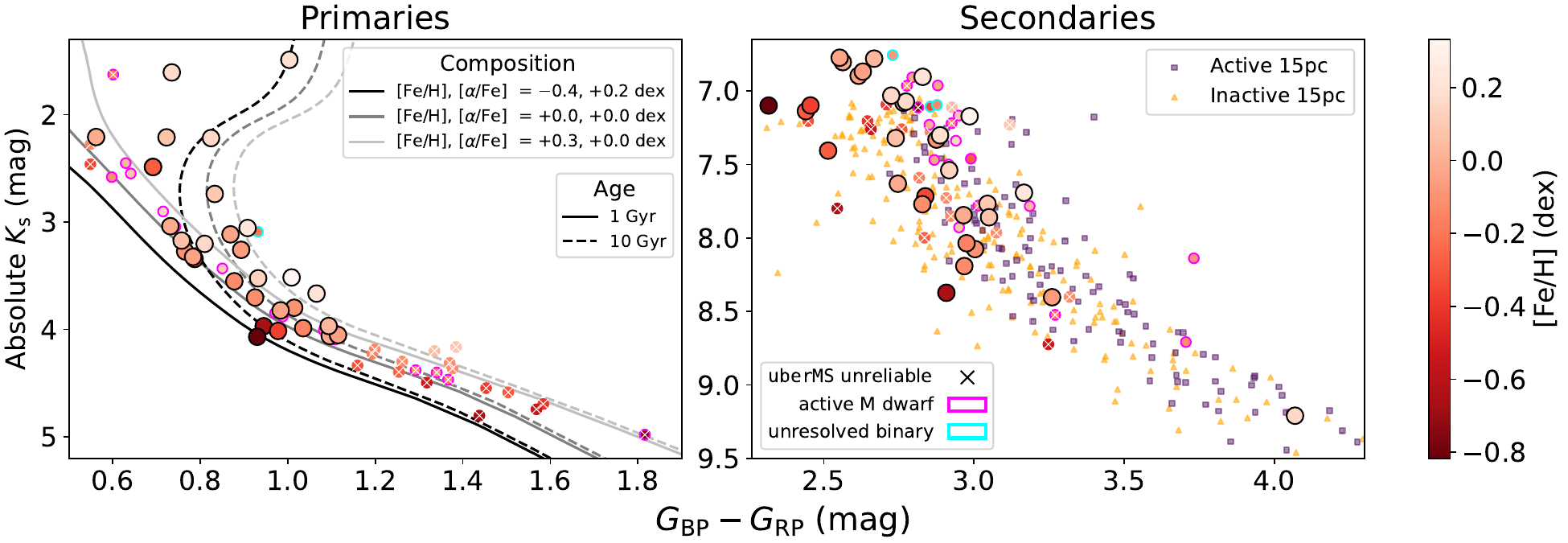}  %cmd_metallicity_paper.py
    \caption{Color-magnitude diagrams for the primaries (left) and secondaries (right). The colorbar indicates the metallicity of the primary from \texttt{uberMS}. Cyan halos indicate components that are unresolved binaries, pink halos indicate systems in which the M dwarf is active, and white crosses indicate primaries with parameters outside the range over which \texttt{uberMS} is reliable. To contextualize the sample, the lefthand plot also includes MIST isochrones \citep{Dotter2026}, while the righthand plot includes the volume-complete sample of single, 0.1--0.3M$_\odot$ M dwarfs within 15pc from \citet{Pass2023_inactive, Pass2023_active}, color-coded by H\textalpha\ activity. A variety of properties can cause a star to fall on the rightmost edge of the population, including evolution (in the case of the primaries), overluminosity due to youth (in the case of the secondaries), unresolved binarity, high metallicity, or high starspot coverage \citep[e.g.,][]{Covey2016}.}
    \label{fig:cmd}
\end{figure*}

\subsection{Stellar properties}

Upon visual inspection of the spectra, we found HD~68638 to be a double-lined spectroscopic binary (SB2) and removed this target from the sample; this result has also been reported elsewhere in the literature \citep{Busa2007, Holl2023, Marcussen2023}. We analyze the remaining primaries using TRES--\texttt{uberMS} \hbox{(\citealt{Pass2025};} P.\ A.\ Cargile et al.\ in prep), an inference code that estimates stellar properties such as effective temperature, surface gravity, $v$sin$i$, metallicity, and alpha enrichment using \texttt{ATLAS-12} model atmospheres \citep{Kurucz2005}, the \texttt{SYNTHE} radiative transfer code \citep{Kurucz1993}, and the \texttt{MIST} stellar isochrone grid \citep{Choi2016}. Specifically, TRES--\texttt{uberMS} uses the Mg b triplet region (5150--5300\AA) of the TRES spectra, as well as photometry from 2MASS \citep{Skrutskie2006} and Gaia DR3 \citep{Gaia2023}. The results of our TRES--\texttt{uberMS} analysis are given in Table~\ref{tab:stellar_parameters}.

We plot the chemistry of the sample in Figure~\ref{fig:sample}, along with kinematics determined from Gaia DR3; both kinematics and chemistry are shared between the members of each pair. The most kinematically and/or chemically interesting targets are labeled in both plots. We remind the reader that while both dynamically hot and chemically \textalpha-enhanced stars are often associated with the Galactic thick disk, neither concept is synonymous with membership of a geometrically defined disk, and neither criterion uniquely predicts the other \citep[e.g.,][]{Hayden2017}.

In Figure~\ref{fig:cmd}, we further contextualize the sample by providing color-magnitude diagrams (CMDs) of the primaries and secondaries. We see that some of the thickness in the CMD is caused by metallicity, with lower metallicity stars having bluer colors at constant luminosity. This is the principle behind photometric determinations of metallicity \citep[e.g.,][]{Dittmann2016}, although such methods generally produce larger uncertainties than spectroscopy-based methods.

\begin{figure*}[t]
    \includegraphics[width=1.0\textwidth]{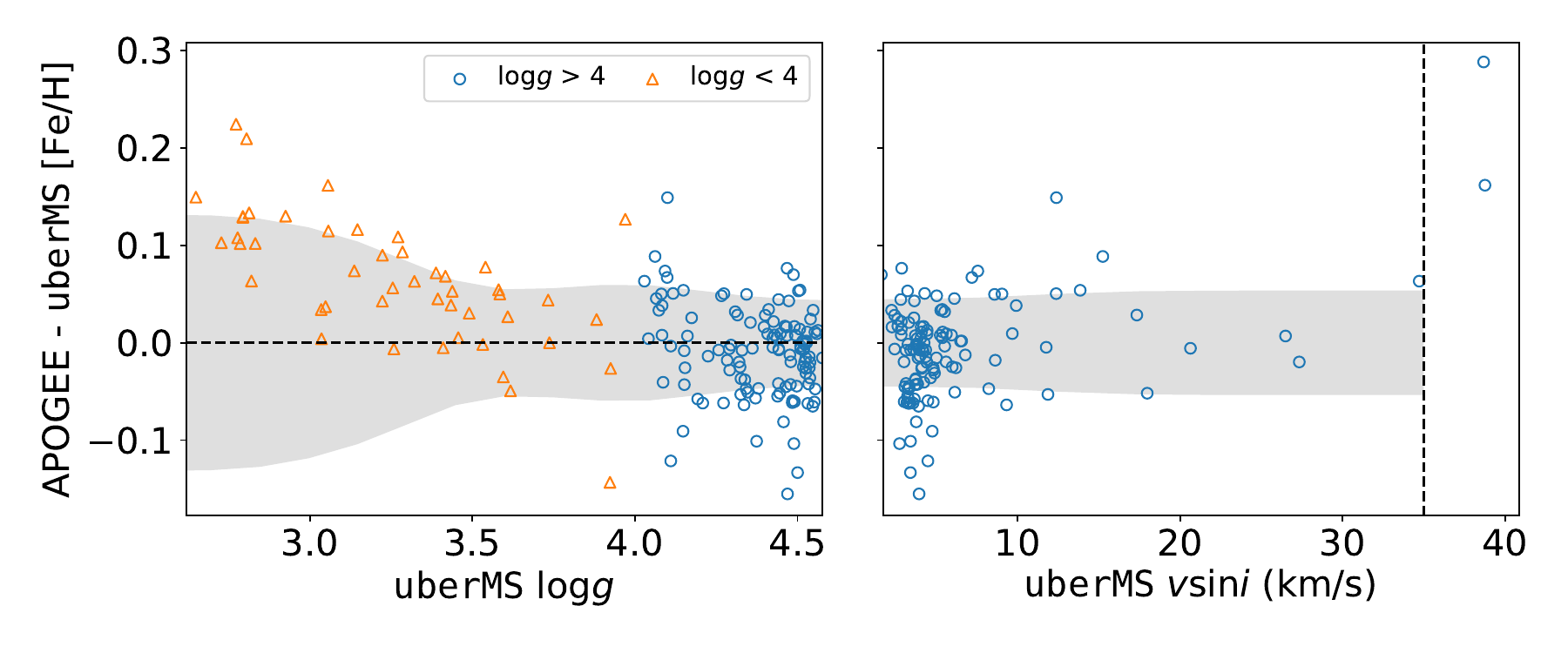}\vspace{-0.6cm}
    \caption{Two additional views of the APOGEE vs TRES--\texttt{uberMS} comparison from \citet{Pass2025}. In the left panel, we show the difference in [Fe/H] as a function of log$g$. The two high $v$sin$i$ targets from the right panel are omitted here. The gray band shows a smoothed, rolling 68\%-ile of the absolute difference in [Fe/H]; i.e., it provides an illustration of the uncertainty as a function of log$g$. By considering bins of log$g$ < 4 and log$g$ > 4~dex, the previous work determined characteristic uncertainties of $\pm$0.04~dex for dwarfs and $\pm$0.10~dex for evolved stars; by probing the log$g$ dependence in greater detail, we see that the errors are only marginally larger at 4.0~dex but worsen for log$g$ < 3.5~dex. The mildly evolved stars in the present work (3.8--4.2~dex) therefore have [Fe/H] uncertainties of roughly $\pm$0.06~dex. In the right figure, we show the same dwarf sample as a function of our measured rotational broadening. High levels of rotational broadening only inflate the [Fe/H] errors for $v$sin$i$ > 35kms$^{-1}$.}
    \label{fig:uberms}
\end{figure*}

As benchmarked in \citet{Pass2025}, TRES--\texttt{uberMS} produces typical uncertainties of $\pm$100~K in $T_{\rm eff}$, $\pm$0.09~dex in $\log g$, $\pm$0.04~dex in [Fe/H], and $\pm$0.03~dex in [\textalpha/Fe] for dwarf stars, although performance is systematically biased for mid-to-late K dwarfs ($T_{\rm eff} < 4800$~K). The performance is also worse for evolved stars, defined in that work as log$g$ < 4.0~dex. As our sample includes a handful of stars that are slightly more evolved than this threshold (log$g$ of 3.8--4.0~dex), in Figure~\ref{fig:uberms} we probe the relationship between log$g$ and [Fe/H] uncertainty in greater detail, using the same APOGEE \citep{Majewski2017} comparison sample as described in \citet{Pass2025}. We find that the uncertainties for stars with log$g$ slightly above and below 4.0~dex are quite similar: that is, marginally worse than the higher log$g$ stars but significantly better than more evolved stars. We therefore adopt [Fe/H] errors of $\pm$0.06~dex for mildly evolved stars with 3.5 < log$g$ < 4.2~dex. In the right panel of this figure, we also show the dependence of [Fe/H] uncertainty on $v$sin$i$; the error is relatively constant except for stars with the highest levels of broadening, motivating the $v$sin$i$ < 35~kms$^{-1}$ threshold adopted in other TRES--\texttt{uberMS} analyses \citep{RodriguezMartinez2026, Ditomasso2026}. This $v$sin$i$ threshold removes the three hottest primaries from our sample, while the 4800~K cut removes the coolest nineteen; these stars are highlighted in red in Table~\ref{tab:stellar_parameters}.

\subsubsection{Newly discovered triples/quadruples}
\label{sec:multiples}
We identify three previously unknown close binaries among our secondaries. The first two are conclusive detections of new companions, with our TRES observations revealing that the M-dwarf components are SB2s. HD 90089 AB are a pair at 23pc, consisting of an F dwarf and an M dwarf separated by 13"; the F dwarf itself also has a close companion \citep{Hartman2026}, making this a quadruple system. HIP 83630 \& 2MASS J17052850+3101533 are a pair at 48pc, consisting of a G dwarf and an M dwarf separated by 24". We follow the method of \citet{Winters2020} to identify and characterize SB2s from TRES spectra, which is based on the \texttt{TODCOR} technique \hbox{\citep{Zucker1994}}. The analysis uses TRES order 41, corresponding to wavelengths 7065--7165\AA, which is dominated by TiO bandhead features for M dwarfs. As in \citet{Winters2020}, we use a spectrum of Barnard's Star as our cross-correlation template.

\begin{deluxetable}{lCCCC}[t]
\tabletypesize{\footnotesize}
\tablecolumns{5}
\tablewidth{0pt}
\tablecaption{\texttt{TODCOR} results for HD 90089 B \& D and 2MASS J17052850+3101533 A \& B \label{tab:TODCOR}}
\tablehead{
 \colhead{\textbf{BJD}} & 
 \colhead{\textbf{RV\bm{$_1$}}} &
 \colhead{\textbf{RV\bm{$_2$}}} &
 \colhead{\vspace{-0.1cm}\bm{$h$}} &  
 \colhead{\bm{$t_{\rm exp}$}} \\[-0.2cm]
 \colhead{[d]} &
 \colhead{[km s$^{-1}$]} &
 \colhead{[km s$^{-1}$]} &
 \colhead{} &
 \colhead{[s]}
}
\startdata
\tightcutinhead{\textbf{HD 90089 B \& D}}
2459265.8442 & -65.060 &  85.624 & 0.77 & 3600 \\
2460460.7087 & -45.429 &  63.788 & 0.84 & 1800 \\
2460461.7379 & -42.239 &  60.204 & 0.82 & 1800 \\
2460463.6706 &  77.517 & -76.678 & 0.89 & 1800 \\
2460464.6790 &  14.396 &  -4.193 & 0.89 & 1800 \\
2460476.6632 &  58.308 & -54.508 & 0.87 & 1980 \\
2460478.6663 & -29.324 &  45.461 & 0.87 & 2040 \\
\tightcutinhead{\textbf{2MASS J17052850+3101533 A \& B}}
2460473.9006 & -95.240 &  18.610 & 0.78 & 3600 \\
2460475.8434 & -91.832 &  15.847 & 0.59 & 1800 \\
2460478.7496 &   4.009 & -81.713 & 0.76 & 3000 \\
2460479.7431 & -79.065 &   2.665 & 0.66 & 1800 \\
2460489.7474 & -71.603 &  -5.656 & 0.51 & 1800 \\
2460491.7216 & -66.059 & -11.257 & 0.45 & 1800 \\
2460494.9430 &  14.492 & -91.267 & 0.65 & 2100
\enddata
\tablecomments{$h$ is the cross-correlation coefficient.}
\end{deluxetable}

We collect seven TRES observations of each of HD 90089 BD and 2MASS J17052850+3101533 AB, with our \texttt{TODCOR}-determined RVs given in Table~\ref{tab:TODCOR}. We measure spectroscopic light ratios of \hbox{$F_{\rm D}/F_{\rm B}=0.75$} for HD 90089 BD and \hbox{$F_{\rm B}/F_{\rm A}=0.94$} for 2MASS J1705+3101 AB. We also obtain $v$sin$i$ estimates of 13kms$^{-1}$, 12kms$^{-1}$, 5.9kms$^{-1}$, and 5.9kms$^{-1}$ for HD 90089 B, HD 90089 D, 2MASS J1705+3101 A, and 2MASS J1705+3101 B, respectively.

\begin{figure*}\centering
    \hspace{-0.3cm}\includegraphics[width=0.5\textwidth]{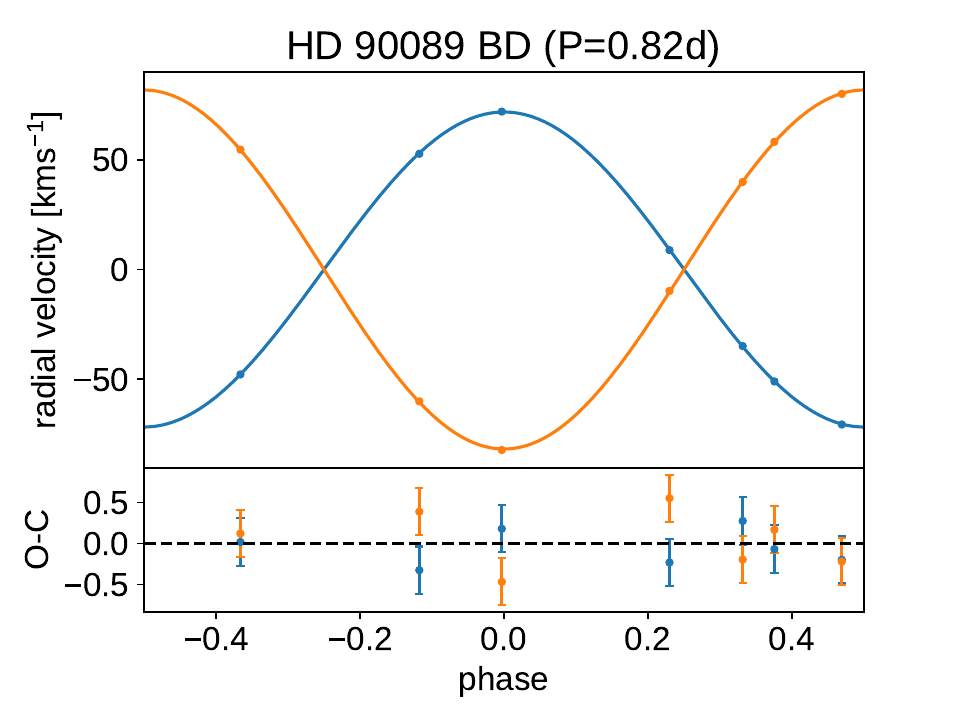}\hspace{-0.5cm}
    \includegraphics[width=0.5\textwidth]{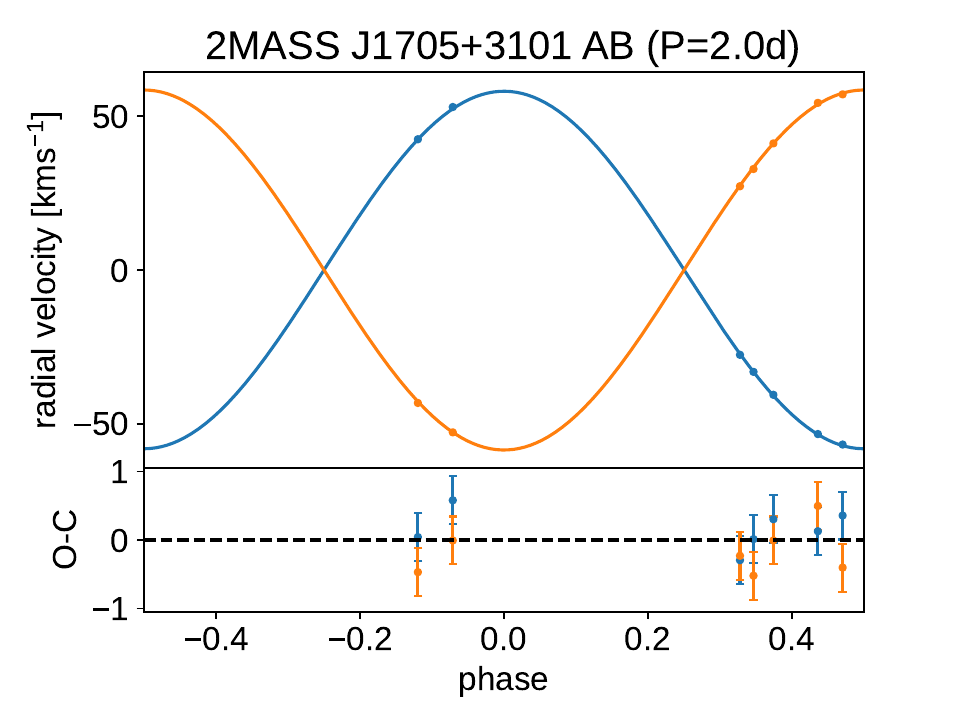}\vspace{-0.3cm}
    \caption{Our MAP orbital solutions for the two newly discovered spectroscopic binaries. HD~90089~BD has a period of 0.82 days and a mass ratio of 0.88, while 2MASS~J1075+3101~AB has a period of 2.0 days and a mass ratio of 0.99. The more massive components are shown in blue.}
    \label{fig:sb2}
\end{figure*}

\begin{deluxetable*}{lllllll}[t]
\vspace{-0.3cm}
\tabletypesize{\footnotesize}
\tablecolumns{7}
\tablewidth{0pt}
 \tablecaption{Orbital fits for HD 90089 BD and  2MASS J1705+3101 AB \label{tab:orbit}}
 \tablehead{
 \colhead{\textbf{Parameter}} & 
 \colhead{\textbf{HD 90089 BD}} &
 \colhead{} &
  \colhead{\textbf{2MASS J1705+3101 AB}} &
 \colhead{} &
 \colhead{\textbf{Unit}}}
\startdata
$P$ & 0.817923 &  $\pm$0.000012 & 2.01199 & $\pm$0.00041 & days\\
$T_{\rm peri}$ & 2459266.234 & $\pm$0.018 & 2460474.9662 & $\pm$0.0037 & BJD\\
$\gamma$ & 5.57 & $\pm$0.50$^{\rm{*}}$ & $-$38.49 & $\pm$0.50$^{*}$ & kms$^{-1}$ \\
$K_{\rm prim}$ & 71.84 & $\pm$0.21 & 58.11 & $\pm$0.39& kms$^{-1}$\\
$K_{\rm sec}$ & 81.88 & $\pm$0.21 & 58.52 &$\pm$0.39 &kms$^{-1}$\\
$q$ & 0.8775& $\pm$0.0030 & 0.9931 & $\pm$0.0065 &--- \\
$\sigma$ & 0.34 & $\pm$0.10 & 0.50 & $\pm$0.15 & kms$^{-1}$
\enddata
\tablecomments{*This error is dominated by the uncertainty in the zero point of our RV scale, resulting from the absolute RV error in the comparison spectrum of Barnard's Star \citep{Winters2026}. Relative RVs are known to much greater precision. \newline}
\end{deluxetable*}
    
To estimate component masses, we assume the TRES light ratio approximates the light ratio in $R$ band. We use the deblending ratio from \citet{Riedel2014} to convert to a $K$--band light ratio, deblend the observed $K$--band magnitude into its components, and apply the \citet{Benedict2016} relation to obtain masses, yielding 0.20M$_\odot$ and 0.18M$_\odot$ for HD 90089 B and D and 0.19M$_\odot$ for both 2MASS J1705+3101 A and B.

We use \texttt{exoplanet} \citep{ForemanMackey2021} to fit an orbital solution to our \texttt{TODCOR} radial velocities (Figure~\ref{fig:sb2}); for each pair, we allow a constant RV uncertainty to be a free parameter of the fit. We identify the maximum a posteriori (MAP) solution with \texttt{exoplanet}, then use \texttt{PyMC3} \hbox{\citep{Salvatier2016}} to sample the posterior, starting from the MAP solution and using two chains each with a 1500 draw burn-in and 2000 draws. The orbital parameters inferred from this analysis are given in Table~\ref{tab:orbit}. We find that both orbits are consistent with circular, with a period of $P=0.81792\pm0.00001$~days for HD 90089 BD and $P=2.0124\pm0.0005$~days for 2MASS J1705+3101 AB.

The final presumed multiple is \hbox{LP 4-248}, which we initially suspected to be a binary due to its high Gaia RUWE of 4.6. This star is quiescent in H\textalpha\ with an equivalent width of $-0.03$\AA, although we measure substantial rotational broadening with a $v$sin$i$ of 6.7kms$^{-1}$. Such rapid rotation would be surprising for an inactive mid-M dwarf \citep{Pass2023_active}, suggesting that this measurement may instead be an artifact of binarity, with two sets of marginally resolved lines mimicking rotational broadening. Our three RV observations taken over a 30-day baseline are mildly inconsistent when analyzed using the mid-to-late M dwarf pipeline from \citet{Pass2023_inactive}, with $P(\chi^2)=0.15$\% and a sample standard deviation of 180ms$^{-1}$. The combination of the Gaia and TRES data therefore suggests that LP~4-248 is an unresolved binary with a multi-year orbital period, such that we see only modest RV evolution over the 30-day observational baseline.

These three systems are noted with daggers in Table~\ref{tab:stellar_parameters} and cyan halos in the right panel of Figure~\ref{fig:cmd}. They have bright absolute magnitudes compared to other M dwarfs of the same color, as expected for binaries, although their binarity cannot be unambiguously identified from photometry alone; there are also single stars (particularly high-metallicity or active systems) that fall at a similar location in the CMD.

\subsubsection{Active stars}
By measuring the equivalent width of the 6563\AA\ H\textalpha\ feature \citep[following][]{Medina2020}, our spectroscopic observations allow us to divide our M-dwarf sample into active and inactive stars. We presented a subset of these data in \citet{Pass2022}, in which ten active M dwarfs in FGK--M pairs were used to help constrain the spindown process of these stars. Our observations since the publication of that work have yielded eight more M dwarfs with H\textalpha\ emission: LP 12-90, 2MASS J14430110$-$0539099, 2MASS J05280096+3345399, LP 16-395, HD 21962 B, 2MASS J04201318$-$0902042, 2MASS J07251770+5634002, and 2MASS J07393210+5031077. The unresolved binaries HD~90089~BD and 2MASS~J1705+3101~AB also show double-lined H\textalpha\ emission.

The primaries of these active M dwarfs are noted with stars in Table~\ref{tab:stellar_parameters}. These systems are also highlighted with pink halos in Figures~\ref{fig:sample} and \ref{fig:cmd}; their positions on the Toomre and Tinsley-Wallerstein diagrams are generally consistent with youth, with the exception of the pair HD~26965/GJ~166~C, which has a chemistry consistent with the Galactic thick disk (age $\sim$ 10~Gyr). We discuss the unusual behavior of this system in \citet{Pass2022}: the pair also has a widely separated white-dwarf primary with a cooling age of only 0.13~Gyr, hinting that the M dwarf's activity may be the result of mass/angular momentum
transfer from the recently transformed white dwarf \citep{Fuhrmann2014}.

\begin{figure} %FGK_calibrators.py
    \hspace{-0.7cm} \includegraphics[width=1.05\columnwidth]{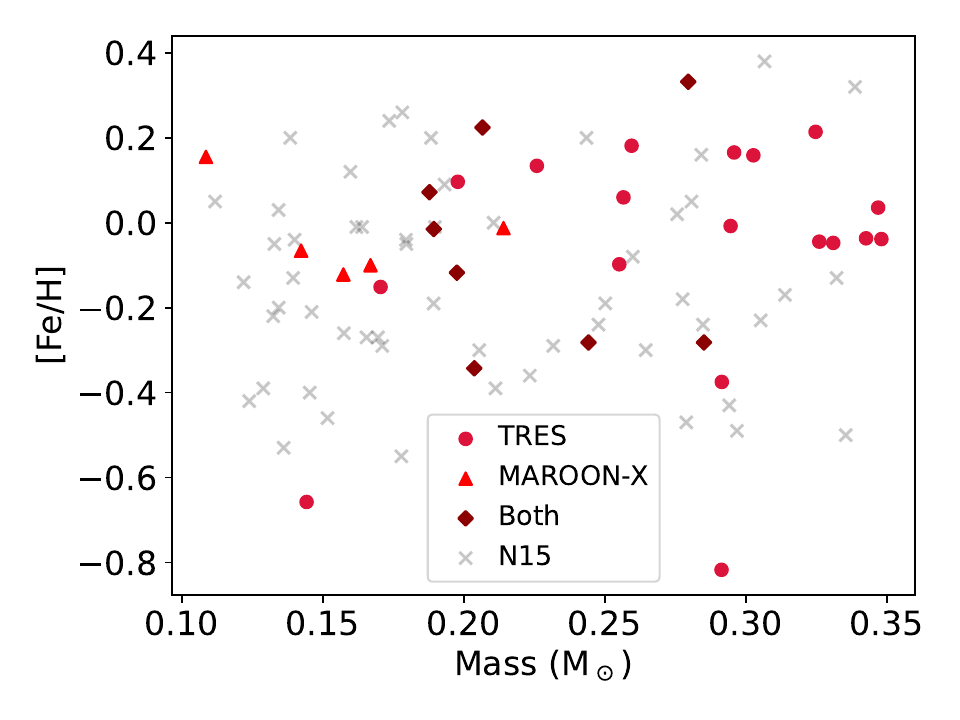}\vspace{-0.4cm}
    \caption{Masses and metallicities of M dwarfs in our prime sample, with the symbol indicating whether the target was observed with TRES, MAROON-X, or both. We also note the stars in the 15pc TRES survey of nearby mid-to-late M dwarfs with [Fe/H] measurements from \citet{Newton2015}, which uses the M+ $K$--band relations.}
    \label{fig:mass-met}
\end{figure}

Neglecting active M dwarfs, unresolved binaries, failed observations, and high-$v$sin$i$ or low-$T_{\rm eff}$ primaries, our sample comprises 31 pairs spanning a metallicity range of $-0.82$ to $+0.33$~dex and an M-dwarf mass range of 0.11 to 0.35M$_\odot$. These targets are shown in Figure~\ref{fig:mass-met}.

\begin{figure} %plotcaii_expanded.py
    \vspace{-0.2cm}\includegraphics[width=0.93\columnwidth]{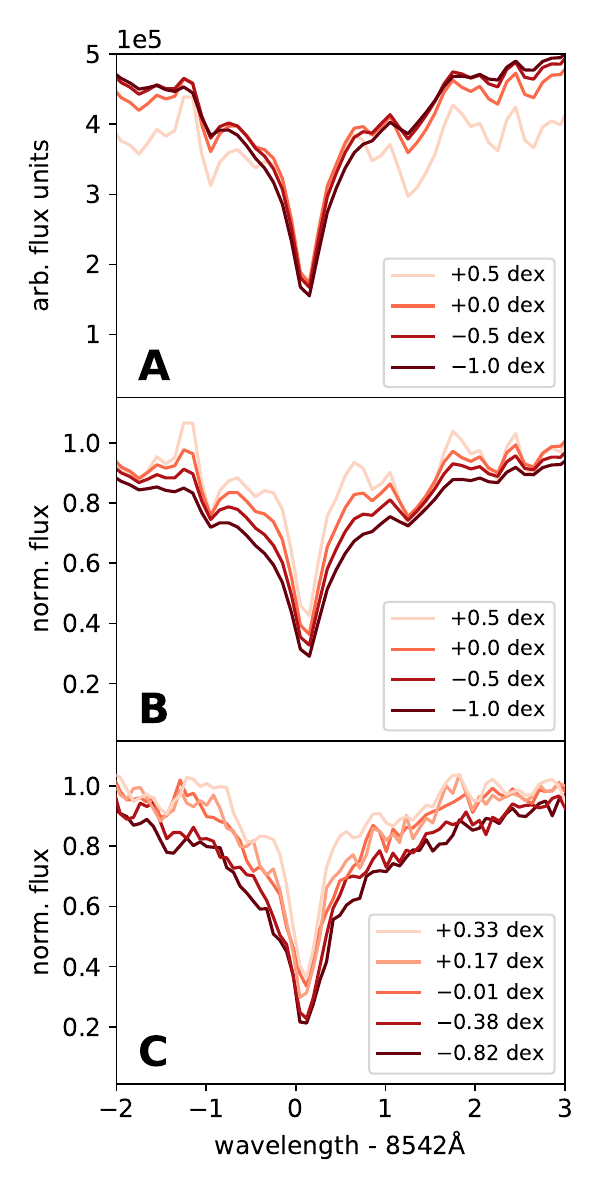}\vspace{-0.5cm}
    \caption{This figure explains the metallicity sensitivity of the CaT feature. \textbf{A}: BT-Settl models \citep{Allard2012} of the 8542\AA\ line for metallicities [Fe/H] = $-1.0$, $-0.5$, 0, and 0.5 dex, convolved to a resolution of 44,000. Temperature and surface gravity are fixed at $T_{\rm eff}=3300$~K and log$g=5.0$~dex. The $y$-axis is in absolute flux units, which is not readily attainable from echelle spectroscopy. As expected, pseudocontinuum features are deeper for enhanced metallicity. \textbf{B}: The same models, now normalized to the echelle order's median flux. The differing depths of the pseudocontinuum combined with the constant strength of the CaT feature generates a metallicity-dependent wing shape; more pronounced wings correlate with lower metallicity, as the weaker wings of the feature can be seen due to the weaker pseudocontinuum. \textbf{C}: TRES observations of GJ~324~B, 2MASS~J11482384$-$1019048, 2MASS~J02580617+2040016, 2MASS~J06165075+4442383, and HD~192031~B, M-dwarf secondaries with similar masses and FGK primaries of known metallicity. These observations display the metallicity-dependent wing variation seen in the models.}
    \label{fig:caii}
\end{figure}

\begin{figure} %comp_flare.py
    \hspace{-0.35cm} \includegraphics[width=1.1\columnwidth]{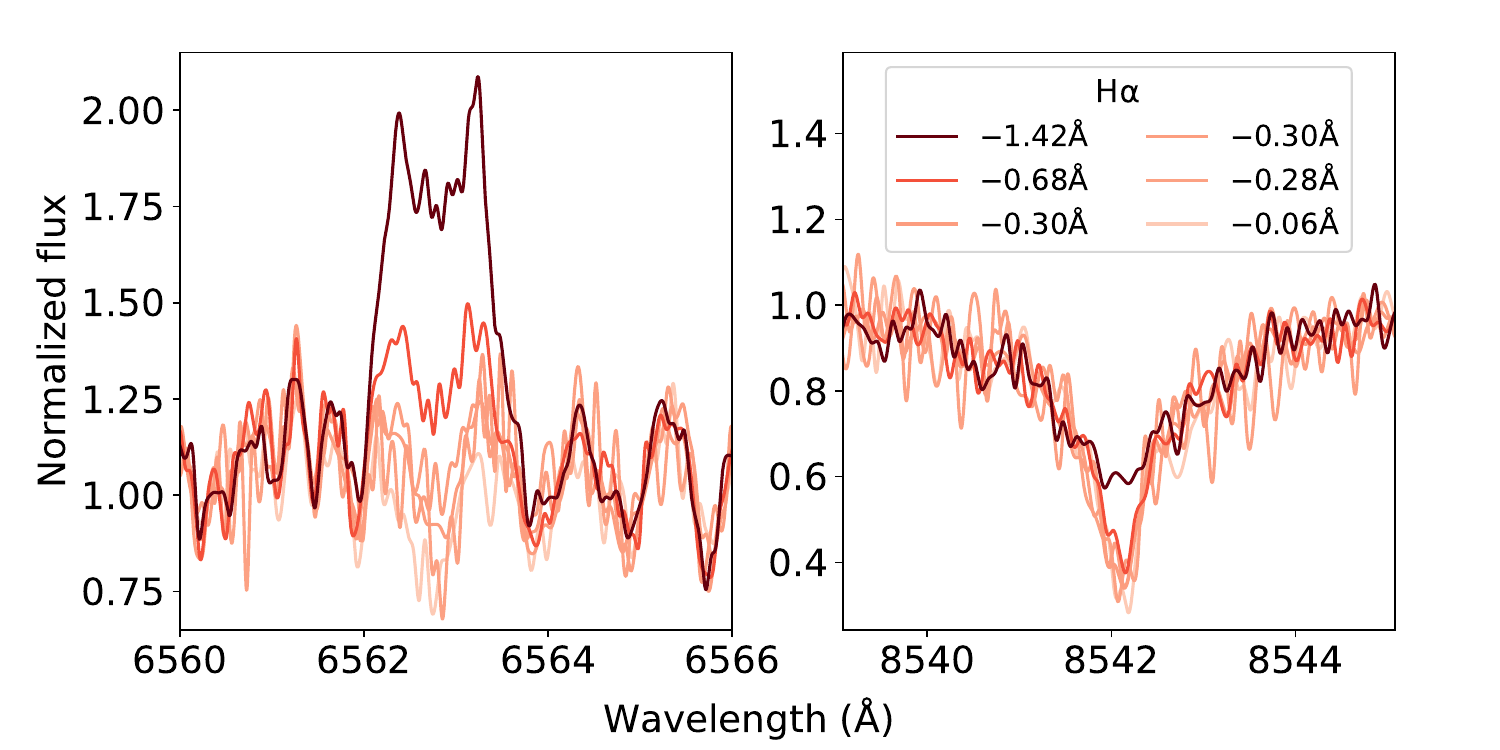}
    \caption{Chromospheric activity causes Ca II line cores to be filled in, as can be seen by comparing our six observations of 2MASS~J19355445+5658586. In the left panel, we show H\textalpha\ and in the right we show the 8542\AA\ Ca II line, with each spectrum color-coded by the equivalent width of the H\textalpha\ feature. During the flare shown in burgundy, H\textalpha\ emission spikes to $-1.42$\AA\ and the 8542\AA\ line core is filled in. For the other five spectra, the 8542\AA\ equivalent width is unchanged within observational uncertainties, although the line core \hbox{does appear to be slightly filled in the H\textalpha\ = $-0.68$\AA\ case.}}
    \label{fig:ha}
\end{figure}

\section{Methods}
\label{sec:calib}
\subsection{Feature selection and properties}
We perform an initial search for metallicity-sensitive features with a by-eye comparison of TRES M-dwarf spectra across the instrument's full wavelength range. We complement this comparison with BT-Settl models \citep{Allard2012} to identify features whose observed variance may potentially be explained by metallicity effects. The Ca II feature at 8542\AA\ emerged as a promising candidate, with significant theoretical variation over the range of metallicities relevant for our stars, as well as significant observed variation (Figure~\ref{fig:caii}). As this strong feature is embedded in the molecular bands of the pseudocontinuum, its wings are quite sensitive to metallicity in the temperature range of interest: for higher metallicity M dwarfs with a stronger pseudocontinuum, less of the wings are visible and the shape and strength of the feature changes.

\begin{figure*} \centering
    \includegraphics[width=0.95\textwidth]{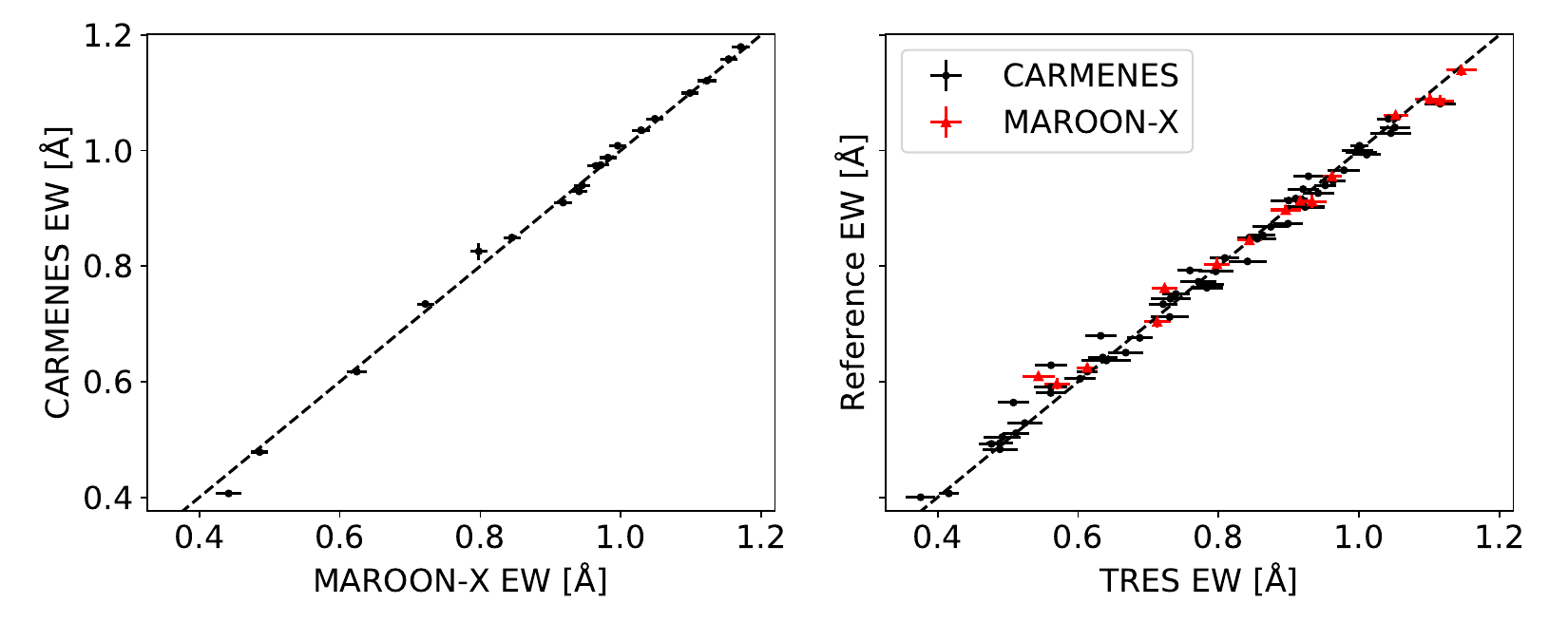}\vspace{-0.4cm}
    \caption{The equivalent width of the 8542\AA\ CaT feature for overlap samples of M dwarfs observed with TRES, CARMENES, and MAROON-X. On the left, we show the comparison between MAROON-X and CARMENES; the \textit{x}-axis error bar has been inflated by the 0.012\AA\ systematic error. On the right, we compare TRES to CARMENES (black circles) and MAROON-X (red triangles), with the TRES uncertainties inflated by a systematic error of 0.012\AA\ in quadrature and a scaling factor of 1.1.}
    \label{fig:ew_calib}
\end{figure*}

In addition to this pseudocontinuum effect, the near-IR Ca II triplet, of which the 8542\AA\ feature is the central line, is known to be triparametric in metallicity, surface gravity, and stellar activity \citep{Mallik1994, Mallik1997}. While CaT strength has been used as a metallicity indicator for stellar populations \citep[e.g.,][]{Sakari2016}, it is better known as an indicator of chromospheric activity \citep[e.g.,][]{Busa2007, Jackson2010, Mittag2017, Martin2017, Fuhrmeister2023}. While the feature's activity sensitivity represents a potential limitation for CaT-based metallicity determination, we note that the MCAL method is similarly activity limited \citep{Neves2014}, and other spectroscopic methods are likely also influenced by activity to some extent, despite these effects not having been quantified in their calibration papers. In Figure~\ref{fig:ha}, we quantify the impact of activity on the present method, comparing six observations of 2MASS~J19355445+5658586. While this star is generally quiescent (H\textalpha\ measurements ranging from $-0.06$\AA\ to $-0.30$\AA), one epoch caught the star during a flare, with H\textalpha\ spiking to EW=$-1.42$\AA\ in emission. During this event, the core of the 8542\AA\ line is observed to fill in, although the wing shape remains unchanged. A second epoch shows a small enhancement in H\textalpha\ emission (EW=$-0.68$\AA); in this observation, the 8542\AA\ line core can be seen to fill in slightly, although the effect is not large enough to change the equivalent width in excess of observational uncertainties. To be conservative, we do not apply our CaT method to stars with H\textalpha\ emission stronger than $-0.5$\AA, and we exclude epochs with enhanced H\textalpha\ emission from our coaddition of each star. This is a slightly more conservative cut than the $-1$\AA\ active/inactive threshold adopted in \citet{Pass2023_inactive, Pass2023_active}, although we note that no M dwarfs in our FGK--M calibration sample have median H\textalpha\ equivalent widths between $-1$\AA\ and $-0.5$\AA; this choice therefore does not influence our sample size.

\subsection{Equivalent width calibration}
\label{sec:ew_calib}
Equivalent widths are challenging to measure in M-dwarf spectra due to the presence of ubiquitous molecular lines. Features cannot be compared to a line-free continuum, removing one of the key advantages of equivalent width---that is, that equivalent width is unaffected by convolution. The blends of features in M-dwarf spectra mean that convolution can change the combination of lines present in the feature or pseudocontinuum windows. These difficulties are further confounded by challenges like differences in spectrograph resolution, instrumental effects like blaze and fringing, and treatment of systematics like tellurics and skylines. Care must be taken to ensure that the equivalent widths that we report are consistent between instruments. This is particularly pertinent given that our calibration sample is split over two instruments, TRES and MAROON-X.

To this end, we consider the telluric-corrected, coadded template spectra from the CARMENES spectrograph \citep{Quirrenbach2014} as published in \citet{Nagel2023}. We build three overlap samples: stars observed by both TRES and MAROON-X, stars observed by both TRES and CARMENES, and stars observed by both CARMENES and MAROON-X, neglecting active stars from all of these samples. In addition to the MAROON-X data of overlapping FGK--M calibrators collected as part of our program, we also consider M-dwarf spectra that were collected as part of MAROON-X's ongoing planet-search work and that were shared by the instrument team for the purposes of this calibration (J.\ Bean, private communication).

We measure the equivalent width using a window with width 4\AA\ centered on 8542.09\AA\ and define the pseudocontinuum as the average value across two neighboring windows: 8530 to 8537\AA\ and 8548 to 8555\AA. In the case of masked skylines in the MAROON-X data, we replace the masked points using linear interpolation and adopt an uncertainty for these points equal to the standard deviation in the pseudocontinuum windows.

The MAROON-X and CARMENES data in these overlap samples are generally of sufficiently high SNR that observational uncertainties have a negligible impact on our observed equivalent widths; the differences between the two measurements reflect either systematics or true astrophysical variation.\footnote{There is one special case where a masked skyline falls near the center of the feature and we only have one MAROON-X observation, making the observational uncertainty non-negligible. This is the lower-left point in the left panel of Figure~\ref{fig:ew_calib}.} As we see in the left panel of Figure~\ref{fig:ew_calib}, our equivalent width measurements agree quite well between the samples, with an RMS error of $\pm$0.012\AA. We therefore add this uncertainty in quadrature to all EW observations.

We next turn to the question of whether our TRES observational uncertainties are appropriately estimated. As TRES is on a smaller telescope, the TRES data generally have lower SNR than their MAROON-X and CARMENES counterparts, making observational uncertainties a more pertinent consideration. For this test, we do not coadd our individual TRES epochs. Instead, we measure a separate equivalent width for each individual spectrum. We consider both our M-dwarf secondaries and the volume-complete 15pc M-dwarf sample in this fashion, again including only inactive stars. We measure the equivalent widths as described above, including the addition of a $\pm$0.012\AA\ systematic error in quadrature. This procedure yields multiple EW measurements per star, with a mode of four and typically spanning years. In each case, we calculate the error scaling factor required to achieve a reduced chi-squared of $\chi^2_{\rm{red}}=1$. We average these scaling factors across our entire sample, finding a median scaling factor of 1.1. We therefore inflate our TRES EW uncertainties by a factor of 1.1 in our subsequent analysis.

In the right panel of Figure~\ref{fig:ew_calib}, we compare TRES to MAROON-X and CARMENES. Our nominally measured TRES equivalent width uncertainties have been inflated by the two factors described above, a 0.012\AA\ systematic error added in quadrature and an error scaling of 1.1. With these two factors applied, we find the expected level of scatter between the TRES and MAROON-X/CARMENES observations, with $\chi^2_{\rm{red}}=1.0$. Our modified errors thus appear to appropriately reflect the uncertainty in 8542\AA\ equivalent width.

\subsection{Preliminary calibration using literature sample}
Before we gathered our FGK--M sample, we first investigated a preliminary calibration of the CaT feature using literature M-dwarf metallicities. While this literature-calibrated relation is not the focus of this paper, it provides useful insight into the limitations of literature methods in the mid-to-late M-dwarf regime and is therefore discussed in this section. Our finalized relation using the FGK--M sample is the topic of Section~\ref{sec:final}.

For the preliminary calibration, we consider the sample of overlapping stars between the 15pc TRES survey and other works that estimate metallicities for large numbers of M dwarfs. One of the largest overlaps is with \citet{Newton2015}: 45 inactive, single stars belong to both samples. We increase this number to 55 by considering similar stars that were observed by TRES but ultimately rejected from the 15pc sample as a result of being slightly too massive or distant (see Table 4 of \citealt{Winters2021}). These 55 overlap targets are noted by crosses in Figure~\ref{fig:mass-met} and span a range of $-0.55$ to $+0.38$~dex in [Fe/H] and 0.11 to 0.33~M$_\odot$ in mass.

We observe a clear trend when plotting the 8542\AA\ equivalent width ($W$) versus literature metallicity, with significant dispersion that correlates with mass ($M$). This motivates a two-parameter fit using a function of the form [Fe/H$] = A + B\times M + C \times W$ and the \texttt{scipy} \texttt{curve\_fit} routine, yielding a best-fit solution of [Fe/H] $= 0.324+ 2.413M -1.211W$. Adding second-order terms does not lead to any notable improvement in the fit.

The 15pc TRES sample also shares some overlapping stars with other works, although generally in more modest numbers than with \citet[][N15 for short]{Newton2015}: 22 with \citet[][R12]{RojasAyala2012}; 35 with \citet[][G14]{Gaidos2014_conch}; 21 with \citet[][N14]{Neves2014}; 25 with \citet[][M15]{Mann2015}; 50 with \citet[][T15]{Terrien2015} 16 with \citet[][F18]{Fouque2018}. These relations vary with respect to their quoted uncertainties, ranging between $\pm$0.08 and $\pm$0.14~dex. N14 and F18 use the optical MCAL method, R12 use an original $K$--band relation based on Ca I and Na I lines, M15 and T15 use the M+ IR relations, and G14 use the M+ optical relation.

\begin{figure}%mit/compnew-firstplot.py
    \includegraphics[width=0.95\columnwidth]{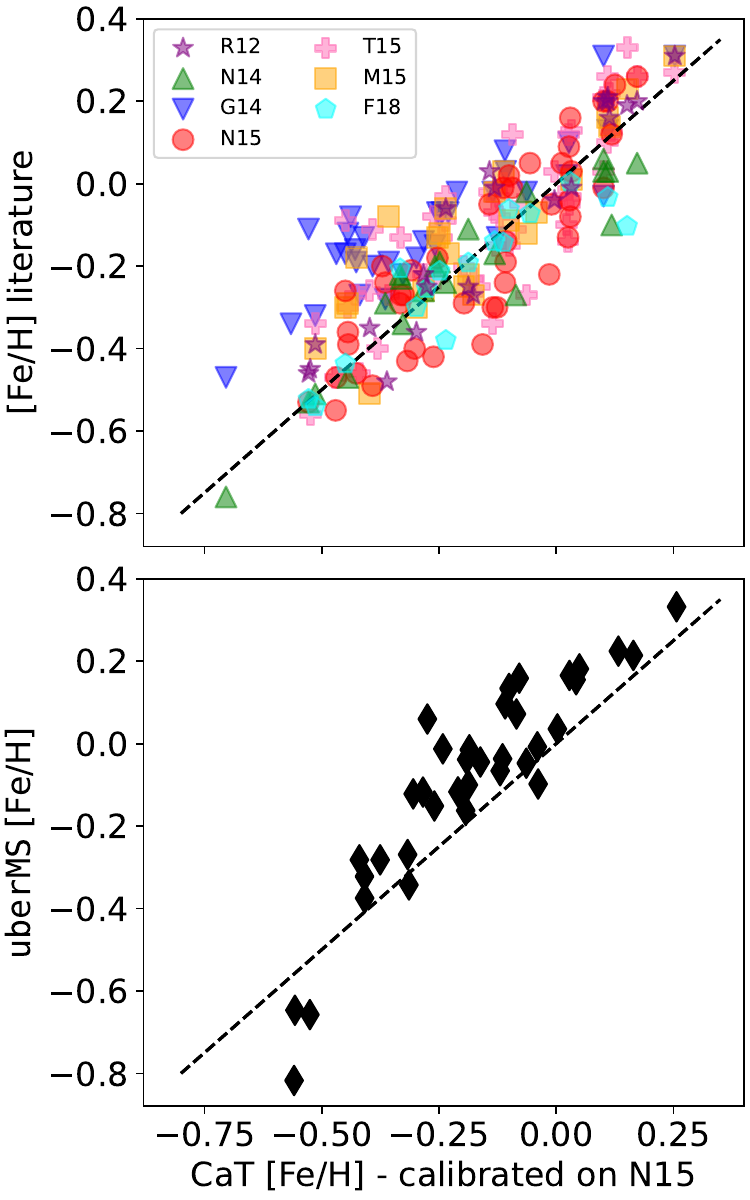}
    \caption{While a relation calibrated using literature metallicities from \citet[][N15]{Newton2015} achieves good agreement with literature measurements---partially by construction---it performs poorly when compared to our gold-standard FGK--M binary sample, which has the most reliable metallicities. The upper plot compares the metallicities we obtain from the N15-calibrated relation to overlap samples with other literature relations: \citet[][R12]{RojasAyala2012}; \citet[][G14]{Gaidos2014_conch}; \citet[][N14]{Neves2014}; \citet[][T15]{Terrien2015}; \citet[][M15]{Mann2015}; \citet[][F18]{Fouque2018}. The lower panel shows the relation applied to the M dwarfs in our FGK--M benchmark sample, with the $y$-axis metallicities determined from our \texttt{uberMS} analysis of their primaries.} 
    \label{fig:n15_calib}
\end{figure}

For stars that have been observed in multiple overlap samples, the literature metallicities can vary substantially. For example, the most discrepant of the points in the overlap sample with \citet{Fouque2018} is GJ~905, with an F18 metallicity of $-0.10$~dex. Others of the overlap samples place it at $+0.33$ (T15), $+0.19$ (R12), and $+0.23$ (M15). The latter three are in much closer agreement with $+0.15$, the metallicity returned by the N15-calibrated relation. By nature of the dispersion in literature estimates, this star will necessarily appear as an outlier in at least one of the overlap samples, regardless of the metallicity returned by our relation. Similarly, the strongest outlier in the comparison with \citet{Neves2014} is GJ~1057. That work reports [Fe/H]$=-0.10$~dex, while \citet{Newton2015} report $+0.12$~dex and \citet{Terrien2015} report $+0.13$~dex. The relation returns $+0.12$~dex, in line with the latter two. We also note that there can be significant discrepancies among the literature, even among works that nominally use the same method: for GJ~232, N15 report $-0.46$~dex while M15 report $-0.18$~dex. Perhaps unsurprisingly, the N15-calibrated relation returns $-0.43$~dex, similar to N15.

The upper panel of Figure~\ref{fig:n15_calib} compares the literature metallicities in these overlap samples to the metallicities we estimate by applying the N15-calibrated relation to the TRES observations. The overlap samples are generally in reasonable agreement with the TRES results considering the uncertainties reported in the original works and assuming a $\pm$0.07~dex error in the N15-calibrated relation, with the exception of G14 (as shown in Figure~\ref{fig:g14}, the metallicities from that work are strongly discrepant with N15, and thus a disagreement with the N15-calibrated relation is unsurprising).

However, the lower panel of Figure~\ref{fig:n15_calib} paints a different picture. In this plot, we compare the N15-calibrated relation with our FGK--M binary sample. For all but the lowest-metallicity stars, the N15-calibrated metallicities are systematically too low, with a typical underestimation of roughly 0.15~dex at solar metallicity. The discrepancy is unlikely to be caused by systematic offsets in the metallicities of the FGK primaries: the primary-star metallicities in M+ were calibrated to the SPOCS \citep{Valenti2005} metallicity scale, and \citet{Pass2025} found SPOCS and TRES-\texttt{uberMS} produce consistent metallicities with an RMS error of $\pm$0.07~dex and no evidence of the systematic offset observed here. Inspection of the specific primaries that overlap between this work and \citet{Valenti2005} similarly shows good agreement between the estimates and no evidence of systematic bias.

This result suggests that the limitations in literature calibrations for mid-to-late M dwarfs may run deeper than just the discrepancy highlighted in Figure~\ref{fig:g14}. To avoid these issues, we should calibrate our relation directly from our sample of FGK--M wide binaries, rather than relying on literature samples that are a step removed from their original FGK--M calibrators (and moreover, we should not supplement our FGK--M sample using single stars with literature estimates). This calibration is the focus of the next section.

\begin{figure*}%mit/compare-new.py
    \includegraphics[width=1.25\columnwidth]{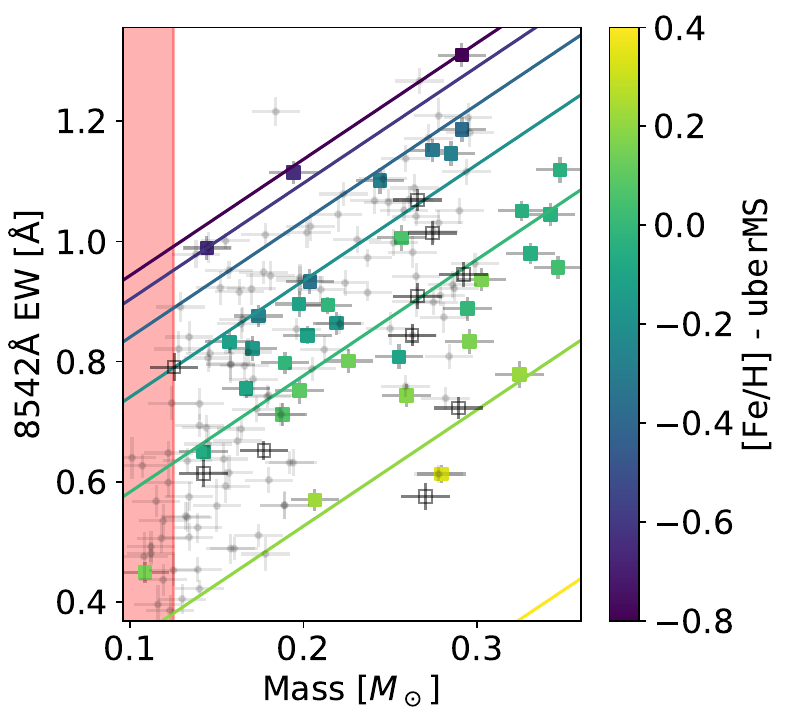}
    \includegraphics[width=0.75\columnwidth]{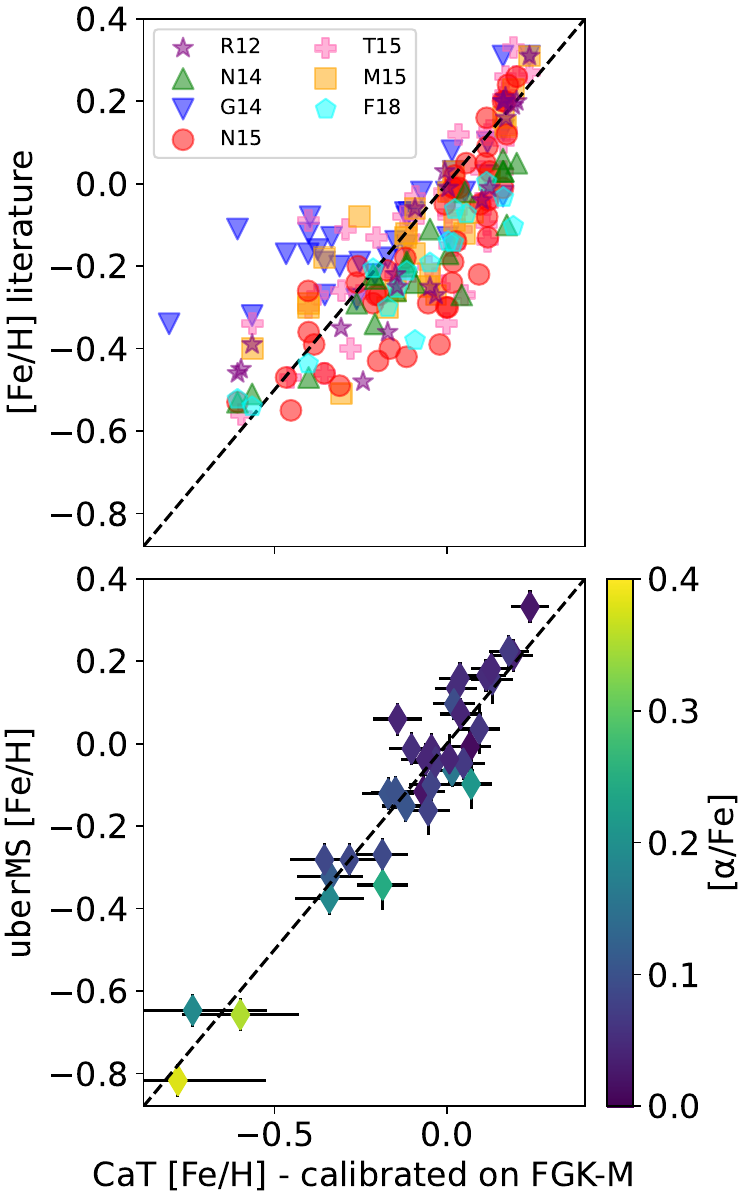}\vspace{-0.15cm}
    \caption{We calibrate our CaT relation using a sample of 36 fully convective M dwarfs with FGK primaries. This sample and calibration are shown in the lefthand plot, with a best-fit function of [Fe/H] $= \log_{10}(1.912+ 4.516M -2.337W$). The 36 calibrators are shown as colored squares; open squares indicate stars in the low-$T_{\rm eff}$ sample that are omitted from the calibration because we lack reliable primary-star metallicities. Translucent dots show the inactive, single stars in the 15pc TRES sample of mid-to-late M dwarfs for which we aim to estimate [Fe/H]. These stars generally fall within the range informed by our calibration, with the exception of LHS~1481 (upper outlier). This star appears to be more metal poor than any of our calibrators. We also note a lack of subsolar metallicities among the 15pc sample for stars below 0.125M$_\odot$; as such a distribution is statistically unlikely and we lack any low-metallicity calibrators in this portion of mass space, we consider our relation to be potentially unreliable in this regime (red shaded region). The upper righthand plot compares the metallicities we obtain from the FGK--M-calibrated relation to overlap samples with the literature: the agreement is necessarily poorer than in Figure~\ref{fig:n15_calib} to resolve the systematic bias in the FGK--M comparison, which is shown in the lower right panel. The colorbar in this figure indicates alpha enrichment; we see that at intermediate metallicities, some of the strongest outliers are alpha enriched.} 
    \label{fig:fgkm_feh_calib}
\end{figure*}

\begin{table*}[]
\centering
\caption{Metallicity, alpha enrichment, and 8542\AA\ equivalent width for the 31 M dwarfs in our prime FGK--M binary sample (upper portion) and 5 M dwarfs with reliable metallicities from the subsample with low-$T_{\rm eff}$ primaries (lower portion)}
\label{tab:ew}\vspace{-0.3cm}
\footnotesize
\begin{tabular}{llMMMMM}
\textbf{M Dwarf} & \textbf{Primary} & \textbf{Mass (M$_\odot$)} & \textbf{[Fe/H] (dex)} & \textbf{[\textalpha/Fe] (dex)} & \textbf{TRES (\AA)} & \textbf{MAROON-X  (\AA)} \\
\hline
HD~192031~B & HD~192031 & 0.291  \pm 0.014 & -0.82  \pm 0.04 & 0.38  \pm 0.03 & 1.310  \pm 0.020 & ... \\
GJ~611~B & HD~144579 & 0.144  \pm 0.014 & -0.66  \pm 0.04 & 0.35  \pm 0.03 & 0.989  \pm 0.020 & ... \\
2MA0616+44 & HD~43147 & 0.291  \pm 0.014 & -0.37  \pm 0.04 & 0.19  \pm 0.03 & 1.186  \pm 0.021 & ... \\
2MA1805+04 & HD~165401 & 0.204  \pm 0.014 & -0.34  \pm 0.04 & 0.25  \pm 0.03 & 0.933  \pm 0.021 & 0.912  \pm 0.016 \\
LEP0355+52 & HD~24421 & 0.244  \pm 0.014 & -0.28  \pm 0.06 & 0.09  \pm 0.03 & 1.101  \pm 0.022 & 1.089  \pm 0.016 \\
HD~24421~B & HD~24421 & 0.285  \pm 0.014 & -0.28  \pm 0.06 & 0.09  \pm 0.03 & 1.146  \pm 0.021 & 1.139  \pm 0.015 \\
GJ~544~B & HD~125455 & 0.170  \pm 0.014 & -0.15  \pm 0.04 & 0.11  \pm 0.03 & 0.822  \pm 0.021 & ... \\
TIC~122574948 & HD~16714 & 0.157  \pm 0.014 & -0.12  \pm 0.04 & 0.10  \pm 0.03 & ... & 0.833  \pm 0.019 \\
LP~587-63 & HD~8828 & 0.198  \pm 0.014 & -0.12  \pm 0.04 & 0.10  \pm 0.03 & 0.896  \pm 0.022 & 0.897  \pm 0.015 \\
TIC~154234106 & HD~67458 & 0.167  \pm 0.014 & -0.10  \pm 0.04 & 0.08  \pm 0.03 & ... & 0.755  \pm 0.015 \\
GJ~783.2~B & HD~191785 & 0.255  \pm 0.014 & -0.10  \pm 0.04 & 0.21  \pm 0.03 & 0.808  \pm 0.021 & ... \\
TIC~39543069 & HD~40397 & 0.142  \pm 0.014 & -0.07  \pm 0.04 & 0.16  \pm 0.03 & ... & 0.650  \pm 0.015 \\
G~66-6 & HD~127871 & 0.331  \pm 0.014 & -0.05  \pm 0.04 & 0.07  \pm 0.03 & 0.980  \pm 0.018 & ... \\
2MA1936+57 & HD~185414 & 0.326  \pm 0.014 & -0.04  \pm 0.04 & 0.06  \pm 0.03 & 1.051  \pm 0.017 & ... \\
HD~141272~B & HD~141272 & 0.348  \pm 0.014 & -0.04  \pm 0.04 & 0.04  \pm 0.03 & 1.119  \pm 0.022 & ... \\
HD~24409~C & HD~24409 & 0.343  \pm 0.014 & -0.04  \pm 0.04 & 0.04  \pm 0.03 & 1.045  \pm 0.019 & ... \\
2MA2000+22 & HD~189733 & 0.189  \pm 0.014 & -0.01  \pm 0.04 & 0.04  \pm 0.03 & 0.798  \pm 0.018 & 0.803  \pm 0.017 \\
TIC~206680308 & HD~44985 & 0.214  \pm 0.014 & -0.01  \pm 0.04 & 0.01  \pm 0.03 & ... & 0.894  \pm 0.014 \\
2MA0258+20 & HD~18404 & 0.295  \pm 0.014 & -0.01  \pm 0.06 & 0.05  \pm 0.03 & 0.888  \pm 0.020 & ... \\
BD+30206~B & HIP~6339 & 0.347  \pm 0.014 & 0.04  \pm 0.04 & 0.06  \pm 0.03 & 0.957  \pm 0.019 & ... \\
NLTT~20915 & HD~77599 & 0.257  \pm 0.014 & 0.06  \pm 0.04 & 0.04  \pm 0.03 & 1.006  \pm 0.022 & ... \\
GJ~61B & HD~9826 & 0.188  \pm 0.014 & 0.07  \pm 0.06 & 0.04  \pm 0.03 & 0.712  \pm 0.019 & 0.704  \pm 0.016 \\
2MA1248+12 & HD~111398 & 0.198  \pm 0.014 & 0.10  \pm 0.04 & 0.09  \pm 0.03 & 0.752  \pm 0.022 & ... \\
2MA1109+02 & HD~96937 & 0.226  \pm 0.014 & 0.13  \pm 0.04 & 0.05  \pm 0.03 & 0.801  \pm 0.020 & ... \\
TIC~122555697 & HD~16417 & 0.109  \pm 0.014 & 0.16  \pm 0.06 & 0.05  \pm 0.03 & ... & 0.449  \pm 0.018 \\
LP~95-334 & HD~110869 & 0.303  \pm 0.014 & 0.16  \pm 0.04 & 0.06  \pm 0.03 & 0.936  \pm 0.019 & ... \\
2MA1148-10 & HD~102574 & 0.296  \pm 0.014 & 0.17  \pm 0.06 & 0.05  \pm 0.03 & 0.834  \pm 0.021 & ... \\
LP~141-7 & HD~175225 & 0.259  \pm 0.014 & 0.18  \pm 0.06 & 0.04  \pm 0.03 & 0.744  \pm 0.019 & ... \\
GJ~118.2~C & HD~18143 & 0.325  \pm 0.014 & 0.21  \pm 0.04 & 0.04  \pm 0.03 & 0.779  \pm 0.022 & ... \\
GJ~777~B & HD~190360 & 0.207  \pm 0.014 & 0.22  \pm 0.04 & 0.07  \pm 0.03 & 0.570  \pm 0.018 & 0.596  \pm 0.015 \\
GJ~324~B & HD~75732 & 0.280  \pm 0.014 & 0.33  \pm 0.04 & 0.02  \pm 0.03 & 0.613  \pm 0.015 & 0.624  \pm 0.014 \\
\hline
HD~93380~B	& HD~93380	& 0.194\pm0.014	& -0.65\pm0.04	& 0.19\pm0.03	& 1.115\pm0.019	&... \\
2MA1645+33	& HIP~82042	& 0.274\pm0.014	& -0.32\pm0.04	& 0.12\pm0.03	& 1.152\pm0.019	& ... \\
GJ~838.1~B & HD~207491& 0.174\pm0.014 &-0.27\pm0.04 & 0.09\pm0.03	& 0.876\pm0.019	& ... \\
LP~138-36	& HD~156985	& 0.219\pm0.014	& -0.16\pm0.04	& 0.08\pm0.03	& 0.864\pm0.019	& ... \\
TIC~14307740	& HD~191902	& 0.202\pm0.014	& -0.12\pm0.04	& 0.02\pm0.03	& ... & 0.844\pm0.015
\\ \hline
\end{tabular}
\end{table*}

\begin{figure*}%mit/massmet_alpha.py
    \includegraphics[width=1.1\columnwidth,trim={1cm, 0.5cm, 0, 0}, clip=true]{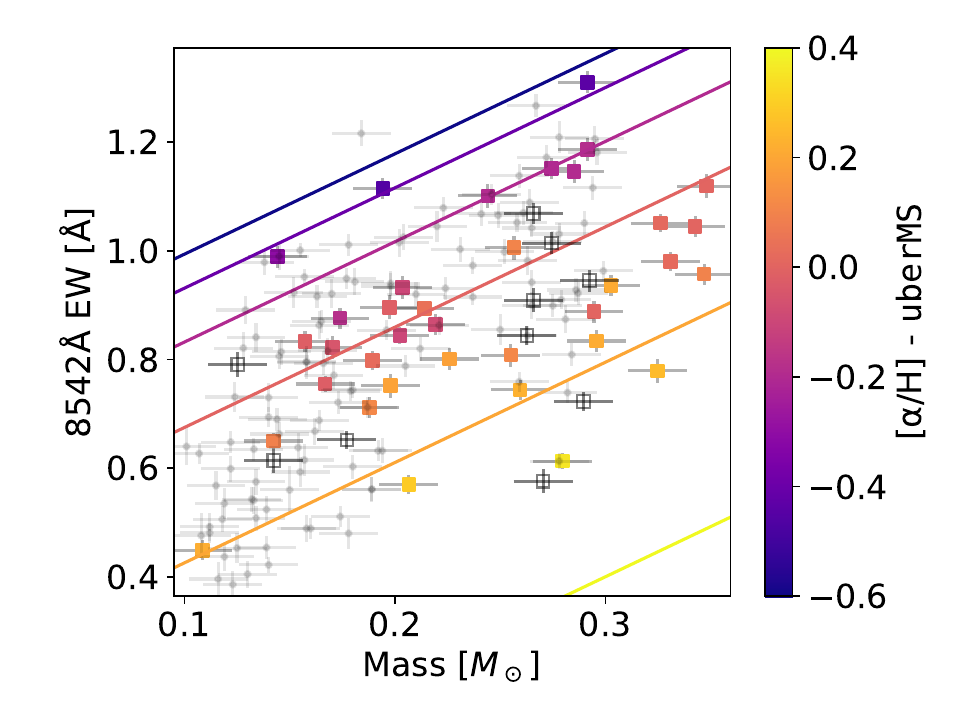}
    \includegraphics[width=0.8\columnwidth, trim={0, 0, 3.5cm, 0}]{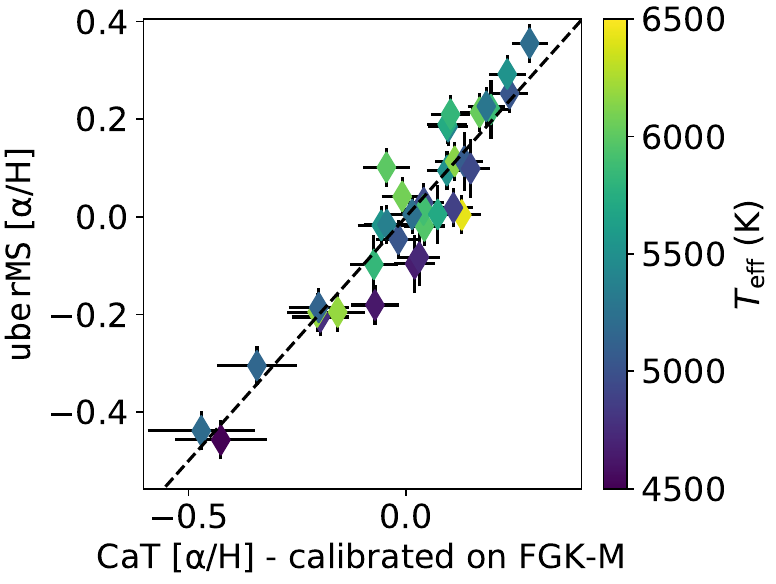}\vspace{-0.15cm}
    \caption{The lefthand plot is the same as in Figure~\ref{fig:fgkm_feh_calib}, but now using [\textalpha/H] instead of [Fe/H]. The best-fit function is [\textalpha/H] $= \log_{10}(2.151+ 4.335M -2.348W$). The righthand plot is the same as the lower righthand plot of Figure~\ref{fig:fgkm_feh_calib}, with the colorbar now showing primary temperature. The strongest outliers tend to have the most extreme primary temperatures.} 
    \label{fig:fgkm_ah_calib}
\end{figure*}

\subsection{Final calibration using FGK--M sample}
\label{sec:final}
We repeat the analysis from the previous section, this time using the FGK--M binary sample to calibrate the relation. We initially consider the same [Fe/H$] = A + B\times M + C \times W$ functional form as before but find that the FGK--M sample favors a relation in which the contours get closer together at low metallicity, motivating a change to [Fe/H$] = \log_{10}(A + B\times M + C \times W)$. This form produces smaller RMS residuals than one with additional polynomial terms, despite containing the same number of free parameters as the previous relation.

To begin, we consider only the 31 M dwarfs in our prime FGK--M sample, as shown in Figure~\ref{fig:mass-met} and the upper portion of Table~\ref{tab:ew}. This prime sample excludes close binaries, active M dwarfs, and primaries with unreliable metallicities due to rapid rotation or low temperature. However, our adopted low-temperature cut is potentially too severe: it is the temperature at which \texttt{uberMS} was found to return biased metallicities for stars in the metal-rich Hyades cluster in \citet{Pass2025}. That work notes that the threshold temperature is likely metallicity dependent: both increasing metallicity and decreasing temperature increase the complexity of the stellar spectrum, and when the spectrum becomes sufficiently complex, stellar models perform poorly. One might therefore expect the low-temperature threshold to be somewhat lower for stars with lower metallicity. However, it is not immediately obvious which of our low-$T_{\rm eff}$ primaries are truly low metallicity versus which have been affected by systematic bias.

To address this question, we take an iterative approach: we first calibrate our relation using only the 31 stars in the prime sample and use this relation to estimate the metallicities of the stars in the low-$T_{\rm eff}$ sample. We then compare the estimated and \texttt{uberMS} metallicities and look for trends as a function of stellar temperature and our CaT-estimated metallicity. Primaries with $4600 < T_{\rm eff} < 4800$~K and [Fe/H$] < 0$~dex appear to be reasonably estimated by \texttt{uberMS}. We also find that the \texttt{uberMS}-determined metallicity of HD~93380 is appropriate despite its temperature being below 4500~K, likely resulting from its very low metallicity ([Fe/H]=$-$0.65~dex). Following this analysis, we readd 5 stars from the low-$T_{\rm eff}$ sample into our FGK--M calibration sample: HD~93380~B, 2MASS~J16453262+3349414, GJ~838.1~B, LP~138-36, and TIC~14307740. These targets are listed in the lower portion of Table~\ref{tab:ew}.

We then recalibrate the relation once more. To evaluate the error in our [Fe/H] estimates that results from uncertainty in the relation's coefficients, we conduct a Monte Carlo analysis: for 2000 iterations, we perturb the masses, equivalent widths, and metallicities within their reported errors and recalculate the fit. This analysis allows us to determine the covariance matrix between the coefficients. When applying the relation to estimate the metallicity of an arbitrary star, the uncertainty has three components: the error in the coefficients ($\sigma_{\rm coef}$), the uncertainty in mass and equivalent width for that star ($\sigma_M$, $\sigma_W$), and the intrinsic scatter in the relation ($\sigma_{\rm int}$). Our final equations are therefore given by:

\begin{equation}
\label{eq:feh}
f(M,W) = 1.912+ 4.516M -2.337W,
\end{equation}
\begin{equation}
[\mathrm{Fe/H}] = \log_{10}(f),
\end{equation}
\begin{align}
\sigma_{\mathrm{coef}}^2
&= 0.02268 + 0.2623M^2 \nonumber\\
&\quad + 0.02411W^2 - 0.06363M \nonumber\\
&\quad - 0.03136W - 0.06204MW,
\end{align}
\begin{align}
\sigma_{\mathrm{[Fe/H]}}
&=
\left[
\sigma_{\mathrm{int}}^2
+
\frac{
\sigma_{\mathrm{coef}}^2
+(4.516\sigma_M)^2
+(2.337\sigma_W)^2
}{
(\ln 10)^2
(f)^2
}
\right]^{1/2}.
\label{eq:error}
\end{align}
Because the metallicity contours converge towards lower metallicities, the same observational errors produce larger uncertainties in [Fe/H] at lower metallicity.

To estimate $\sigma_{\mathrm{int}}$, we consider the residuals in the fit to the calibration sample but only propagate the errors from the input mass, equivalent width, and metallically, not the errors in the polynomial coefficients. We then determine the value of $\sigma_{\mathrm{int}}$ that must be added in quadrature to obtain $\chi^2_{\rm{red}}=1$. We find $\sigma_{\mathrm{int}} = 0.05$~dex.

As can be seen in the right panel of Figure~\ref{fig:fgkm_feh_calib}, many of the strongest outliers are \textalpha-enriched. This is perhaps unsurprising considering that the metallicity sensitivity of the CaT triplet comes from its embeddedness in the pseudocontinuum, which contains a mix of metals. Moreover, this feature is specifically embedded in a TiO band \citep[e.g., see Figure 7 of][]{Sanchez2012}, and Ti is an \textalpha\ element.

In light of this dependency on \textalpha-enrichment, we also consider a version of our fit in which we link CaT equivalent width to [\textalpha/H] instead of [Fe/H]. These results are shown in Figure~\ref{fig:fgkm_ah_calib} and yield
\begin{equation}
\label{eq:ah}
f(M,W) = 2.151+ 4.335M -2.348W,
\end{equation}
\begin{align}
\sigma_{\mathrm{coef}}^2
&= 0.01712 + 0.1772M^2 \nonumber\\
&\quad + 0.02214W^2 - 0.02701M \nonumber\\
&\quad - 0.02834W - 0.05909MW,
\end{align}
\begin{align}
\sigma_{\mathrm{[\alpha/H]}}
&=
\left[
\sigma_{\mathrm{int}}^2
+
\frac{
\sigma_{\mathrm{coef}}^2
+(4.335\sigma_M)^2
+(2.348\sigma_W)^2
}{
(\ln 10)^2
(f)^2
}
\right]^{1/2},
\label{eq:error2}
\end{align}
\noindent and $\sigma_{\mathrm{int}} = 0.035$~dex.

As can be seen in the righthand panel of Figure~\ref{fig:fgkm_ah_calib}, calibrating the CaT feature on [\textalpha/H] reduces scatter, although it also reduces the dynamic range. High-\textalpha\ stars are no longer outliers in this fit; instead, the strongest outliers are some of the cooler stars added in the previous step, with residuals at the 0.11--0.12~dex level: GJ~838.1~B, LP~138-36, TIC~14307740, as well as 2MA~0258+20, which orbits the hottest and most rotationally broadened primary in the sample. That is, these targets are at the extremes of the primary-star sample and their abundance estimates may thus be slightly biased, although not to such a significant extent that they are wholly uninformative for our calibration. At 4800~K in the Hyades, \citet{Pass2025} found offsets of roughly 0.1~dex; we are likely seeing a similar effect here, but pushed to lower temperatures at lower metallicity. We note that NLTT~20915 is also one of the strongest outliers in both the [Fe/H] and [\textalpha/H] views, but we are unable to identify a clear reason for this behavior.

The 15pc sample contains two wide binaries where both stars are inactive M dwarfs. These systems provide a test for our calibrations: given our expectation that the components formed from the same molecular cloud, we expect both components to have the same metallicity. For GJ 1194 AB (masses of 0.30M$_\odot$ and 0.20M$_\odot$), we obtain [Fe/H] = $-0.37\pm0.11$~dex for A and $-0.36\pm0.10$~dex B. For GJ 589 AB (masses of 0.30M$_\odot$ and 0.13M$_\odot$), we obtain [Fe/H] = $-0.31\pm0.10$~dex for A and [Fe/H] = $-0.39\pm0.12$~dex for B. The [\textalpha/Fe]-calibrated relation gives $-0.22\pm0.08$~dex, $-0.20\pm0.07$~dex, $-0.18\pm0.07$~dex, and $-0.21\pm0.08$~dex, respectively. The components consistently agree within uncertainties.

\begin{figure*}
    \hspace{-0.7cm}\includegraphics[width=1.05\textwidth]{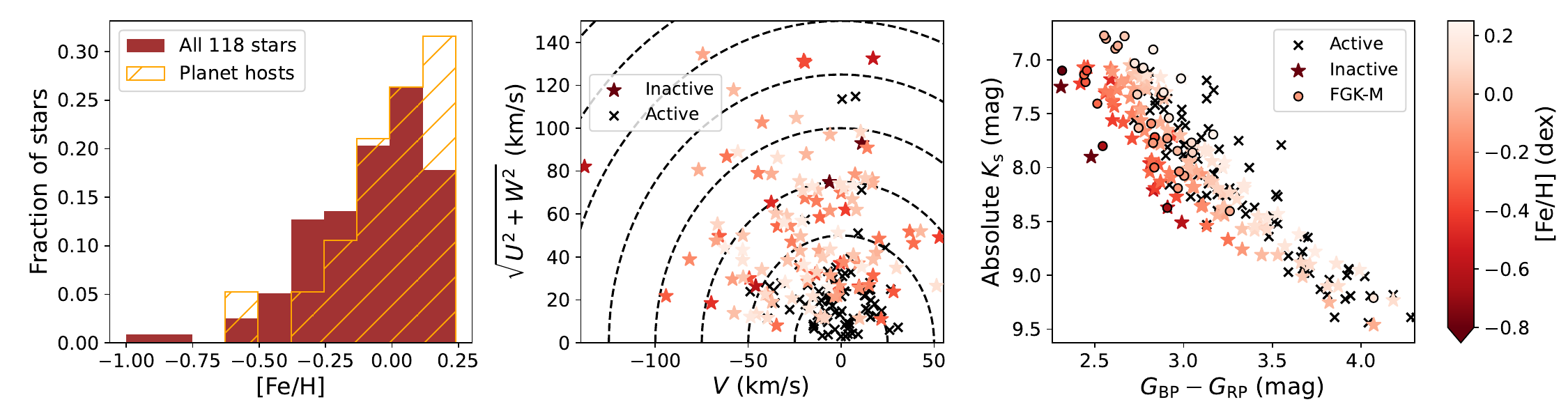}\vspace{-0.3cm}
    \caption{Our method of [Fe/H] estimation applied to the 118 single, inactive M dwarfs observed by TRES in the volume-complete survey of 0.1--0.3M$_\odot$ stars within 15pc \citep{Winters2026}. The left panel shows a histogram of the resulting [Fe/H] estimates, the middle panel shows galactic kinematics, and the right panel shows a color-magnitude diagram. In the middle and right panels, the excluded active stars are plotted as black Xs to help contextualize the sample.}
    \label{fig:results}
\end{figure*}

\section{Application to M-Dwarf Survey}
\label{sec:met}
In Table~\ref{tab:results}, we apply Equations~\ref{eq:feh} and \ref{eq:ah} to estimate metallicities for the volume-complete sample of 118 single, inactive M dwarfs observed by TRES in \citet{Winters2026}; wide binaries are also included, as they are sufficiently separated for TRES spectra to be obtained of each component. While we provide these estimates as both [Fe/H] and [\textalpha/H], both quantities are determined from the same equivalent width. They are \textit{not} independently measured, and therefore it would be meaningless to combine the two to estimate [\textalpha/Fe].

Uncertainties vary from star to star following Equations~\ref{eq:error} and \ref{eq:error2}, but we find a median [Fe/H] error of $\pm$0.07~dex and a median [\textalpha/H] error of $\pm$0.05~dex.

\subsection{Population-level statistics}
\label{sec:pop}
Our TRES requirement limits us to the northern and equatorial targets from the volume-complete sample. Eventually, we would like to also apply our method to observations from the CTIO HIgh ResolutiON (CHIRON; $R=80$,000; \citealt{Tokovinin2013}) spectrograph at the Cerro Tololo Inter-American Observatory \hbox{1.5 m} telescope, as this instrument was used to observe targets in the 15pc sample that were too far south to be accessible with TRES ($\delta \lesssim -15\degree$); however, this extension is left for future work since additional calibration is still needed for the CHIRON data. That said, the 15pc sample is expected to be isotropic and so this selection is not likely to bias our population-level statistics.

Meanwhile, our cut on activity means that our sample is missing the youngest stars, and this selection effect is mass dependent: at 0.3M$_\odot$, M dwarfs retain their H\textalpha\ emission for 1.3~Gyr on average, while 0.1M$_\odot$ stars do so for 4.4~Gyr \citep{Pass2024}. However, this cut does not necessarily induce a substantial bias: the age-metallicity relation in the solar neighborhood is characterized by a large intrinsic dispersion, such that there is not a clear trend between older stars and systematically lower metallicities \citep[e.g.,][]{Rebassa-Mansergas2026}. Indeed, we find a mean and standard deviation of [Fe/H]=$-0.09\pm0.22$~dex for our sample, similar to the \citet{Buder2019} $-0.04\pm0.26$~dex distribution for stars in the solar neighborhood from GALAH \citep{DeSilva2015}. Our distribution is shown in the left panel of Figure~\ref{fig:results}. As expected, it includes a large tail to low metallicities, with a median of $-0.04$~dex and a 68th-percentile interval of $-0.31$ to 0.12~dex.

In the middle and right panels of Figure~\ref{fig:results}, we plot the TRES sample on Toomre and color-magnitude diagrams, analogous to those plotted for the calibration sample in Figures~\ref{fig:sample} and \ref{fig:cmd}. To place this sample in context, we also show the active stars that have been excluded from our analysis; active stars tend to fall near the origin on the Toomre diagram and to the upper right of the CMD. Among the inactive stars, there is no clear trend between metallicity and position on the Toomre diagram; this is in line with the previously stated argument that there is not a strong age-metallicity relation in the solar neighborhood. We see the expected metallicity dependence in the CMD, where bluer stars have lower metallicity at constant absolute $K$--band magnitude.

For most of our parameter space, we do not see a correlation between stellar mass and metallicity. However, there appears to be a pileup at super-solar metallicities for the lowest stellar masses: of the 15 stars with masses $\leq 0.125$M$_\odot$, the lowest metallicity is [Fe/H] = $-0.12$~dex, which would still place it in the 39th percentile of the sample. This result is statistically unlikely and may suggest that our linear scaling breaks down at the lowest stellar masses probed. As we only have a single FGK--M calibrator in the range 0.1--0.125~M$_\odot$ and it is itself super-solar, our calibration sample does not allow us to map this potential curvature. The agreement between GJ~589~A and B (0.296 and 0.129M$_\odot$) disfavors an abrupt downturn that affects stars at 0.13M$_\odot$, as such a behavior would disrupt the agreement between the measurements by moving B to a lower metallicity; if a departure from linearity exists, only the 0.1--0.125~M$_\odot$ range is affected. While such a departure is a likely explanation (particularly considering the large range in the CMD covered by these low masses), it is also plausible that a portion of this trend could be astrophysical (e.g., due to the metallicity dependence of the stellar initial mass function; \citealt{Li2023}). In any case, we flag the stars with masses $\leq 0.125$M$_\odot$ in Table~\ref{tab:results} and caution that additional reference stars  are needed in this regime to validate their metallicities.

\subsection{Planet-hosting stars}
While the correlation between metallicity and giant-planet occurrence is well established for Sun-like stars \citep[e.g.,][]{Santos2001, Fischer2005, Fulton2021, Rosenthal2022}, the picture is more complicated for M dwarfs, particularly mid-to-late ones \citep{Gaidos2014, Gan2025}. The nature of the correlation is also more debated for small planets \citep[e.g.,][]{Petigura2018, Lu2020, Kutra2021}. To investigate this topic, we crossmatch our 118-star sample with the Exoplanet Archive \citep{Christiansen2025} to produce the list of known planet hosts among our targets. There are 4 hosts of transiting planets in the sample: GJ~1214 \citep{Charbonneau2009}, LHS~1140 \citep{Dittmann2017, Ment2019}, LTT 1445~A \citep{Winters2019, Winters2022}, and GJ~12 \citep{Dholakia2024}. There are a further 15 hosts of known non-transiting planets: LHS~1723 \citep{AstudilloDefru2017}, GJ 447 \citep{Bonfils2018b}, GJ~1265 \& LHS~350 \citep{Luque2018}, GJ~1057 \citep{Bauer2020}, GJ~585 \citep{Harakawa2022}, GJ~1151 \citep{BlancoPozo2023}, GJ~1002 \citep{SuarezMascareno2023}, GJ~1253 \citep{Kossakowski2023}, LHS~3262 \citep{Gorrini2023}, Barnard's Star \citep{GonzalezHernandez2024}, LHS~1809 \citep{Kaminski2025}, GJ~324~B \citep{Moutou2026}, GJ~725~B \citep{OuldElhkim2026}, and LHS~1805 \citep{Robertson2026}, for a total of 19 planet hosts. All of the known planets are smaller than Neptune.

We can then ask: are the known planets preferentially found around metal-rich hosts? Or more specifically, we can test the null hypothesis that the known planet hosts are drawn from the same distribution as the other stars. We conduct a one-sided Kolmogorov–Smirnov (KS) test to compare the cumulative distribution of [Fe/H] for the planet-hosting stars against the same distribution for the other 99 stars in the sample, which may or may not host planets. We find a p-value of 0.22, representing tentative but not statistically significant evidence that the known planets preferentially orbit more metal-rich stars than the overall distribution. A one-sided Mann–Whitney U (MWU) test results in a p-value of 0.056, representing slightly stronger yet still tentative support of the planet hosts being anomalously metal rich. The MWU is arguably the more relevant metric for this specific science question, as we are most interested in a potential difference in distribution median. This result is not biased by the possible overestimation of metallicities for very low-mass stars that we discussed above: if we consider only the 103 stars with masses $>0.125$M$_\odot$, the MWU further strengthens to 0.023.

Nonetheless, we caution that there are likely many yet-unknown planets within the sample. Future investigation into the selection effects of the heterogeneous radial velocity surveys that have produced the known planet population is thus critical before claiming that there exists a statistically significant trend.

\section{Conclusions}
\label{sec:conclusion}
We develop a novel spectroscopic metallicity indicator for inactive, mid-to-late M dwarfs that leverages the metallicity sensitivity of the calcium infrared triplet (Figure~\ref{fig:caii}). Our relation is calibrated using a sample of 36 FGK--M wide binaries observed with the TRES and MAROON-X spectrographs (Table~\ref{tab:ew}), spanning a [Fe/H] range of $-0.82$ to $+0.33$~dex and an M-dwarf mass range of 0.11 to 0.35M$_\odot$. Due to a limited number of low-mass calibrators, metallicity estimates may be unreliable for stellar masses below 0.125M$_\odot$. Primary-star metallicities are homogeneously measured using the TRES-\texttt{uberMS} code \citep{Pass2025}. We provide equations that link CaT strength to either [Fe/H] (Equation~\ref{eq:feh}; median uncertainty of $\pm$0.07~dex) or [\textalpha/H] (Equation~\ref{eq:ah}; median uncertainty of $\pm$0.05~dex). We stress that these relations are calibrated on the same feature. They are not independent measurements and therefore should not be combined to create an estimate of [\textalpha/Fe]. We also caveat that our relation is specifically designed for inactive M dwarfs, as strong chromospheric activity can fill in the CaT line core (Figure~\ref{fig:ha}).

We show that there are disagreements among various literature relations when applied to mid-to-late M dwarfs (Figure~\ref{fig:g14}), and that these relations produce systematically biased [Fe/H] estimates when applied to our sample of FGK--M calibrators (Figure~\ref{fig:n15_calib}), motivating the use of this novel calibration.

We apply our relation to the volume-complete sample of mid-to-late M dwarfs observed in \citet{Winters2026}, or specifically, the 118 of these stars that are single, inactive, and observed with the TRES spectrograph (Table~\ref{tab:results}). We find that the distribution of metallicities for these mid-to-late M dwarfs is comparable to findings for the solar neighborhood at large from surveys like GALAH \citep{Buder2019}. We also recover the expected large dispersion in the age-metallicity relation and the strong correlation between CMD position and metallicity (Figure~\ref{fig:results}). We compare the 19 known planet-hosting stars against the overall metallicity distribution, finding a tentative correlation between high metallicity and planet occurrence (94\% confidence in a Mann-Whitney U test). However, we stress that while our host-star metallicities have been homogeneously determined, planet-search efforts are strongly heterogeneous; there are likely many yet-undiscovered planets in this sample, and therefore the tentative correlation could plausibly be sculpted by selection effects in the various radial-velocity search efforts.

Lastly, we report the discovery and orbital characterization of two new double-lined spectroscopic binaries, HD 90089 BD ($P=0.82$~days) and 2MASS J1075+3101 AB ($P=2.0$~days) that are members of higher-order multiples (Figure~\ref{fig:sb2}). LP~4-248 is also likely an unresolved binary in a hierarchical triple, as inferred based on its high Gaia RUWE and our measurement of substantial spectroscopic broadening that is inconsistent with its observed H\textalpha\ quiescence.

\section*{Acknowledgements}
E.P.\ is supported by a Juan Carlos Torres Postdoctoral Fellowship at the Massachusetts Institute of Technology and A.V.\ is supported in part by a Sloan Research Fellowship. We thank Jonathan Irwin and Amber Medina for their work developing codes to analyze TRES spectra of M dwarfs. We are also grateful to the MAROON-X team for their support, particularly Jacob Bean, Andreas Seifahrt, and Nina Brown.

This work is based on observations obtained at the international Gemini Observatory, a program of NSF’s NOIRLab, which is managed by the Association of Universities for Research in Astronomy (AURA) under a cooperative agreement with the National Science Foundation on behalf of the Gemini Observatory partnership: the National Science Foundation (United States), National Research Council (Canada), Agencia Nacional de Investigaci\'{o}n y Desarrollo (Chile), Ministerio de Ciencia, Tecnolog\'{i}a e Innovaci\'{o}n (Argentina), Minist\'{e}rio da Ci\^{e}ncia, Tecnologia, Inova\c{c}\~{o}es e Comunica\c{c}\~{o}es (Brazil), and Korea Astronomy and Space Science Institute (Republic of Korea). These observations were obtained under programs GN-2022B-Q-323 and GN-2023B-Q-325.

This work has made use of data from the European Space Agency (ESA) mission Gaia (\url{https://www.cosmos.esa.int/gaia}), processed by the Gaia
Data Processing and Analysis Consortium (DPAC, \url{https://www.cosmos.esa.int/web/gaia/dpac/consortium}). Funding for the DPAC
has been provided by national institutions, in particular the institutions
participating in the Gaia Multilateral Agreement. It also uses data products from the Two Micron All Sky Survey, which is a joint project of the University of Massachusetts and the Infrared Processing and Analysis Center/California Institute of Technology, funded by the National Aeronautics and Space Administration and the National Science Foundation. This research also uses the SIMBAD database, operated at CDS, Strasbourg, France.

\facilities{FLWO:1.5m (TRES), Gemini:Gillett (MAROON-X)}
\software{\texttt{exoplanet} \citep{ForemanMackey2021}, \texttt{Matplotlib} \citep{Hunter2007}, \texttt{NumPy} \citep{Harris2020}, \texttt{pandas} \citep{Reback2021}, \texttt{PyMC3} \citep{Salvatier2016}, \texttt{SciPy} \citep{Scipy2020}}

\startlongtable
\begin{deluxetable*}{llccrr}
\tablecaption{Metallicities for single, inactive stars in the 15pc TRES mid-to-late M-dwarf sample \label{tab:results}}
\tablewidth{0pt}
\tabletypesize{\footnotesize}
\tablehead{
\colhead{Name} &
\colhead{2MASS ID} &
\colhead{Mass (M$_\odot$)} &
\colhead{EW (\AA)} &
\colhead{[Fe/H] (dex)} &
\colhead{[\textalpha/H] (dex)}
}
\startdata
GJ 1002$^{*\ddag}$ & 2MASSJ00064325$-$0732147 & $0.115$ & $0.568 \pm 0.027$ & $0.04 \pm 0.07$ & $0.12 \pm 0.05$ \\
LEP0011+59$^\ddag$ & 2MASSJ00113182+5908400 & $0.107$ & $0.627 \pm 0.020$ & $-0.03 \pm 0.07$ & $0.06 \pm 0.05$ \\
GJ 12$^{*}$ & 2MASSJ00154919+1333218 & $0.249$ & $1.065 \pm 0.023$ & $-0.26 \pm 0.08$ & $-0.14 \pm 0.06$ \\
GJ 15~B & 2MASSJ00182549+4401376 & $0.157$ & $0.952 \pm 0.015$ & $-0.40 \pm 0.11$ & $-0.22 \pm 0.07$ \\
LHS 112$^\ddag$ & 2MASSJ00202922+3305081 & $0.119$ & $0.437 \pm 0.027$ & $0.15 \pm 0.06$ & $0.22 \pm 0.05$ \\
GJ 1013 & 2MASSJ00313539$-$0552115 & $0.275$ & $1.017 \pm 0.019$ & $-0.11 \pm 0.07$ & $-0.02 \pm 0.05$ \\
GJ 1014 & 2MASSJ00355557+1028352 & $0.132$ & $0.543 \pm 0.023$ & $0.09 \pm 0.06$ & $0.16 \pm 0.05$ \\
LHS 1140$^{*}$ & 2MASSJ00445930-1516166 & $0.179$ & $0.742 \pm 0.019$ & $-0.01 \pm 0.06$ & $0.07 \pm 0.05$ \\
GDR20049+65 & 2MASSJ00492565+6518038 & $0.140$ & $0.694 \pm 0.019$ & $-0.04 \pm 0.07$ & $0.05 \pm 0.05$ \\
GJ 1025 & 2MASSJ01005643$-$0426561 & $0.198$ & $0.940 \pm 0.023$ & $-0.22 \pm 0.08$ & $-0.10 \pm 0.06$ \\
GJ 1035 & 2MASSJ01195227+8409327 & $0.155$ & $0.593 \pm 0.028$ & $0.09 \pm 0.06$ & $0.16 \pm 0.05$ \\
GJ 61~B & 2MASSJ01365042+4123325 & $0.187$ & $0.712 \pm 0.019$ & $0.04 \pm 0.06$ & $0.11 \pm 0.05$ \\
LHS 1326$^\ddag$ & 2MASSJ02021620+1020136 & $0.112$ & $0.481 \pm 0.019$ & $0.11 \pm 0.06$ & $0.18 \pm 0.05$ \\
GJ 105~B & 2MASSJ02361535+0652191 & $0.263$ & $0.844 \pm 0.018$ & $0.05 \pm 0.06$ & $0.12 \pm 0.04$ \\
LHS 1443$^\ddag$ & 2MASSJ02463486+1625115 & $0.101$ & $0.640 \pm 0.035$ & $-0.06 \pm 0.08$ & $0.04 \pm 0.06$ \\
LHS 1481$^\dag$ & 2MASSJ02581021$-$1253066 & $0.184$ & $1.216 \pm 0.024$ & $\lesssim -1$ & $-1.03 \pm 0.47$ \\
LTT 1445~A$^{*}$ & 2MASSJ03015142$-$1635356 & $0.258$ & $1.052 \pm 0.018$ & $-0.21 \pm 0.08$ & $-0.10 \pm 0.06$ \\
GJ 1055 & 2MASSJ03090015+1001257 & $0.133$ & $0.635 \pm 0.021$ & $0.01 \pm 0.07$ & $0.09 \pm 0.05$ \\
GJ 1053 & 2MASSJ03105861+7346189 & $0.128$ & $0.821 \pm 0.030$ & $-0.24 \pm 0.10$ & $-0.11 \pm 0.07$ \\
GJ 1057$^{*}$ & 2MASSJ03132299+0446293 & $0.160$ & $0.489 \pm 0.019$ & $0.17 \pm 0.06$ & $0.23 \pm 0.04$ \\
GJ 1059$^\ddag$ & 2MASSJ03230175+4200269 & $0.124$ & $0.731 \pm 0.028$ & $-0.12 \pm 0.08$ & $-0.01 \pm 0.06$ \\
LHS 176$^\ddag$ & 2MASSJ03353849$-$0829223 & $0.119$ & $0.535 \pm 0.029$ & $0.08 \pm 0.07$ & $0.15 \pm 0.05$ \\
LHS 1593 & 2MASSJ03472091+0841464 & $0.134$ & $0.841 \pm 0.027$ & $-0.26 \pm 0.10$ & $-0.12 \pm 0.07$ \\
GJ 1065 & 2MASSJ03504432$-$0605400 & $0.198$ & $0.935 \pm 0.024$ & $-0.21 \pm 0.08$ & $-0.09 \pm 0.06$ \\
LP 357-56$^\ddag$ & 2MASSJ03544620+2416246 & $0.116$ & $0.396 \pm 0.032$ & $0.18 \pm 0.06$ & $0.24 \pm 0.05$ \\
2MA0406-05 & 2MASSJ04060688$-$0534444 & $0.224$ & $0.931 \pm 0.021$ & $-0.13 \pm 0.07$ & $-0.03 \pm 0.05$ \\
LHS 1629 & 2MASSJ04063732+7916012 & $0.134$ & $0.575 \pm 0.024$ & $0.07 \pm 0.06$ & $0.14 \pm 0.05$ \\
GJ 1072 & 2MASSJ04505083+2207224 & $0.139$ & $0.454 \pm 0.021$ & $0.17 \pm 0.06$ & $0.23 \pm 0.04$ \\
GJ 1073 & 2MASSJ04523448+4042255 & $0.212$ & $0.820 \pm 0.022$ & $-0.02 \pm 0.06$ & $0.06 \pm 0.05$ \\
LHS 1723$^{*}$ & 2MASSJ05015746$-$0656459 & $0.166$ & $0.793 \pm 0.022$ & $-0.09 \pm 0.07$ & $0.00 \pm 0.05$ \\
LHS 1731 & 2MASSJ05032009$-$1722245 & $0.290$ & $1.051 \pm 0.022$ & $-0.12 \pm 0.07$ & $-0.03 \pm 0.05$ \\
UPM0505+44 & 2MASSJ05050591+4414037 & $0.145$ & $0.663 \pm 0.022$ & $0.01 \pm 0.07$ & $0.09 \pm 0.05$ \\
G 86-28 & 2MASSJ05103956+2946479 & $0.261$ & $1.070 \pm 0.022$ & $-0.23 \pm 0.08$ & $-0.11 \pm 0.06$ \\
GJ 203 & 2MASSJ05280015+0938382 & $0.231$ & $1.003 \pm 0.023$ & $-0.21 \pm 0.08$ & $-0.10 \pm 0.06$ \\
LHS 5108 & 2MASSJ05325194+3349474 & $0.202$ & $1.015 \pm 0.026$ & $-0.35 \pm 0.10$ & $-0.19 \pm 0.07$ \\
GJ 213 & 2MASSJ05420897+1229252 & $0.221$ & $0.863 \pm 0.017$ & $-0.05 \pm 0.06$ & $0.04 \pm 0.05$ \\
GJ 2045$^\ddag$ & 2MASSJ05421271$-$0527567 & $0.122$ & $0.648 \pm 0.020$ & $-0.02 \pm 0.07$ & $0.06 \pm 0.05$ \\
LEP0556+11$^\ddag$ & 2MASSJ05565722+1144333 & $0.125$ & $0.453 \pm 0.027$ & $0.15 \pm 0.06$ & $0.21 \pm 0.05$ \\
LHS 1805$^{*}$ & 2MASSJ06011106+5935508 & $0.275$ & $0.899 \pm 0.021$ & $0.02 \pm 0.06$ & $0.09 \pm 0.05$ \\
LHS 1809$^{*}$ & 2MASSJ06022918+4951561 & $0.134$ & $0.508 \pm 0.023$ & $0.12 \pm 0.06$ & $0.19 \pm 0.05$ \\
G 192-22 & 2MASSJ06140240+5140081 & $0.264$ & $1.089 \pm 0.024$ & $-0.25 \pm 0.09$ & $-0.13 \pm 0.06$ \\
GJ 232 & 2MASSJ06244132+2325585 & $0.152$ & $0.923 \pm 0.028$ & $-0.36 \pm 0.11$ & $-0.19 \pm 0.08$ \\
GJ 1092 & 2MASSJ06490542+3706533 & $0.177$ & $0.949 \pm 0.019$ & $-0.31 \pm 0.09$ & $-0.16 \pm 0.06$ \\
GJ 1093$^\ddag$ & 2MASSJ06592868+1920577 & $0.123$ & $0.386 \pm 0.025$ & $0.19 \pm 0.06$ & $0.25 \pm 0.05$ \\
G 107-48 & 2MASSJ07073776+4841138 & $0.163$ & $0.916 \pm 0.021$ & $-0.29 \pm 0.09$ & $-0.15 \pm 0.06$ \\
LHS 1950 & 2MASSJ07515138+0532572 & $0.154$ & $0.638 \pm 0.026$ & $0.05 \pm 0.06$ & $0.12 \pm 0.05$ \\
GJ 1103 & 2MASSJ07515465$-$0000117 & $0.192$ & $0.632 \pm 0.019$ & $0.11 \pm 0.06$ & $0.18 \pm 0.04$ \\
GJ 1105 & 2MASSJ07581269+4118134 & $0.284$ & $0.809 \pm 0.021$ & $0.12 \pm 0.06$ & $0.17 \pm 0.04$ \\
GJ 299 & 2MASSJ08115757+0846220 & $0.138$ & $0.979 \pm 0.023$ & $-0.61 \pm 0.18$ & $-0.35 \pm 0.10$ \\
LHS 2025 & 2MASSJ08313011+7303459 & $0.279$ & $0.910 \pm 0.017$ & $0.02 \pm 0.06$ & $0.09 \pm 0.05$ \\
GJ 2070 & 2MASSJ08342587$-$0108391 & $0.246$ & $1.104 \pm 0.019$ & $-0.35 \pm 0.10$ & $-0.20 \pm 0.07$ \\
LEP0840+31 & 2MASSJ08401597+3127068 & $0.299$ & $0.963 \pm 0.018$ & $0.01 \pm 0.06$ & $0.07 \pm 0.05$ \\
GJ 324~B$^{*}$ & 2MASSJ08524084+2818589 & $0.278$ & $0.613 \pm 0.015$ & $0.24 \pm 0.06$ & $0.28 \pm 0.04$ \\
LHS 2088 & 2MASSJ08595604+7257364 & $0.158$ & $0.796 \pm 0.025$ & $-0.12 \pm 0.07$ & $-0.02 \pm 0.05$ \\
LP 60-179 & 2MASSJ09025284+6803464 & $0.265$ & $0.942 \pm 0.022$ & $-0.04 \pm 0.06$ & $0.04 \pm 0.05$ \\
LEP0921$-$02 & 2MASSJ09214812$-$0219433 & $0.272$ & $1.172 \pm 0.023$ & $-0.40 \pm 0.11$ & $-0.24 \pm 0.08$ \\
GJ 359 & 2MASSJ09410199+2201291 & $0.145$ & $0.806 \pm 0.024$ & $-0.17 \pm 0.08$ & $-0.05 \pm 0.06$ \\
LHS 2224 & 2MASSJ10092996+5117197 & $0.181$ & $0.943 \pm 0.025$ & $-0.28 \pm 0.09$ & $-0.14 \pm 0.06$ \\
LEP1015+17 & 2MASSJ10155390+1729271 & $0.287$ & $0.922 \pm 0.021$ & $0.02 \pm 0.06$ & $0.09 \pm 0.05$ \\
GJ 1134 & 2MASSJ10413809+3736397 & $0.205$ & $0.788 \pm 0.020$ & $0.00 \pm 0.06$ & $0.08 \pm 0.05$ \\
GJ 402 & 2MASSJ10505201+0648292 & $0.282$ & $0.739 \pm 0.019$ & $0.16 \pm 0.06$ & $0.21 \pm 0.04$ \\
GJ 403 & 2MASSJ10520440+1359509 & $0.278$ & $1.031 \pm 0.021$ & $-0.12 \pm 0.07$ & $-0.03 \pm 0.05$ \\
LHS 296 & 2MASSJ11011965+0300171 & $0.164$ & $0.863 \pm 0.026$ & $-0.20 \pm 0.08$ & $-0.08 \pm 0.06$ \\
LHS 2395$^\ddag$ & 2MASSJ11193058+4641437 & $0.112$ & $0.492 \pm 0.026$ & $0.10 \pm 0.06$ & $0.17 \pm 0.05$ \\
LHS 2415 & 2MASSJ11285624+1010395 & $0.286$ & $0.928 \pm 0.022$ & $0.01 \pm 0.06$ & $0.08 \pm 0.05$ \\
LHS 306 & 2MASSJ11310835$-$1457201 & $0.157$ & $0.615 \pm 0.029$ & $0.07 \pm 0.06$ & $0.14 \pm 0.05$ \\
GJ 445 & 2MASSJ11474143+7841283 & $0.252$ & $0.998 \pm 0.023$ & $-0.14 \pm 0.07$ & $-0.05 \pm 0.05$ \\
GJ 447$^{*}$ & 2MASSJ11474440+0048164 & $0.173$ & $0.721 \pm 0.021$ & $0.00 \pm 0.06$ & $0.08 \pm 0.05$ \\
GJ 1151$^{*}$ & 2MASSJ11505787+4822395 & $0.164$ & $0.688 \pm 0.019$ & $0.02 \pm 0.06$ & $0.10 \pm 0.05$ \\
LHS 2674 & 2MASSJ13065025+3050549 & $0.133$ & $0.541 \pm 0.025$ & $0.10 \pm 0.06$ & $0.16 \pm 0.05$ \\
LHS 350$^{*}$ & 2MASSJ13225673+2428034 & $0.265$ & $0.900 \pm 0.025$ & $0.00 \pm 0.06$ & $0.07 \pm 0.05$ \\
GJ 1171 & 2MASSJ13303106+1909340 & $0.146$ & $0.814 \pm 0.024$ & $-0.17 \pm 0.08$ & $-0.06 \pm 0.06$ \\
LHS 2784 & 2MASSJ13424328+3317255 & $0.281$ & $0.874 \pm 0.026$ & $0.06 \pm 0.06$ & $0.12 \pm 0.05$ \\
GJ 1179~A$^\ddag$ & 2MASSJ13481341+2336486 & $0.122$ & $0.599 \pm 0.024$ & $0.03 \pm 0.07$ & $0.11 \pm 0.05$ \\
LHS 2830 & 2MASSJ13581392+1234438 & $0.220$ & $1.045 \pm 0.029$ & $-0.33 \pm 0.10$ & $-0.19 \pm 0.07$ \\
G 165-58 & 2MASSJ14155637+3616368 & $0.241$ & $1.068 \pm 0.022$ & $-0.30 \pm 0.09$ & $-0.16 \pm 0.06$ \\
GJ 545 & 2MASSJ14200739$-$0937127 & $0.263$ & $0.982 \pm 0.027$ & $-0.09 \pm 0.07$ & $-0.01 \pm 0.05$ \\
LHS 2899 & 2MASSJ14211512$-$0107199 & $0.237$ & $0.915 \pm 0.014$ & $-0.07 \pm 0.06$ & $0.01 \pm 0.05$ \\
LEP1455+30 & 2MASSJ14551146+3006454 & $0.170$ & $0.921 \pm 0.027$ & $-0.28 \pm 0.09$ & $-0.14 \pm 0.07$ \\
GJ 585$^{*}$ & 2MASSJ15235112+1727569 & $0.171$ & $0.771 \pm 0.025$ & $-0.05 \pm 0.07$ & $0.03 \pm 0.05$ \\
GJ 589~B & 2MASSJ15352039+1743045 & $0.129$ & $0.891 \pm 0.025$ & $-0.39 \pm 0.12$ & $-0.21 \pm 0.08$ \\
GJ 589~A & 2MASSJ15352059+1742470 & $0.296$ & $1.181 \pm 0.024$ & $-0.31 \pm 0.10$ & $-0.18 \pm 0.07$ \\
GJ 1194~A & 2MASSJ15400352+4329396 & $0.295$ & $1.206 \pm 0.024$ & $-0.37 \pm 0.11$ & $-0.22 \pm 0.08$ \\
GJ 1194~B & 2MASSJ15400374+4329355 & $0.204$ & $1.025 \pm 0.024$ & $-0.36 \pm 0.10$ & $-0.20 \pm 0.07$ \\
GJ 609 & 2MASSJ16025098+2035218 & $0.250$ & $0.855 \pm 0.028$ & $0.02 \pm 0.06$ & $0.09 \pm 0.05$ \\
GJ 611~B & 2MASSJ16045093+3909359 & $0.145$ & $0.990 \pm 0.026$ & $-0.60 \pm 0.18$ & $-0.34 \pm 0.10$ \\
LHS 3241$^\ddag$ & 2MASSJ16463154+3434554 & $0.108$ & $0.476 \pm 0.041$ & $0.11 \pm 0.07$ & $0.18 \pm 0.05$ \\
GJ 643~D & 2MASSJ16552527$-$0819207 & $0.210$ & $0.920 \pm 0.022$ & $-0.15 \pm 0.07$ & $-0.05 \pm 0.05$ \\
LHS 3262$^{*}$ & 2MASSJ17032384+5124219 & $0.180$ & $0.603 \pm 0.022$ & $0.12 \pm 0.06$ & $0.18 \pm 0.04$ \\
GJ 1214$^{*}$ & 2MASSJ17151894+0457496 & $0.178$ & $0.480 \pm 0.029$ & $0.20 \pm 0.06$ & $0.25 \pm 0.04$ \\
GJ 1220 & 2MASSJ17311725+8205198 & $0.158$ & $0.795 \pm 0.030$ & $-0.12 \pm 0.08$ & $-0.01 \pm 0.06$ \\
LHS 3324 & 2MASSJ17460465+2439049 & $0.278$ & $1.209 \pm 0.029$ & $-0.47 \pm 0.14$ & $-0.29 \pm 0.09$ \\
Barnard's Star$^{*}$ & 2MASSJ17574849+0441405 & $0.155$ & $1.001 \pm 0.013$ & $-0.56 \pm 0.14$ & $-0.33 \pm 0.08$ \\
GJ 1223 & 2MASSJ18024624+3731048 & $0.139$ & $0.524 \pm 0.025$ & $0.12 \pm 0.06$ & $0.18 \pm 0.05$ \\
LP 449-10 & 2MASSJ18064856+1720472 & $0.196$ & $0.854 \pm 0.018$ & $-0.10 \pm 0.07$ & $0.00 \pm 0.05$ \\
G 140-51 & 2MASSJ18163154+0452456 & $0.174$ & $0.511 \pm 0.019$ & $0.18 \pm 0.06$ & $0.23 \pm 0.04$ \\
LHS 461~B & 2MASSJ18180345+3846359 & $0.165$ & $0.870 \pm 0.021$ & $-0.21 \pm 0.08$ & $-0.08 \pm 0.06$ \\
GJ 712 & 2MASSJ18220671+0620376 & $0.294$ & $1.116 \pm 0.022$ & $-0.20 \pm 0.08$ & $-0.09 \pm 0.06$ \\
GJ 1227 & 2MASSJ18222719+6203025 & $0.162$ & $0.668 \pm 0.025$ & $0.03 \pm 0.06$ & $0.11 \pm 0.05$ \\
GJ 725~B$^{*}$ & 2MASSJ18424688+5937374 & $0.265$ & $1.042 \pm 0.016$ & $-0.17 \pm 0.07$ & $-0.07 \pm 0.05$ \\
GJ 1232 & 2MASSJ19095098+1740074 & $0.189$ & $0.561 \pm 0.023$ & $0.16 \pm 0.06$ & $0.22 \pm 0.04$ \\
LEP1916+84 & 2MASSJ19162483+8413411 & $0.150$ & $0.560 \pm 0.044$ & $0.11 \pm 0.07$ & $0.17 \pm 0.05$ \\
GJ 754.1~B & 2MASSJ19203346$-$0739435 & $0.259$ & $1.138 \pm 0.022$ & $-0.37 \pm 0.10$ & $-0.22 \pm 0.07$ \\
GJ 1235 & 2MASSJ19213867+2052028 & $0.194$ & $0.632 \pm 0.023$ & $0.12 \pm 0.06$ & $0.18 \pm 0.04$ \\
GJ 1236 & 2MASSJ19220206+0702310 & $0.223$ & $1.079 \pm 0.020$ & $-0.40 \pm 0.10$ & $-0.23 \pm 0.07$ \\
G 125-34 & 2MASSJ19484080+3555178 & $0.237$ & $0.973 \pm 0.015$ & $-0.15 \pm 0.07$ & $-0.05 \pm 0.05$ \\
GJ 1248 & 2MASSJ20035098+0559440 & $0.267$ & $1.267 \pm 0.021$ & $-0.81 \pm 0.27$ & $-0.48 \pm 0.12$ \\
GJ 1253$^{*}$ & 2MASSJ20260528+5834224 & $0.158$ & $0.489 \pm 0.025$ & $0.17 \pm 0.06$ & $0.23 \pm 0.04$ \\
GJ 1256 & 2MASSJ20403364+1529572 & $0.189$ & $0.561 \pm 0.022$ & $0.16 \pm 0.06$ & $0.22 \pm 0.04$ \\
LHS 3593 & 2MASSJ20533304+1037020 & $0.180$ & $0.744 \pm 0.015$ & $-0.01 \pm 0.06$ & $0.07 \pm 0.05$ \\
GJ 810~B & 2MASSJ20553706$-$1403545 & $0.140$ & $0.730 \pm 0.027$ & $-0.08 \pm 0.07$ & $0.02 \pm 0.05$ \\
LEP2124+40 & 2MASSJ21243234+4003599 & $0.130$ & $0.405 \pm 0.025$ & $0.19 \pm 0.06$ & $0.25 \pm 0.05$ \\
LEP2146+38 & 2MASSJ21462206+3813047 & $0.178$ & $1.011 \pm 0.020$ & $-0.45 \pm 0.12$ & $-0.26 \pm 0.08$ \\
LHS 516 & 2MASSJ21565513$-$0154100 & $0.144$ & $0.690 \pm 0.021$ & $-0.02 \pm 0.07$ & $0.06 \pm 0.05$ \\
GJ 1265$^{*}$ & 2MASSJ22134277$-$1741081 & $0.168$ & $0.816 \pm 0.024$ & $-0.12 \pm 0.07$ & $-0.02 \pm 0.05$ \\
GJ 1270 & 2MASSJ22294885+4128479 & $0.259$ & $0.759 \pm 0.018$ & $0.12 \pm 0.06$ & $0.17 \pm 0.04$ \\
GJ 1286$^\ddag$ & 2MASSJ23351050$-$0223214 & $0.118$ & $0.506 \pm 0.024$ & $0.10 \pm 0.06$ & $0.17 \pm 0.05$ \\
GJ 905 & 2MASSJ23415498+4410407 & $0.140$ & $0.422 \pm 0.014$ & $0.19 \pm 0.06$ & $0.25 \pm 0.04$ \\
\enddata
\tablecomments{$^{\dag}$LHS~1481 is outside of the range of metallicities defined by our calibration. $^\ddag$ Metallicities may be overestimated for stellar masses below 0.125M$_\odot$ (see Section~\ref{sec:pop}).  $^{*}$Known planet host.}
\end{deluxetable*}
\bibliography{metals}{}

@Misc{Reback2021,
  author       = {{Reback}, Jeff and {Jbrockmendel} and {McKinney}, Wes and {Van Den Bossche}, Joris and {Augspurger}, Tom and {Cloud}, Phillip and {Hawkins}, Simon and {Gfyoung} and {Sinhrks} and {Roeschke}, Matthew and {Klein}, Adam and {Petersen}, Terji and {Tratner}, Jeff and {She}, Chang and {Ayd}, William and {Hoefler}, Patrick and {Naveh}, Shahar and {Garcia}, Marc and {Schendel}, Jeremy and {Hayden}, Andy and {Saxton}, Daniel and {Gorelli}, Marco Edward and {Shadrach}, Richard and {Jancauskas}, Vytautas and {McMaster}, Ali and {Li}, Fangchen and {Battiston}, Pietro and {Seabold}, Skipper and {Attack68} and {Dong}, Kaiqi},
  howpublished = {Zenodo},
  month        = jul,
  note         = {doi:10.5281/zenodo.3509134},
  title        = {{pandas-dev/pandas: Pandas 1.3.0}},
  year         = {2021},
  adsurl       = {https://ui.adsabs.harvard.edu/abs/2020zndo...3509134R},
  doi          = {10.5281/zenodo.3509134},
  eid          = {10.5281/zenodo.3509134},
  publisher    = {Zenodo},
  version      = {v1.3.0},
}

@Article{Gaia2023,
  author        = {{Gaia Collaboration} and {Vallenari}, A. and {Brown}, A.~G.~A. and {Prusti}, T. and {de Bruijne}, J.~H.~J. and {Arenou}, F. and {Babusiaux}, C. and {Biermann}, M. and {Creevey}, O.~L. and {Ducourant}, C. and {Evans}, D.~W. and {Eyer}, L. and {Guerra}, R. and {Hutton}, A. and {Jordi}, C. and {Klioner}, S.~A. and {Lammers}, U.~L. and {Lindegren}, L. and {Luri}, X. and {Mignard}, F. and {Panem}, C. and {Pourbaix}, D. and {Randich}, S. and {Sartoretti}, P. and {Soubiran}, C. and {Tanga}, P. and {Walton}, N.~A. and {Bailer-Jones}, C.~A.~L. and {Bastian}, U. and {Drimmel}, R. and {Jansen}, F. and {Katz}, D. and {Lattanzi}, M.~G. and {van Leeuwen}, F. and {Bakker}, J. and {Cacciari}, C. and {Casta{\~n}eda}, J. and {De Angeli}, F. and {Fabricius}, C. and {Fouesneau}, M. and {Fr{\'e}mat}, Y. and {Galluccio}, L. and {Guerrier}, A. and {Heiter}, U. and {Masana}, E. and {Messineo}, R. and {Mowlavi}, N. and {Nicolas}, C. and {Nienartowicz}, K. and {Pailler}, F. and {Panuzzo}, P. and {Riclet}, F. and {Roux}, W. and {Seabroke}, G.~M. and {Sordo}, R. and {Th{\'e}venin}, F. and {Gracia-Abril}, G. and {Portell}, J. and {Teyssier}, D. and {Altmann}, M. and {Andrae}, R. and {Audard}, M. and {Bellas-Velidis}, I. and {Benson}, K. and {Berthier}, J. and {Blomme}, R. and {Burgess}, P.~W. and {Busonero}, D. and {Busso}, G. and {C{\'a}novas}, H. and {Carry}, B. and {Cellino}, A. and {Cheek}, N. and {Clementini}, G. and {Damerdji}, Y. and {Davidson}, M. and {de Teodoro}, P. and {Nu{\~n}ez Campos}, M. and {Delchambre}, L. and {Dell'Oro}, A. and {Esquej}, P. and {Fern{\'a}ndez-Hern{\'a}ndez}, J. and {Fraile}, E. and {Garabato}, D. and {Garc{\'\i}a-Lario}, P. and {Gosset}, E. and {Haigron}, R. and {Halbwachs}, J. -L. and {Hambly}, N.~C. and {Harrison}, D.~L. and {Hern{\'a}ndez}, J. and {Hestroffer}, D. and {Hodgkin}, S.~T. and {Holl}, B. and {Jan{\ss}en}, K. and {Jevardat de Fombelle}, G. and {Jordan}, S. and {Krone-Martins}, A. and {Lanzafame}, A.~C. and {L{\"o}ffler}, W. and {Marchal}, O. and {Marrese}, P.~M. and {Moitinho}, A. and {Muinonen}, K. and {Osborne}, P. and {Pancino}, E. and {Pauwels}, T. and {Recio-Blanco}, A. and {Reyl{\'e}}, C. and {Riello}, M. and {Rimoldini}, L. and {Roegiers}, T. and {Rybizki}, J. and {Sarro}, L.~M. and {Siopis}, C. and {Smith}, M. and {Sozzetti}, A. and {Utrilla}, E. and {van Leeuwen}, M. and {Abbas}, U. and {{\'A}brah{\'a}m}, P. and {Abreu Aramburu}, A. and {Aerts}, C. and {Aguado}, J.~J. and {Ajaj}, M. and {Aldea-Montero}, F. and {Altavilla}, G. and {{\'A}lvarez}, M.~A. and {Alves}, J. and {Anders}, F. and {Anderson}, R.~I. and {Anglada Varela}, E. and {Antoja}, T. and {Baines}, D. and {Baker}, S.~G. and {Balaguer-N{\'u}{\~n}ez}, L. and {Balbinot}, E. and {Balog}, Z. and {Barache}, C. and {Barbato}, D. and {Barros}, M. and {Barstow}, M.~A. and {Bartolom{\'e}}, S. and {Bassilana}, J. -L. and {Bauchet}, N. and {Becciani}, U. and {Bellazzini}, M. and {Berihuete}, A. and {Bernet}, M. and {Bertone}, S. and {Bianchi}, L. and {Binnenfeld}, A. and {Blanco-Cuaresma}, S. and {Blazere}, A. and {Boch}, T. and {Bombrun}, A. and {Bossini}, D. and {Bouquillon}, S. and {Bragaglia}, A. and {Bramante}, L. and {Breedt}, E. and {Bressan}, A. and {Brouillet}, N. and {Brugaletta}, E. and {Bucciarelli}, B. and {Burlacu}, A. and {Butkevich}, A.~G. and {Buzzi}, R. and {Caffau}, E. and {Cancelliere}, R. and {Cantat-Gaudin}, T. and {Carballo}, R. and {Carlucci}, T. and {Carnerero}, M.~I. and {Carrasco}, J.~M. and {Casamiquela}, L. and {Castellani}, M. and {Castro-Ginard}, A. and {Chaoul}, L. and {Charlot}, P. and {Chemin}, L. and {Chiaramida}, V. and {Chiavassa}, A. and {Chornay}, N. and {Comoretto}, G. and {Contursi}, G. and {Cooper}, W.~J. and {Cornez}, T. and {Cowell}, S. and {Crifo}, F. and {Cropper}, M. and {Crosta}, M. and {Crowley}, C. and {Dafonte}, C. and {Dapergolas}, A. and {David}, M. and {David}, P. and {de Laverny}, P. and {De Luise}, F. and {De March}, R. and {De Ridder}, J. and {de Souza}, R. and {de Torres}, A. and {del Peloso}, E.~F. and {del Pozo}, E. and {Delbo}, M. and {Delgado}, A. and {Delisle}, J. -B. and {Demouchy}, C. and {Dharmawardena}, T.~E. and {Di Matteo}, P. and {Diakite}, S. and {Diener}, C. and {Distefano}, E. and {Dolding}, C. and {Edvardsson}, B. and {Enke}, H. and {Fabre}, C. and {Fabrizio}, M. and {Faigler}, S. and {Fedorets}, G. and {Fernique}, P. and {Fienga}, A. and {Figueras}, F. and {Fournier}, Y. and {Fouron}, C. and {Fragkoudi}, F. and {Gai}, M. and {Garcia-Gutierrez}, A. and {Garcia-Reinaldos}, M. and {Garc{\'\i}a-Torres}, M. and {Garofalo}, A. and {Gavel}, A. and {Gavras}, P. and {Gerlach}, E. and {Geyer}, R. and {Giacobbe}, P. and {Gilmore}, G. and {Girona}, S. and {Giuffrida}, G. and {Gomel}, R. and {Gomez}, A. and {Gonz{\'a}lez-N{\'u}{\~n}ez}, J. and {Gonz{\'a}lez-Santamar{\'\i}a}, I. and {Gonz{\'a}lez-Vidal}, J.~J. and {Granvik}, M. and {Guillout}, P. and {Guiraud}, J. and {Guti{\'e}rrez-S{\'a}nchez}, R. and {Guy}, L.~P. and {Hatzidimitriou}, D. and {Hauser}, M. and {Haywood}, M. and {Helmer}, A. and {Helmi}, A. and {Sarmiento}, M.~H. and {Hidalgo}, S.~L. and {Hilger}, T. and {H{\l}adczuk}, N. and {Hobbs}, D. and {Holland}, G. and {Huckle}, H.~E. and {Jardine}, K. and {Jasniewicz}, G. and {Jean-Antoine Piccolo}, A. and {Jim{\'e}nez-Arranz}, {\'O}. and {Jorissen}, A. and {Juaristi Campillo}, J. and {Julbe}, F. and {Karbevska}, L. and {Kervella}, P. and {Khanna}, S. and {Kontizas}, M. and {Kordopatis}, G. and {Korn}, A.~J. and {K{\'o}sp{\'a}l}, {\'A}. and {Kostrzewa-Rutkowska}, Z. and {Kruszy{\'n}ska}, K. and {Kun}, M. and {Laizeau}, P. and {Lambert}, S. and {Lanza}, A.~F. and {Lasne}, Y. and {Le Campion}, J. -F. and {Lebreton}, Y. and {Lebzelter}, T. and {Leccia}, S. and {Leclerc}, N. and {Lecoeur-Taibi}, I. and {Liao}, S. and {Licata}, E.~L. and {Lindstr{\o}m}, H.~E.~P. and {Lister}, T.~A. and {Livanou}, E. and {Lobel}, A. and {Lorca}, A. and {Loup}, C. and {Madrero Pardo}, P. and {Magdaleno Romeo}, A. and {Managau}, S. and {Mann}, R.~G. and {Manteiga}, M. and {Marchant}, J.~M. and {Marconi}, M. and {Marcos}, J. and {Marcos Santos}, M.~M.~S. and {Mar{\'\i}n Pina}, D. and {Marinoni}, S. and {Marocco}, F. and {Marshall}, D.~J. and {Martin Polo}, L. and {Mart{\'\i}n-Fleitas}, J.~M. and {Marton}, G. and {Mary}, N. and {Masip}, A. and {Massari}, D. and {Mastrobuono-Battisti}, A. and {Mazeh}, T. and {McMillan}, P.~J. and {Messina}, S. and {Michalik}, D. and {Millar}, N.~R. and {Mints}, A. and {Molina}, D. and {Molinaro}, R. and {Moln{\'a}r}, L. and {Monari}, G. and {Mongui{\'o}}, M. and {Montegriffo}, P. and {Montero}, A. and {Mor}, R. and {Mora}, A. and {Morbidelli}, R. and {Morel}, T. and {Morris}, D. and {Muraveva}, T. and {Murphy}, C.~P. and {Musella}, I. and {Nagy}, Z. and {Noval}, L. and {Oca{\~n}a}, F. and {Ogden}, A. and {Ordenovic}, C. and {Osinde}, J.~O. and {Pagani}, C. and {Pagano}, I. and {Palaversa}, L. and {Palicio}, P.~A. and {Pallas-Quintela}, L. and {Panahi}, A. and {Payne-Wardenaar}, S. and {Pe{\~n}alosa Esteller}, X. and {Penttil{\"a}}, A. and {Pichon}, B. and {Piersimoni}, A.~M. and {Pineau}, F. -X. and {Plachy}, E. and {Plum}, G. and {Poggio}, E. and {Pr{\v{s}}a}, A. and {Pulone}, L. and {Racero}, E. and {Ragaini}, S. and {Rainer}, M. and {Raiteri}, C.~M. and {Rambaux}, N. and {Ramos}, P. and {Ramos-Lerate}, M. and {Re Fiorentin}, P. and {Regibo}, S. and {Richards}, P.~J. and {Rios Diaz}, C. and {Ripepi}, V. and {Riva}, A. and {Rix}, H. -W. and {Rixon}, G. and {Robichon}, N. and {Robin}, A.~C. and {Robin}, C. and {Roelens}, M. and {Rogues}, H.~R.~O. and {Rohrbasser}, L. and {Romero-G{\'o}mez}, M. and {Rowell}, N. and {Royer}, F. and {Ruz Mieres}, D. and {Rybicki}, K.~A. and {Sadowski}, G. and {S{\'a}ez N{\'u}{\~n}ez}, A. and {Sagrist{\`a} Sell{\'e}s}, A. and {Sahlmann}, J. and {Salguero}, E. and {Samaras}, N. and {Sanchez Gimenez}, V. and {Sanna}, N. and {Santove{\~n}a}, R. and {Sarasso}, M. and {Schultheis}, M. and {Sciacca}, E. and {Segol}, M. and {Segovia}, J.~C. and {S{\'e}gransan}, D. and {Semeux}, D. and {Shahaf}, S. and {Siddiqui}, H.~I. and {Siebert}, A. and {Siltala}, L. and {Silvelo}, A. and {Slezak}, E. and {Slezak}, I. and {Smart}, R.~L. and {Snaith}, O.~N. and {Solano}, E. and {Solitro}, F. and {Souami}, D. and {Souchay}, J. and {Spagna}, A. and {Spina}, L. and {Spoto}, F. and {Steele}, I.~A. and {Steidelm{\"u}ller}, H. and {Stephenson}, C.~A. and {S{\"u}veges}, M. and {Surdej}, J. and {Szabados}, L. and {Szegedi-Elek}, E. and {Taris}, F. and {Taylor}, M.~B. and {Teixeira}, R. and {Tolomei}, L. and {Tonello}, N. and {Torra}, F. and {Torra}, J. and {Torralba Elipe}, G. and {Trabucchi}, M. and {Tsounis}, A.~T. and {Turon}, C. and {Ulla}, A. and {Unger}, N. and {Vaillant}, M.~V. and {van Dillen}, E. and {van Reeven}, W. and {Vanel}, O. and {Vecchiato}, A. and {Viala}, Y. and {Vicente}, D. and {Voutsinas}, S. and {Weiler}, M. and {Wevers}, T. and {Wyrzykowski}, {\L}. and {Yoldas}, A. and {Yvard}, P. and {Zhao}, H. and {Zorec}, J. and {Zucker}, S. and {Zwitter}, T.},
  journal       = {\aap},
  title         = {{Gaia Data Release 3. Summary of the content and survey properties}},
  year          = {2023},
  month         = jun,
  pages         = {A1},
  volume        = {674},
  adsurl        = {https://ui.adsabs.harvard.edu/abs/2023A&A...674A...1G},
  archiveprefix = {arXiv},
  doi           = {10.1051/0004-6361/202243940},
  eid           = {A1},
  eprint        = {2208.00211},
  primaryclass  = {astro-ph.GA},
}

@Article{Charbonneau2007,
  author        = {{Charbonneau}, David and {Deming}, Drake},
  title         = {{The Dynamics-Based Approach to Studying Terrestrial Exoplanets}},
  journal       = {arXiv e-prints},
  year          = {2007},
  pages         = {arXiv:0706.1047},
  month         = jun,
  adsurl        = {https://ui.adsabs.harvard.edu/abs/2007arXiv0706.1047C},
  archiveprefix = {arXiv},
  eid           = {arXiv:0706.1047},
  eprint        = {0706.1047},
  primaryclass  = {astro-ph},
}

@Article{Fischer2005,
  author   = {{Fischer}, Debra A. and {Valenti}, Jeff},
  title    = {{The Planet-Metallicity Correlation}},
  journal  = {\apj},
  year     = {2005},
  volume   = {622},
  number   = {2},
  pages    = {1102-1117},
  month    = apr,
  adsurl   = {https://ui.adsabs.harvard.edu/abs/2005ApJ...622.1102F},
  doi      = {10.1086/428383},
}

@Article{Fulton2021,
  author        = {{Fulton}, Benjamin J. and {Rosenthal}, Lee J. and {Hirsch}, Lea A. and {Isaacson}, Howard and {Howard}, Andrew W. and {Dedrick}, Cayla M. and {Sherstyuk}, Ilya A. and {Blunt}, Sarah C. and {Petigura}, Erik A. and {Knutson}, Heather A. and {Behmard}, Aida and {Chontos}, Ashley and {Crepp}, Justin R. and {Crossfield}, Ian J.~M. and {Dalba}, Paul A. and {Fischer}, Debra A. and {Henry}, Gregory W. and {Kane}, Stephen R. and {Kosiarek}, Molly and {Marcy}, Geoffrey W. and {Rubenzahl}, Ryan A. and {Weiss}, Lauren M. and {Wright}, Jason T.},
  title         = {{California Legacy Survey. II. Occurrence of Giant Planets beyond the Ice Line}},
  journal       = {\apjs},
  year          = {2021},
  volume        = {255},
  number        = {1},
  pages         = {14},
  month         = jul,
  adsurl        = {https://ui.adsabs.harvard.edu/abs/2021ApJS..255...14F},
  archiveprefix = {arXiv},
  doi           = {10.3847/1538-4365/abfcc1},
  eid           = {14},
  eprint        = {2105.11584},
  primaryclass  = {astro-ph.EP},
}

@Article{Charbonneau2009,
  author        = {{Charbonneau}, David and {Berta}, Zachory K. and {Irwin}, Jonathan and {Burke}, Christopher J. and {Nutzman}, Philip and {Buchhave}, Lars A. and {Lovis}, Christophe and {Bonfils}, Xavier and {Latham}, David W. and {Udry}, St{\'e}phane and {Murray-Clay}, Ruth A. and {Holman}, Matthew J. and {Falco}, Emilio E. and {Winn}, Joshua N. and {Queloz}, Didier and {Pepe}, Francesco and {Mayor}, Michel and {Delfosse}, Xavier and {Forveille}, Thierry},
  title         = {{A super-Earth transiting a nearby low-mass star}},
  journal       = {\nat},
  year          = {2009},
  volume        = {462},
  number        = {7275},
  pages         = {891-894},
  month         = dec,
  adsurl        = {https://ui.adsabs.harvard.edu/abs/2009Natur.462..891C},
  archiveprefix = {arXiv},
  doi           = {10.1038/nature08679},
  eprint        = {0912.3229},
  primaryclass  = {astro-ph.EP},
}

@Article{AstudilloDefru2017,
  author        = {{Astudillo-Defru}, N. and {D{\'\i}az}, R.~F. and {Bonfils}, X. and {Almenara}, J.~M. and {Delisle}, J. -B. and {Bouchy}, F. and {Delfosse}, X. and {Forveille}, T. and {Lovis}, C. and {Mayor}, M. and {Murgas}, F. and {Pepe}, F. and {Santos}, N.~C. and {S{\'e}gransan}, D. and {Udry}, S. and {W{\"u}nsche}, A.},
  title         = {{The HARPS search for southern extra-solar planets. XLII. A system of Earth-mass planets around the nearby M dwarf YZ Ceti}},
  journal       = {\aap},
  year          = {2017},
  volume        = {605},
  pages         = {L11},
  month         = sep,
  adsurl        = {https://ui.adsabs.harvard.edu/abs/2017A&A...605L..11A},
  archiveprefix = {arXiv},
  doi           = {10.1051/0004-6361/201731581},
  eid           = {L11},
  eprint        = {1708.03336},
  primaryclass  = {astro-ph.EP},
}

@Article{Newton2015,
  author        = {{Newton}, Elisabeth R. and {Charbonneau}, David and {Irwin}, Jonathan and {Mann}, Andrew W.},
  title         = {{An Empirical Calibration to Estimate Cool Dwarf Fundamental Parameters from H-band Spectra}},
  journal       = {\apj},
  year          = {2015},
  volume        = {800},
  number        = {2},
  pages         = {85},
  month         = feb,
  adsurl        = {https://ui.adsabs.harvard.edu/abs/2015ApJ...800...85N},
  archiveprefix = {arXiv},
  doi           = {10.1088/0004-637X/800/2/85},
  eid           = {85},
  eprint        = {1412.2758},
  primaryclass  = {astro-ph.SR},
}

@Article{Rosenthal2022,
  author        = {{Rosenthal}, Lee J. and {Knutson}, Heather A. and {Chachan}, Yayaati and {Dai}, Fei and {Howard}, Andrew W. and {Fulton}, Benjamin J. and {Chontos}, Ashley and {Crepp}, Justin R. and {Dalba}, Paul A. and {Henry}, Gregory W. and {Kane}, Stephen R. and {Petigura}, Erik A. and {Weiss}, Lauren M. and {Wright}, Jason T.},
  journal       = {\apjs},
  title         = {{The California Legacy Survey. III. On the Shoulders of (Some) Giants: The Relationship between Inner Small Planets and Outer Massive Planets}},
  year          = {2022},
  month         = sep,
  number        = {1},
  pages         = {1},
  volume        = {262},
  adsurl        = {https://ui.adsabs.harvard.edu/abs/2022ApJS..262....1R},
  archiveprefix = {arXiv},
  doi           = {10.3847/1538-4365/ac7230},
  eid           = {1},
  eprint        = {2112.03399},
  primaryclass  = {astro-ph.EP},
}

@Article{Ment2019,
  author        = {{Ment}, Kristo and {Dittmann}, Jason A. and {Astudillo-Defru}, Nicola and {Charbonneau}, David and {Irwin}, Jonathan and {Bonfils}, Xavier and {Murgas}, Felipe and {Almenara}, Jose-Manuel and {Forveille}, Thierry and {Agol}, Eric and {Ballard}, Sarah and {Berta-Thompson}, Zachory K. and {Bouchy}, Fran{\c{c}}ois and {Cloutier}, Ryan and {Delfosse}, Xavier and {Doyon}, Ren{\'e} and {Dressing}, Courtney D. and {Esquerdo}, Gilbert A. and {Haywood}, Rapha{\"e}lle D. and {Kipping}, David M. and {Latham}, David W. and {Lovis}, Christophe and {Newton}, Elisabeth R. and {Pepe}, Francesco and {Rodriguez}, Joseph E. and {Santos}, Nuno C. and {Tan}, Thiam-Guan and {Udry}, Stephane and {Winters}, Jennifer G. and {W{\"u}nsche}, Ana{\"e}l},
  journal       = {\aj},
  title         = {{A Second Terrestrial Planet Orbiting the Nearby M Dwarf LHS 1140}},
  year          = {2019},
  month         = jan,
  number        = {1},
  pages         = {32},
  volume        = {157},
  adsurl        = {https://ui.adsabs.harvard.edu/abs/2019AJ....157...32M},
  archiveprefix = {arXiv},
  doi           = {10.3847/1538-3881/aaf1b1},
  eid           = {32},
  eprint        = {1808.00485},
  primaryclass  = {astro-ph.EP},
}

@Article{Tayar2022,
  author        = {{Tayar}, Jamie and {Claytor}, Zachary R. and {Huber}, Daniel and {van Saders}, Jennifer},
  journal       = {\apj},
  title         = {{A Guide to Realistic Uncertainties on the Fundamental Properties of Solar-type Exoplanet Host Stars}},
  year          = {2022},
  month         = mar,
  number        = {1},
  pages         = {31},
  volume        = {927},
  adsurl        = {https://ui.adsabs.harvard.edu/abs/2022ApJ...927...31T},
  archiveprefix = {arXiv},
  doi           = {10.3847/1538-4357/ac4bbc},
  eid           = {31},
  eprint        = {2012.07957},
  primaryclass  = {astro-ph.EP},
}

@InProceedings{Seifahrt2016,
  author        = {{Seifahrt}, Andreas and {Bean}, Jacob L. and {St{\"u}rmer}, Julian and {Gers}, Luke and {Grobler}, Deon S. and {Reed}, Tony and {Jones}, Damien J.},
  booktitle     = {Ground-based and Airborne Instrumentation for Astronomy VI},
  title         = {{Development and construction of MAROON-X}},
  year          = {2016},
  editor        = {{Evans}, Christopher J. and {Simard}, Luc and {Takami}, Hideki},
  month         = aug,
  pages         = {990818},
  series        = {Society of Photo-Optical Instrumentation Engineers (SPIE) Conference Series},
  volume        = {9908},
  adsurl        = {https://ui.adsabs.harvard.edu/abs/2016SPIE.9908E..18S},
  archiveprefix = {arXiv},
  doi           = {10.1117/12.2232069},
  eid           = {990818},
  eprint        = {1606.07140},
  primaryclass  = {astro-ph.IM},
}

@InProceedings{Seifahrt2020,
  author        = {{Seifahrt}, Andreas and {Bean}, Jacob L. and {St{\"u}rmer}, Julian and {Kasper}, David and {Gers}, Luke and {Schwab}, Christian and {Zechmeister}, Mathias and {Stef{\'a}nsson}, Gudmundur and {Montet}, Ben and {Dos Santos}, Leonardo A. and {Peck}, Alison and {White}, John and {Tapia}, Eduardo},
  booktitle     = {Ground-based and Airborne Instrumentation for Astronomy VIII},
  title         = {{On-sky commissioning of MAROON-X: a new precision radial velocity spectrograph for Gemini North}},
  year          = {2020},
  editor        = {{Evans}, Christopher J. and {Bryant}, Julia J. and {Motohara}, Kentaro},
  month         = dec,
  pages         = {114471F},
  series        = {Society of Photo-Optical Instrumentation Engineers (SPIE) Conference Series},
  volume        = {11447},
  adsurl        = {https://ui.adsabs.harvard.edu/abs/2020SPIE11447E..1FS},
  archiveprefix = {arXiv},
  doi           = {10.1117/12.2561564},
  eid           = {114471F},
  eprint        = {2106.02157},
  primaryclass  = {astro-ph.IM},
}

@Article{Pass2023_inactive,
  author        = {{Pass}, Emily K. and {Winters}, Jennifer G. and {Charbonneau}, David and {Irwin}, Jonathan M. and {Latham}, David W. and {Berlind}, Perry and {Calkins}, Michael L. and {Esquerdo}, Gilbert A. and {Mink}, Jessica},
  journal       = {\aj},
  title         = {{Mid-to-late M Dwarfs Lack Jupiter Analogs}},
  year          = {2023},
  month         = jul,
  number        = {1},
  pages         = {11},
  volume        = {166},
  adsurl        = {https://ui.adsabs.harvard.edu/abs/2023AJ....166...11P},
  archiveprefix = {arXiv},
  doi           = {10.3847/1538-3881/acd349},
  eid           = {11},
  eprint        = {2305.19357},
  primaryclass  = {astro-ph.EP},
}

@Article{ForemanMackey2021,
  author        = {{Foreman-Mackey}, Daniel and {Luger}, Rodrigo and {Agol}, Eric and {Barclay}, Thomas and {Bouma}, Luke and {Brandt}, Timothy and {Czekala}, Ian and {David}, Trevor and {Dong}, Jiayin and {Gilbert}, Emily and {Gordon}, Tyler and {Hedges}, Christina and {Hey}, Daniel and {Morris}, Brett and {Price-Whelan}, Adrian and {Savel}, Arjun},
  journal       = {JOSS},
  title         = {{exoplanet: Gradient-based probabilistic inference for exoplanet data \& other astronomical time series}},
  year          = {2021},
  month         = jun,
  number        = {62},
  pages         = {3285},
  volume        = {6},
  adsurl        = {https://ui.adsabs.harvard.edu/abs/2021JOSS....6.3285F},
  archiveprefix = {arXiv},
  doi           = {10.21105/joss.03285},
  eid           = {3285},
  eprint        = {2105.01994},
  primaryclass  = {astro-ph.IM},
}

@Article{Desidera2006,
  author   = {{Desidera}, S. and {Gratton}, R.~G. and {Lucatello}, S. and {Claudi}, R.~U.},
  journal  = {\aap},
  title    = {{Abundance difference between components of wide binaries. II. The southern sample}},
  year     = {2006},
  month    = aug,
  number   = {2},
  pages    = {581-593},
  volume   = {454},
  adsurl   = {https://ui.adsabs.harvard.edu/abs/2006A&A...454..581D},
  doi      = {10.1051/0004-6361:20064896},
}

@Article{Desidera2004,
  author        = {{Desidera}, S. and {Gratton}, R.~G. and {Scuderi}, S. and {Claudi}, R.~U. and {Cosentino}, R. and {Barbieri}, M. and {Bonanno}, G. and {Carretta}, E. and {Endl}, M. and {Lucatello}, S. and {Martinez Fiorenzano}, A.~F. and {Marzari}, F.},
  journal       = {\aap},
  title         = {{Abundance difference between components of wide binaries}},
  year          = {2004},
  month         = jun,
  pages         = {683-697},
  volume        = {420},
  adsurl        = {https://ui.adsabs.harvard.edu/abs/2004A&A...420..683D},
  archiveprefix = {arXiv},
  doi           = {10.1051/0004-6361:20041242},
  eprint        = {astro-ph/0403051},
  primaryclass  = {astro-ph},
}

@Article{Skrutskie2006,
  author   = {{Skrutskie}, M.~F. and {Cutri}, R.~M. and {Stiening}, R. and {Weinberg}, M.~D. and {Schneider}, S. and {Carpenter}, J.~M. and {Beichman}, C. and {Capps}, R. and {Chester}, T. and {Elias}, J. and {Huchra}, J. and {Liebert}, J. and {Lonsdale}, C. and {Monet}, D.~G. and {Price}, S. and {Seitzer}, P. and {Jarrett}, T. and {Kirkpatrick}, J.~D. and {Gizis}, J.~E. and {Howard}, E. and {Evans}, T. and {Fowler}, J. and {Fullmer}, L. and {Hurt}, R. and {Light}, R. and {Kopan}, E.~L. and {Marsh}, K.~A. and {McCallon}, H.~L. and {Tam}, R. and {Van Dyk}, S. and {Wheelock}, S.},
  title    = {{The Two Micron All Sky Survey (2MASS)}},
  journal  = {\aj},
  year     = {2006},
  volume   = {131},
  number   = {2},
  pages    = {1163-1183},
  month    = feb,
  adsurl   = {https://ui.adsabs.harvard.edu/abs/2006AJ....131.1163S},
  doi      = {10.1086/498708},
}

@Article{Gaidos2014_conch,
  author        = {{Gaidos}, E. and {Mann}, A.~W. and {L{\'e}pine}, S. and {Buccino}, A. and {James}, D. and {Ansdell}, M. and {Petrucci}, R. and {Mauas}, P. and {Hilton}, E.~J.},
  journal       = {\mnras},
  title         = {{Trumpeting M dwarfs with CONCH-SHELL: a catalogue of nearby cool host-stars for habitable exoplanets and life}},
  year          = {2014},
  month         = sep,
  number        = {3},
  pages         = {2561-2578},
  volume        = {443},
  adsurl        = {https://ui.adsabs.harvard.edu/abs/2014MNRAS.443.2561G},
  archiveprefix = {arXiv},
  doi           = {10.1093/mnras/stu1313},
  eprint        = {1406.7353},
  primaryclass  = {astro-ph.SR},
}

@Article{El-Badry2021,
  author        = {{El-Badry}, Kareem and {Rix}, Hans-Walter and {Heintz}, Tyler M.},
  title         = {{A million binaries from Gaia eDR3: sample selection and validation of Gaia parallax uncertainties}},
  journal       = {\mnras},
  year          = {2021},
  volume        = {506},
  number        = {2},
  pages         = {2269-2295},
  month         = sep,
  adsurl        = {https://ui.adsabs.harvard.edu/abs/2021MNRAS.506.2269E},
  archiveprefix = {arXiv},
  doi           = {10.1093/mnras/stab323},
  eprint        = {2101.05282},
  primaryclass  = {astro-ph.SR},
}

@Article{Harris2020,
  author    = {Charles R. Harris and K. Jarrod Millman and St{\'{e}}fan J. van der Walt and Ralf Gommers and Pauli Virtanen and David Cournapeau and Eric Wieser and Julian Taylor and Sebastian Berg and Nathaniel J. Smith and Robert Kern and Matti Picus and Stephan Hoyer and Marten H. van Kerkwijk and Matthew Brett and Allan Haldane and Jaime Fern{\'{a}}ndez del R{\'{i}}o and Mark Wiebe and Pearu Peterson and Pierre G{\'{e}}rard-Marchant and Kevin Sheppard and Tyler Reddy and Warren Weckesser and Hameer Abbasi and Christoph Gohlke and Travis E. Oliphant},
  journal   = {Nature},
  title     = {Array programming with {NumPy}},
  year      = {2020},
  month     = sep,
  number    = {7825},
  pages     = {357--362},
  volume    = {585},
  adsurl    = {https://ui.adsabs.harvard.edu/abs/2020Natur.585..357H},
  doi       = {10.1038/s41586-020-2649-2},
  publisher = {Springer Science and Business Media {LLC}},
  url       = {https://doi.org/10.1038/s41586-020-2649-2},
}

@Article{Fouque2018,
  author        = {{Fouqu{\'e}}, Pascal and {Moutou}, Claire and {Malo}, Lison and {Martioli}, Eder and {Lim}, Olivia and {Rajpurohit}, Arvind and {Artigau}, Etienne and {Delfosse}, Xavier and {Donati}, Jean-Fran{\c{c}}ois and {Forveille}, Thierry and {Morin}, Julien and {Allard}, France and {Delage}, Rapha{\"e}l and {Doyon}, Ren{\'e} and {H{\'e}brard}, Elodie and {Neves}, Vasco},
  title         = {{SPIRou Input Catalogue: global properties of 440 M dwarfs observed with ESPaDOnS at CFHT}},
  journal       = {\mnras},
  year          = {2018},
  volume        = {475},
  number        = {2},
  pages         = {1960-1986},
  month         = apr,
  adsurl        = {https://ui.adsabs.harvard.edu/abs/2018MNRAS.475.1960F},
  archiveprefix = {arXiv},
  doi           = {10.1093/mnras/stx3246},
  eprint        = {1712.04490},
  primaryclass  = {astro-ph.SR},
}

@Article{Ricker2015,
  author  = {{Ricker}, George R. and {Winn}, Joshua N. and {Vanderspek}, Roland and {Latham}, David W. and {Bakos}, G{\'a}sp{\'a}r {\'A}. and {Bean}, Jacob L. and {Berta-Thompson}, Zachory K. and {Brown}, Timothy M. and {Buchhave}, Lars and {Butler}, Nathaniel R. and {Butler}, R. Paul and {Chaplin}, William J. and {Charbonneau}, David and {Christensen-Dalsgaard}, J{\o}rgen and {Clampin}, Mark and {Deming}, Drake and {Doty}, John and {De Lee}, Nathan and {Dressing}, Courtney and {Dunham}, Edward W. and {Endl}, Michael and {Fressin}, Francois and {Ge}, Jian and {Henning}, Thomas and {Holman}, Matthew J. and {Howard}, Andrew W. and {Ida}, Shigeru and {Jenkins}, Jon M. and {Jernigan}, Garrett and {Johnson}, John Asher and {Kaltenegger}, Lisa and {Kawai}, Nobuyuki and {Kjeldsen}, Hans and {Laughlin}, Gregory and {Levine}, Alan M. and {Lin}, Douglas and {Lissauer}, Jack J. and {MacQueen}, Phillip and {Marcy}, Geoffrey and {McCullough}, Peter R. and {Morton}, Timothy D. and {Narita}, Norio and {Paegert}, Martin and {Palle}, Enric and {Pepe}, Francesco and {Pepper}, Joshua and {Quirrenbach}, Andreas and {Rinehart}, Stephen A. and {Sasselov}, Dimitar and {Sato}, Bun'ei and {Seager}, Sara and {Sozzetti}, Alessandro and {Stassun}, Keivan G. and {Sullivan}, Peter and {Szentgyorgyi}, Andrew and {Torres}, Guillermo and {Udry}, Stephane and {Villasenor}, Joel},
  journal = {JATIS},
  title   = {{Transiting Exoplanet Survey Satellite (TESS)}},
  year    = {2015},
  month   = jan,
  pages   = {014003},
  volume  = {1},
  adsurl  = {https://ui.adsabs.harvard.edu/abs/2015JATIS...1a4003R},
  doi     = {10.1117/1.JATIS.1.1.014003},
  eid     = {014003},
}

@Article{Tokovinin2013,
  author    = {Andrei Tokovinin and Debra A. Fischer and Marco Bonati and Matthew J. Giguere and Peter Moore and Christian Schwab and Julien F. P. Spronck and Andrew Szymkowiak},
  journal   = {PASP},
  title     = {{CHIRON}{\textemdash}A Fiber Fed Spectrometer for Precise Radial Velocities},
  year      = {2013},
  month     = {nov},
  number    = {933},
  pages     = {1336--1347},
  volume    = {125},
  adsurl    = {https://ui.adsabs.harvard.edu/abs/2013PASP..125.1336T},
  doi       = {10.1086/674012},
  publisher = {{IOP} Publishing},
}

@Article{Pass2022,
  author        = {{Pass}, Emily K. and {Charbonneau}, David and {Irwin}, Jonathan M. and {Winters}, Jennifer G.},
  journal       = {\apj},
  title         = {{Constraints on the Spindown of Fully Convective M Dwarfs Using Wide Field Binaries}},
  year          = {2022},
  month         = sep,
  number        = {2},
  pages         = {109},
  volume        = {936},
  adsurl        = {https://ui.adsabs.harvard.edu/abs/2022ApJ...936..109P},
  archiveprefix = {arXiv},
  doi           = {10.3847/1538-4357/ac7da8},
  eid           = {109},
  eprint        = {2206.15318},
  primaryclass  = {astro-ph.SR},
}

@Article{Fuhrmann2014,
  author   = {{Fuhrmann}, K. and {Chini}, R. and {Buda}, L. -S. and {Pozo Nu{\~n}ez}, F.},
  journal  = {\apj},
  title    = {{On the Age of Gliese 86}},
  year     = {2014},
  month    = apr,
  number   = {1},
  pages    = {68},
  volume   = {785},
  adsurl   = {https://ui.adsabs.harvard.edu/abs/2014ApJ...785...68F},
  doi      = {10.1088/0004-637X/785/1/68},
  eid      = {68},
}

@Article{Cutri2003,
  author   = {{Cutri}, R.~M. and {Skrutskie}, M.~F. and {van Dyk}, S. and {Beichman}, C.~A. and {Carpenter}, J.~M. and {Chester}, T. and {Cambresy}, L. and {Evans}, T. and {Fowler}, J. and {Gizis}, J. and {Howard}, E. and {Huchra}, J. and {Jarrett}, T. and {Kopan}, E.~L. and {Kirkpatrick}, J.~D. and {Light}, R.~M. and {Marsh}, K.~A. and {McCallon}, H. and {Schneider}, S. and {Stiening}, R. and {Sykes}, M. and {Weinberg}, M. and {Wheaton}, W.~A. and {Wheelock}, S. and {Zacarias}, N.},
  journal  = {VizieR Online Data Catalog},
  title    = {{VizieR Online Data Catalog: 2MASS All-Sky Catalog of Point Sources (Cutri+ 2003)}},
  year     = {2003},
  month    = jun,
  pages    = {II/246},
  adsurl   = {https://ui.adsabs.harvard.edu/abs/2003yCat.2246....0C},
  eid      = {II/246},
}

@Article{Covey2016,
  author        = {{Covey}, Kevin R. and {Ag{\"u}eros}, Marcel A. and {Law}, Nicholas M. and {Liu}, Jiyu and {Ahmadi}, Aida and {Laher}, Russ and {Levitan}, David and {Sesar}, Branimir and {Surace}, Jason},
  journal       = {\apj},
  title         = {{Why Are Rapidly Rotating M Dwarfs in the Pleiades so (Infra)red? New Period Measurements Confirm Rotation-dependent Color Offsets From the Cluster Sequence}},
  year          = {2016},
  month         = may,
  number        = {2},
  pages         = {81},
  volume        = {822},
  adsurl        = {https://ui.adsabs.harvard.edu/abs/2016ApJ...822...81C},
  archiveprefix = {arXiv},
  doi           = {10.3847/0004-637X/822/2/81},
  eid           = {81},
  eprint        = {1601.07237},
  primaryclass  = {astro-ph.SR},
}

@Article{Zucker1994,
  author   = {{Zucker}, S. and {Mazeh}, T.},
  journal  = {\apj},
  title    = {{Study of Spectroscopic Binaries with TODCOR. I. A New Two-dimensional Correlation Algorithm to Derive the Radial Velocities of the Two Components}},
  year     = {1994},
  month    = jan,
  pages    = {806},
  volume   = {420},
  adsurl   = {https://ui.adsabs.harvard.edu/abs/1994ApJ...420..806Z},
  doi      = {10.1086/173605},
}

@Article{Gaia2016,
  author        = {{Gaia Collaboration} and {Prusti}, T. and {de Bruijne}, J.~H.~J. and {Brown}, A.~G.~A. and {Vallenari}, A. and {Babusiaux}, C. and {Bailer-Jones}, C.~A.~L. and {Bastian}, U. and {Biermann}, M. and {Evans}, D.~W. and {Eyer}, L. and {Jansen}, F. and {Jordi}, C. and {Klioner}, S.~A. and {Lammers}, U. and {Lindegren}, L. and {Luri}, X. and {Mignard}, F. and {Milligan}, D.~J. and {Panem}, C. and {Poinsignon}, V. and {Pourbaix}, D. and {Randich}, S. and {Sarri}, G. and {Sartoretti}, P. and {Siddiqui}, H.~I. and {Soubiran}, C. and {Valette}, V. and {van Leeuwen}, F. and {Walton}, N.~A. and {Aerts}, C. and {Arenou}, F. and {Cropper}, M. and {Drimmel}, R. and {H{\o}g}, E. and {Katz}, D. and {Lattanzi}, M.~G. and {O'Mullane}, W. and {Grebel}, E.~K. and {Holland}, A.~D. and {Huc}, C. and {Passot}, X. and {Bramante}, L. and {Cacciari}, C. and {Casta{\~n}eda}, J. and {Chaoul}, L. and {Cheek}, N. and {De Angeli}, F. and {Fabricius}, C. and {Guerra}, R. and {Hern{\'a}ndez}, J. and {Jean-Antoine-Piccolo}, A. and {Masana}, E. and {Messineo}, R. and {Mowlavi}, N. and {Nienartowicz}, K. and {Ord{\'o}{\~n}ez-Blanco}, D. and {Panuzzo}, P. and {Portell}, J. and {Richards}, P.~J. and {Riello}, M. and {Seabroke}, G.~M. and {Tanga}, P. and {Th{\'e}venin}, F. and {Torra}, J. and {Els}, S.~G. and {Gracia-Abril}, G. and {Comoretto}, G. and {Garcia-Reinaldos}, M. and {Lock}, T. and {Mercier}, E. and {Altmann}, M. and {Andrae}, R. and {Astraatmadja}, T.~L. and {Bellas-Velidis}, I. and {Benson}, K. and {Berthier}, J. and {Blomme}, R. and {Busso}, G. and {Carry}, B. and {Cellino}, A. and {Clementini}, G. and {Cowell}, S. and {Creevey}, O. and {Cuypers}, J. and {Davidson}, M. and {De Ridder}, J. and {de Torres}, A. and {Delchambre}, L. and {Dell'Oro}, A. and {Ducourant}, C. and {Fr{\'e}mat}, Y. and {Garc{\'\i}a-Torres}, M. and {Gosset}, E. and {Halbwachs}, J. -L. and {Hambly}, N.~C. and {Harrison}, D.~L. and {Hauser}, M. and {Hestroffer}, D. and {Hodgkin}, S.~T. and {Huckle}, H.~E. and {Hutton}, A. and {Jasniewicz}, G. and {Jordan}, S. and {Kontizas}, M. and {Korn}, A.~J. and {Lanzafame}, A.~C. and {Manteiga}, M. and {Moitinho}, A. and {Muinonen}, K. and {Osinde}, J. and {Pancino}, E. and {Pauwels}, T. and {Petit}, J. -M. and {Recio-Blanco}, A. and {Robin}, A.~C. and {Sarro}, L.~M. and {Siopis}, C. and {Smith}, M. and {Smith}, K.~W. and {Sozzetti}, A. and {Thuillot}, W. and {van Reeven}, W. and {Viala}, Y. and {Abbas}, U. and {Abreu Aramburu}, A. and {Accart}, S. and {Aguado}, J.~J. and {Allan}, P.~M. and {Allasia}, W. and {Altavilla}, G. and {{\'A}lvarez}, M.~A. and {Alves}, J. and {Anderson}, R.~I. and {Andrei}, A.~H. and {Anglada Varela}, E. and {Antiche}, E. and {Antoja}, T. and {Ant{\'o}n}, S. and {Arcay}, B. and {Atzei}, A. and {Ayache}, L. and {Bach}, N. and {Baker}, S.~G. and {Balaguer-N{\'u}{\~n}ez}, L. and {Barache}, C. and {Barata}, C. and {Barbier}, A. and {Barblan}, F. and {Baroni}, M. and {Barrado y Navascu{\'e}s}, D. and {Barros}, M. and {Barstow}, M.~A. and {Becciani}, U. and {Bellazzini}, M. and {Bellei}, G. and {Bello Garc{\'\i}a}, A. and {Belokurov}, V. and {Bendjoya}, P. and {Berihuete}, A. and {Bianchi}, L. and {Bienaym{\'e}}, O. and {Billebaud}, F. and {Blagorodnova}, N. and {Blanco-Cuaresma}, S. and {Boch}, T. and {Bombrun}, A. and {Borrachero}, R. and {Bouquillon}, S. and {Bourda}, G. and {Bouy}, H. and {Bragaglia}, A. and {Breddels}, M.~A. and {Brouillet}, N. and {Br{\"u}semeister}, T. and {Bucciarelli}, B. and {Budnik}, F. and {Burgess}, P. and {Burgon}, R. and {Burlacu}, A. and {Busonero}, D. and {Buzzi}, R. and {Caffau}, E. and {Cambras}, J. and {Campbell}, H. and {Cancelliere}, R. and {Cantat-Gaudin}, T. and {Carlucci}, T. and {Carrasco}, J.~M. and {Castellani}, M. and {Charlot}, P. and {Charnas}, J. and {Charvet}, P. and {Chassat}, F. and {Chiavassa}, A. and {Clotet}, M. and {Cocozza}, G. and {Collins}, R.~S. and {Collins}, P. and {Costigan}, G. and {Crifo}, F. and {Cross}, N.~J.~G. and {Crosta}, M. and {Crowley}, C. and {Dafonte}, C. and {Damerdji}, Y. and {Dapergolas}, A. and {David}, P. and {David}, M. and {De Cat}, P. and {de Felice}, F. and {de Laverny}, P. and {De Luise}, F. and {De March}, R. and {de Martino}, D. and {de Souza}, R. and {Debosscher}, J. and {del Pozo}, E. and {Delbo}, M. and {Delgado}, A. and {Delgado}, H.~E. and {di Marco}, F. and {Di Matteo}, P. and {Diakite}, S. and {Distefano}, E. and {Dolding}, C. and {Dos Anjos}, S. and {Drazinos}, P. and {Dur{\'a}n}, J. and {Dzigan}, Y. and {Ecale}, E. and {Edvardsson}, B. and {Enke}, H. and {Erdmann}, M. and {Escolar}, D. and {Espina}, M. and {Evans}, N.~W. and {Eynard Bontemps}, G. and {Fabre}, C. and {Fabrizio}, M. and {Faigler}, S. and {Falc{\~a}o}, A.~J. and {Farr{\`a}s Casas}, M. and {Faye}, F. and {Federici}, L. and {Fedorets}, G. and {Fern{\'a}ndez-Hern{\'a}ndez}, J. and {Fernique}, P. and {Fienga}, A. and {Figueras}, F. and {Filippi}, F. and {Findeisen}, K. and {Fonti}, A. and {Fouesneau}, M. and {Fraile}, E. and {Fraser}, M. and {Fuchs}, J. and {Furnell}, R. and {Gai}, M. and {Galleti}, S. and {Galluccio}, L. and {Garabato}, D. and {Garc{\'\i}a-Sedano}, F. and {Gar{\'e}}, P. and {Garofalo}, A. and {Garralda}, N. and {Gavras}, P. and {Gerssen}, J. and {Geyer}, R. and {Gilmore}, G. and {Girona}, S. and {Giuffrida}, G. and {Gomes}, M. and {Gonz{\'a}lez-Marcos}, A. and {Gonz{\'a}lez-N{\'u}{\~n}ez}, J. and {Gonz{\'a}lez-Vidal}, J.~J. and {Granvik}, M. and {Guerrier}, A. and {Guillout}, P. and {Guiraud}, J. and {G{\'u}rpide}, A. and {Guti{\'e}rrez-S{\'a}nchez}, R. and {Guy}, L.~P. and {Haigron}, R. and {Hatzidimitriou}, D. and {Haywood}, M. and {Heiter}, U. and {Helmi}, A. and {Hobbs}, D. and {Hofmann}, W. and {Holl}, B. and {Holland}, G. and {Hunt}, J.~A.~S. and {Hypki}, A. and {Icardi}, V. and {Irwin}, M. and {Jevardat de Fombelle}, G. and {Jofr{\'e}}, P. and {Jonker}, P.~G. and {Jorissen}, A. and {Julbe}, F. and {Karampelas}, A. and {Kochoska}, A. and {Kohley}, R. and {Kolenberg}, K. and {Kontizas}, E. and {Koposov}, S.~E. and {Kordopatis}, G. and {Koubsky}, P. and {Kowalczyk}, A. and {Krone-Martins}, A. and {Kudryashova}, M. and {Kull}, I. and {Bachchan}, R.~K. and {Lacoste-Seris}, F. and {Lanza}, A.~F. and {Lavigne}, J. -B. and {Le Poncin-Lafitte}, C. and {Lebreton}, Y. and {Lebzelter}, T. and {Leccia}, S. and {Leclerc}, N. and {Lecoeur-Taibi}, I. and {Lemaitre}, V. and {Lenhardt}, H. and {Leroux}, F. and {Liao}, S. and {Licata}, E. and {Lindstr{\o}m}, H.~E.~P. and {Lister}, T.~A. and {Livanou}, E. and {Lobel}, A. and {L{\"o}ffler}, W. and {L{\'o}pez}, M. and {Lopez-Lozano}, A. and {Lorenz}, D. and {Loureiro}, T. and {MacDonald}, I. and {Magalh{\~a}es Fernandes}, T. and {Managau}, S. and {Mann}, R.~G. and {Mantelet}, G. and {Marchal}, O. and {Marchant}, J.~M. and {Marconi}, M. and {Marie}, J. and {Marinoni}, S. and {Marrese}, P.~M. and {Marschalk{\'o}}, G. and {Marshall}, D.~J. and {Mart{\'\i}n-Fleitas}, J.~M. and {Martino}, M. and {Mary}, N. and {Matijevi{\v{c}}}, G. and {Mazeh}, T. and {McMillan}, P.~J. and {Messina}, S. and {Mestre}, A. and {Michalik}, D. and {Millar}, N.~R. and {Miranda}, B.~M.~H. and {Molina}, D. and {Molinaro}, R. and {Molinaro}, M. and {Moln{\'a}r}, L. and {Moniez}, M. and {Montegriffo}, P. and {Monteiro}, D. and {Mor}, R. and {Mora}, A. and {Morbidelli}, R. and {Morel}, T. and {Morgenthaler}, S. and {Morley}, T. and {Morris}, D. and {Mulone}, A.~F. and {Muraveva}, T. and {Musella}, I. and {Narbonne}, J. and {Nelemans}, G. and {Nicastro}, L. and {Noval}, L. and {Ord{\'e}novic}, C. and {Ordieres-Mer{\'e}}, J. and {Osborne}, P. and {Pagani}, C. and {Pagano}, I. and {Pailler}, F. and {Palacin}, H. and {Palaversa}, L. and {Parsons}, P. and {Paulsen}, T. and {Pecoraro}, M. and {Pedrosa}, R. and {Pentik{\"a}inen}, H. and {Pereira}, J. and {Pichon}, B. and {Piersimoni}, A.~M. and {Pineau}, F. -X. and {Plachy}, E. and {Plum}, G. and {Poujoulet}, E. and {Pr{\v{s}}a}, A. and {Pulone}, L. and {Ragaini}, S. and {Rago}, S. and {Rambaux}, N. and {Ramos-Lerate}, M. and {Ranalli}, P. and {Rauw}, G. and {Read}, A. and {Regibo}, S. and {Renk}, F. and {Reyl{\'e}}, C. and {Ribeiro}, R.~A. and {Rimoldini}, L. and {Ripepi}, V. and {Riva}, A. and {Rixon}, G. and {Roelens}, M. and {Romero-G{\'o}mez}, M. and {Rowell}, N. and {Royer}, F. and {Rudolph}, A. and {Ruiz-Dern}, L. and {Sadowski}, G. and {Sagrist{\`a} Sell{\'e}s}, T. and {Sahlmann}, J. and {Salgado}, J. and {Salguero}, E. and {Sarasso}, M. and {Savietto}, H. and {Schnorhk}, A. and {Schultheis}, M. and {Sciacca}, E. and {Segol}, M. and {Segovia}, J.~C. and {Segransan}, D. and {Serpell}, E. and {Shih}, I. -C. and {Smareglia}, R. and {Smart}, R.~L. and {Smith}, C. and {Solano}, E. and {Solitro}, F. and {Sordo}, R. and {Soria Nieto}, S. and {Souchay}, J. and {Spagna}, A. and {Spoto}, F. and {Stampa}, U. and {Steele}, I.~A. and {Steidelm{\"u}ller}, H. and {Stephenson}, C.~A. and {Stoev}, H. and {Suess}, F.~F. and {S{\"u}veges}, M. and {Surdej}, J. and {Szabados}, L. and {Szegedi-Elek}, E. and {Tapiador}, D. and {Taris}, F. and {Tauran}, G. and {Taylor}, M.~B. and {Teixeira}, R. and {Terrett}, D. and {Tingley}, B. and {Trager}, S.~C. and {Turon}, C. and {Ulla}, A. and {Utrilla}, E. and {Valentini}, G. and {van Elteren}, A. and {Van Hemelryck}, E. and {van Leeuwen}, M. and {Varadi}, M. and {Vecchiato}, A. and {Veljanoski}, J. and {Via}, T. and {Vicente}, D. and {Vogt}, S. and {Voss}, H. and {Votruba}, V. and {Voutsinas}, S. and {Walmsley}, G. and {Weiler}, M. and {Weingrill}, K. and {Werner}, D. and {Wevers}, T. and {Whitehead}, G. and {Wyrzykowski}, {\L}. and {Yoldas}, A. and {{\v{Z}}erjal}, M. and {Zucker}, S. and {Zurbach}, C. and {Zwitter}, T. and {Alecu}, A. and {Allen}, M. and {Allende Prieto}, C. and {Amorim}, A. and {Anglada-Escud{\'e}}, G. and {Arsenijevic}, V. and {Azaz}, S. and {Balm}, P. and {Beck}, M. and {Bernstein}, H. -H. and {Bigot}, L. and {Bijaoui}, A. and {Blasco}, C. and {Bonfigli}, M. and {Bono}, G. and {Boudreault}, S. and {Bressan}, A. and {Brown}, S. and {Brunet}, P. -M. and {Bunclark}, P. and {Buonanno}, R. and {Butkevich}, A.~G. and {Carret}, C. and {Carrion}, C. and {Chemin}, L. and {Ch{\'e}reau}, F. and {Corcione}, L. and {Darmigny}, E. and {de Boer}, K.~S. and {de Teodoro}, P. and {de Zeeuw}, P.~T. and {Delle Luche}, C. and {Domingues}, C.~D. and {Dubath}, P. and {Fodor}, F. and {Fr{\'e}zouls}, B. and {Fries}, A. and {Fustes}, D. and {Fyfe}, D. and {Gallardo}, E. and {Gallegos}, J. and {Gardiol}, D. and {Gebran}, M. and {Gomboc}, A. and {G{\'o}mez}, A. and {Grux}, E. and {Gueguen}, A. and {Heyrovsky}, A. and {Hoar}, J. and {Iannicola}, G. and {Isasi Parache}, Y. and {Janotto}, A. -M. and {Joliet}, E. and {Jonckheere}, A. and {Keil}, R. and {Kim}, D. -W. and {Klagyivik}, P. and {Klar}, J. and {Knude}, J. and {Kochukhov}, O. and {Kolka}, I. and {Kos}, J. and {Kutka}, A. and {Lainey}, V. and {LeBouquin}, D. and {Liu}, C. and {Loreggia}, D. and {Makarov}, V.~V. and {Marseille}, M.~G. and {Martayan}, C. and {Martinez-Rubi}, O. and {Massart}, B. and {Meynadier}, F. and {Mignot}, S. and {Munari}, U. and {Nguyen}, A. -T. and {Nordlander}, T. and {Ocvirk}, P. and {O'Flaherty}, K.~S. and {Olias Sanz}, A. and {Ortiz}, P. and {Osorio}, J. and {Oszkiewicz}, D. and {Ouzounis}, A. and {Palmer}, M. and {Park}, P. and {Pasquato}, E. and {Peltzer}, C. and {Peralta}, J. and {P{\'e}turaud}, F. and {Pieniluoma}, T. and {Pigozzi}, E. and {Poels}, J. and {Prat}, G. and {Prod'homme}, T. and {Raison}, F. and {Rebordao}, J.~M. and {Risquez}, D. and {Rocca-Volmerange}, B. and {Rosen}, S. and {Ruiz-Fuertes}, M.~I. and {Russo}, F. and {Sembay}, S. and {Serraller Vizcaino}, I. and {Short}, A. and {Siebert}, A. and {Silva}, H. and {Sinachopoulos}, D. and {Slezak}, E. and {Soffel}, M. and {Sosnowska}, D. and {Strai{\v{z}}ys}, V. and {ter Linden}, M. and {Terrell}, D. and {Theil}, S. and {Tiede}, C. and {Troisi}, L. and {Tsalmantza}, P. and {Tur}, D. and {Vaccari}, M. and {Vachier}, F. and {Valles}, P. and {Van Hamme}, W. and {Veltz}, L. and {Virtanen}, J. and {Wallut}, J. -M. and {Wichmann}, R. and {Wilkinson}, M.~I. and {Ziaeepour}, H. and {Zschocke}, S.},
  journal       = {\aap},
  title         = {{The Gaia mission}},
  year          = {2016},
  month         = nov,
  pages         = {A1},
  volume        = {595},
  adsurl        = {https://ui.adsabs.harvard.edu/abs/2016A&A...595A...1G},
  archiveprefix = {arXiv},
  doi           = {10.1051/0004-6361/201629272},
  eid           = {A1},
  eprint        = {1609.04153},
  primaryclass  = {astro-ph.IM},
}

@Article{Riedel2014,
  author        = {{Riedel}, Adric R. and {Finch}, Charlie T. and {Henry}, Todd J. and {Subasavage}, John P. and {Jao}, Wei-Chun and {Malo}, Lison and {Rodriguez}, David R. and {White}, Russel J. and {Gies}, Douglas R. and {Dieterich}, Sergio B. and {Winters}, Jennifer G. and {Davison}, Cassy L. and {Nelan}, Edmund P. and {Blunt}, Sarah C. and {Cruz}, Kelle L. and {Rice}, Emily L. and {Ianna}, Philip A.},
  journal       = {\aj},
  title         = {{The Solar Neighborhood. XXXIII. Parallax Results from the CTIOPI 0.9 m Program: Trigonometric Parallaxes of Nearby Low-mass Active and Young Systems}},
  year          = {2014},
  month         = apr,
  number        = {4},
  pages         = {85},
  volume        = {147},
  adsurl        = {https://ui.adsabs.harvard.edu/abs/2014AJ....147...85R},
  archiveprefix = {arXiv},
  doi           = {10.1088/0004-6256/147/4/85},
  eid           = {85},
  eprint        = {1401.0722},
  primaryclass  = {astro-ph.SR},
}

@InProceedings{Szentgyorgyi2007,
  author    = {{Szentgyorgyi}, A.~H. and {Fur{\'e}sz}, G.},
  booktitle = {Revista Mexicana de Astronomia y Astrofisica Conference Series},
  title     = {{Precision Radial Velocities for the Kepler Era}},
  year      = {2007},
  editor    = {{Kurtz}, Stanley},
  month     = jun,
  pages     = {129},
  volume    = {28},
  adsurl    = {https://ui.adsabs.harvard.edu/abs/2007RMxAC..28..129S},
}

@Article{Choi2016,
  author        = {{Choi}, Jieun and {Dotter}, Aaron and {Conroy}, Charlie and {Cantiello}, Matteo and {Paxton}, Bill and {Johnson}, Benjamin D.},
  journal       = {\apj},
  title         = {{Mesa Isochrones and Stellar Tracks (MIST). I. Solar-scaled Models}},
  year          = {2016},
  month         = jun,
  number        = {2},
  pages         = {102},
  volume        = {823},
  adsurl        = {https://ui.adsabs.harvard.edu/abs/2016ApJ...823..102C},
  archiveprefix = {arXiv},
  doi           = {10.3847/0004-637X/823/2/102},
  eid           = {102},
  eprint        = {1604.08592},
  primaryclass  = {astro-ph.SR},
}

@Article{Allard2012,
  author        = {{Allard}, F. and {Homeier}, D. and {Freytag}, B.},
  journal       = {RSPT},
  title         = {{Models of very-low-mass stars, brown dwarfs and exoplanets}},
  year          = {2012},
  month         = jun,
  number        = {1968},
  pages         = {2765-2777},
  volume        = {370},
  adsurl        = {https://ui.adsabs.harvard.edu/abs/2012RSPTA.370.2765A},
  archiveprefix = {arXiv},
  doi           = {10.1098/rsta.2011.0269},
  eprint        = {1112.3591},
  primaryclass  = {astro-ph.SR},
}

@Article{Mann2015,
  author        = {{Mann}, Andrew W. and {Feiden}, Gregory A. and {Gaidos}, Eric and {Boyajian}, Tabetha and {von Braun}, Kaspar},
  journal       = {\apj},
  title         = {{How to Constrain Your M Dwarf: Measuring Effective Temperature, Bolometric Luminosity, Mass, and Radius}},
  year          = {2015},
  month         = may,
  number        = {1},
  pages         = {64},
  volume        = {804},
  adsurl        = {https://ui.adsabs.harvard.edu/abs/2015ApJ...804...64M},
  archiveprefix = {arXiv},
  doi           = {10.1088/0004-637X/804/1/64},
  eid           = {64},
  eprint        = {1501.01635},
  primaryclass  = {astro-ph.SR},
  publisher     = {American Astronomical Society},
}

@Article{Winters2020,
  author    = {Jennifer G. Winters and Jonathan M. Irwin and David Charbonneau and David W. Latham and Amber M. Medina and Jessica Mink and Gilbert A. Esquerdo and Perry Berlind and Michael L. Calkins and Zachory K. Berta-Thompson},
  journal   = {AJ},
  title     = {{Spectroscopic Orbits of 11 Nearby, Mid-to-late M-dwarf Binaries}},
  year      = {2020},
  month     = {jun},
  number    = {6},
  pages     = {290},
  volume    = {159},
  adsurl    = {https://ui.adsabs.harvard.edu/abs/2020AJ....159..290W},
  doi       = {10.3847/1538-3881/ab8d31},
  publisher = {American Astronomical Society},
}

@Article{Winters2021,
  author        = {{Winters}, Jennifer G. and David Charbonneau and Todd J. Henry and Jonathan M. Irwin and Wei-Chun Jao and Adric R. Riedel and Kenneth Slatten},
  journal       = {AJ},
  title         = {{The Volume-complete Sample of M Dwarfs with Masses 0.1 $\leq$ M/M $\odot$ $\leq$ 0.3 within 15 Parsecs}},
  year          = {2021},
  month         = {jan},
  number        = {2},
  pages         = {63},
  volume        = {161},
  adsurl        = {https://ui.adsabs.harvard.edu/abs/2021AJ....161...63W},
  archiveprefix = {arXiv},
  doi           = {10.3847/1538-3881/abcc74},
  eid           = {63},
  eprint        = {2011.09409},
  primaryclass  = {astro-ph.SR},
  publisher     = {American Astronomical Society},
}

@Article{Benedict2016,
  author    = {G. F. Benedict and T. J. Henry and O. G. Franz and B. E. McArthur and L. H. Wasserman and Wei-Chun Jao and P. A. Cargile and S. B. Dieterich and A. J. Bradley and E. P. Nelan and A. L. Whipple},
  journal   = {AJ},
  title     = {{The solar neighborhood XXXVII: The mass-luminosity relation for main-sequence M dwarfs}},
  year      = {2016},
  month     = {oct},
  number    = {5},
  pages     = {141},
  volume    = {152},
  adsurl    = {https://ui.adsabs.harvard.edu/abs/2016AJ....152..141B},
  doi       = {10.3847/0004-6256/152/5/141},
  publisher = {American Astronomical Society},
}

@Article{Morley2017,
  author        = {Caroline V. Morley and Laura Kreidberg and Zafar Rustamkulov and Tyler Robinson and Jonathan J. Fortney},
  journal       = {ApJ},
  title         = {{Observing the Atmospheres of Known Temperate Earth-sized Planets with JWST}},
  year          = {2017},
  month         = {nov},
  number        = {2},
  pages         = {121},
  volume        = {850},
  adsurl        = {https://ui.adsabs.harvard.edu/abs/2017ApJ...850..121M},
  archiveprefix = {arXiv},
  doi           = {10.3847/1538-4357/aa927b},
  eid           = {121},
  eprint        = {1708.04239},
  primaryclass  = {astro-ph.EP},
  publisher     = {American Astronomical Society},
}

@Article{Medina2020,
  author        = {{Medina}, Amber A. and {Winters}, Jennifer G. and {Irwin}, Jonathan M. and {Charbonneau}, David},
  journal       = {ApJ},
  title         = {{Flare Rates, Rotation Periods, and Spectroscopic Activity Indicators of a Volume-complete Sample of Mid- to Late-M Dwarfs within 15 pc}},
  year          = {2020},
  month         = {dec},
  number        = {2},
  pages         = {107},
  volume        = {905},
  adsurl        = {https://ui.adsabs.harvard.edu/abs/2020ApJ...905..107M},
  archiveprefix = {arXiv},
  doi           = {10.3847/1538-4357/abc686},
  eid           = {107},
  eprint        = {2010.15635},
  primaryclass  = {astro-ph.SR},
  publisher     = {American Astronomical Society},
}

@Article{Scipy2020,
  author  = {Virtanen, Pauli and Gommers, Ralf and Oliphant, Travis E. and Haberland, Matt and Reddy, Tyler and Cournapeau, David and Burovski, Evgeni and Peterson, Pearu and Weckesser, Warren and Bright, Jonathan and {van der Walt}, St{\'e}fan J. and Brett, Matthew and Wilson, Joshua and Millman, K. Jarrod and Mayorov, Nikolay and Nelson, Andrew R. J. and Jones, Eric and Kern, Robert and Larson, Eric and Carey, C J and Polat, {\.I}lhan and Feng, Yu and Moore, Eric W. and {VanderPlas}, Jake and Laxalde, Denis and Perktold, Josef and Cimrman, Robert and Henriksen, Ian and Quintero, E. A. and Harris, Charles R. and Archibald, Anne M. and Ribeiro, Ant{\^o}nio H. and Pedregosa, Fabian and {van Mulbregt}, Paul and {SciPy 1.0 Contributors}},
  journal = {Nature Methods},
  title   = {{{SciPy} 1.0: Fundamental Algorithms for Scientific Computing in Python}},
  year    = {2020},
  pages   = {261--272},
  volume  = {17},
  adsurl  = {https://ui.adsabs.harvard.edu/abs/2020NatMe..17..261V},
  doi     = {10.1038/s41592-019-0686-2},
}

@Article{Hunter2007,
  author    = {Hunter, J. D.},
  journal   = {CSE},
  title     = {Matplotlib: A 2D graphics environment},
  year      = {2007},
  number    = {3},
  pages     = {90--95},
  volume    = {9},
  adsurl    = {https://ui.adsabs.harvard.edu/abs/2007CSE.....9...90H},
  doi       = {10.1109/MCSE.2007.55},
  publisher = {IEEE COMPUTER SOC},
}

@InProceedings{Seifahrt2018,
  author        = {{Seifahrt}, Andreas and {St{\"u}rmer}, Julian and {Bean}, Jacob L. and {Schwab}, Christian},
  booktitle     = {Ground-based and Airborne Instrumentation for Astronomy VII},
  title         = {{MAROON-X: a radial velocity spectrograph for the Gemini Observatory}},
  year          = {2018},
  editor        = {{Evans}, Christopher J. and {Simard}, Luc and {Takami}, Hideki},
  month         = jul,
  pages         = {107026D},
  series        = {Society of Photo-Optical Instrumentation Engineers (SPIE) Conference Series},
  volume        = {10702},
  adsurl        = {https://ui.adsabs.harvard.edu/abs/2018SPIE10702E..6DS},
  archiveprefix = {arXiv},
  doi           = {10.1117/12.2312936},
  eid           = {107026D},
  eprint        = {1805.09276},
  journal       = {Proc. SPIE},
  primaryclass  = {astro-ph.IM},
}

@Article{Winters2022,
  author        = {{Winters}, Jennifer G. and {Cloutier}, Ryan and {Medina}, Amber A. and {Irwin}, Jonathan M. and {Charbonneau}, David and {Astudillo-Defru}, Nicola and {Bonfils}, Xavier and {Howard}, Andrew W. and {Isaacson}, Howard and {Bean}, Jacob L. and {Seifahrt}, Andreas and {Teske}, Johanna K. and {Eastman}, Jason D. and {Twicken}, Joseph D. and {Collins}, Karen A. and {Jensen}, Eric L.~N. and {Quinn}, Samuel N. and {Payne}, Matthew J. and {Kristiansen}, Martti H. and {Spencer}, Alton and {Vanderburg}, Andrew and {Zechmeister}, Mathias and {Weiss}, Lauren M. and {Wang}, Sharon Xuesong and {Wang}, Gavin and {Udry}, St{\'e}phane and {Terentev}, Ivan A. and {St{\"u}rmer}, Julian and {Stef{\'a}nsson}, Gudmundur and {Shporer}, Avi and {Shectman}, Stephen and {Sefako}, Ramotholo and {Schwengeler}, Hans Martin and {Schwarz}, Richard P. and {Scarsdale}, Nicholas and {Rubenzahl}, Ryan A. and {Roy}, Arpita and {Rosenthal}, Lee J. and {Robertson}, Paul and {Petigura}, Erik A. and {Pepe}, Francesco and {Omohundro}, Mark and {Murphy}, Joseph M. Akana and {Murgas}, Felipe and {Mo{\v{c}}nik}, Teo and {Montet}, Benjamin T. and {Mennickent}, Ronald and {Mayo}, Andrew W. and {Massey}, Bob and {Lubin}, Jack and {Lovis}, Christophe and {Lewin}, Pablo and {Kasper}, David and {Kane}, Stephen R. and {Jenkins}, Jon M. and {Huber}, Daniel and {Horne}, Keith and {Hill}, Michelle L. and {Gorrini}, Paula and {Giacalone}, Steven and {Fulton}, Benjamin and {Forveille}, Thierry and {Figueira}, Pedro and {Fetherolf}, Tara and {Dressing}, Courtney and {D{\'\i}az}, Rodrigo F. and {Delfosse}, Xavier and {Dalba}, Paul A. and {Dai}, Fei and {Cort{\'e}s}, C.~C. and {Crossfield}, Ian J.~M. and {Crane}, Jeffrey D. and {Conti}, Dennis M. and {Collins}, Kevin I. and {Chontos}, Ashley and {Butler}, R. Paul and {Brown}, Peyton and {Brady}, Madison and {Behmard}, Aida and {Beard}, Corey and {Batalha}, Natalie M. and {Almenara}, Jose-Manuel},
  journal       = {\aj},
  title         = {{A Second Planet Transiting LTT 1445A and a Determination of the Masses of Both Worlds}},
  year          = {2022},
  month         = apr,
  number        = {4},
  pages         = {168},
  volume        = {163},
  adsurl        = {https://ui.adsabs.harvard.edu/abs/2022AJ....163..168W},
  archiveprefix = {arXiv},
  doi           = {10.3847/1538-3881/ac50a9},
  eid           = {168},
  eprint        = {2107.14737},
  primaryclass  = {astro-ph.EP},
}

@Article{Salvatier2016,
  author    = {Salvatier, John and Wiecki, Thomas V and Fonnesbeck, Christopher},
  journal   = {PeerJ Computer Science},
  title     = {Probabilistic programming in Python using PyMC3},
  year      = {2016},
  pages     = {e55},
  volume    = {2},
  adsurl    = {a},
  doi       = {10.7717/peerj-cs.55},
  publisher = {PeerJ Inc.},
}

@Article{Pass2023_active,
  author        = {{Pass}, Emily K. and {Winters}, Jennifer G. and {Charbonneau}, David and {Irwin}, Jonathan M. and {Medina}, Amber A.},
  journal       = {\aj},
  title         = {{Active Stars in the Spectroscopic Survey of Mid-to-late M Dwarfs within 15 pc}},
  year          = {2023},
  month         = jul,
  number        = {1},
  pages         = {16},
  volume        = {166},
  adsurl        = {https://ui.adsabs.harvard.edu/abs/2023AJ....166...16P},
  archiveprefix = {arXiv},
  doi           = {10.3847/1538-3881/acd6a2},
  eid           = {16},
  eprint        = {2306.00799},
  primaryclass  = {astro-ph.SR},
}

@Article{Montes2018,
  author        = {{Montes}, D. and {Gonz{\'a}lez-Peinado}, R. and {Tabernero}, H.~M. and {Caballero}, J.~A. and {Marfil}, E. and {Alonso-Floriano}, F.~J. and {Cort{\'e}s-Contreras}, M. and {Gonz{\'a}lez Hern{\'a}ndez}, J.~I. and {Klutsch}, A. and {Moreno-J{\'o}dar}, C.},
  journal       = {\mnras},
  title         = {{Calibrating the metallicity of M dwarfs in wide physical binaries with F-, G-, and K-primaries - I: High-resolution spectroscopy with HERMES: stellar parameters, abundances, and kinematics}},
  year          = {2018},
  month         = sep,
  number        = {1},
  pages         = {1332-1382},
  volume        = {479},
  adsurl        = {https://ui.adsabs.harvard.edu/abs/2018MNRAS.479.1332M},
  archiveprefix = {arXiv},
  doi           = {10.1093/mnras/sty1295},
  eprint        = {1805.05394},
  primaryclass  = {astro-ph.SR},
}

@Article{Valenti2005,
  author   = {{Valenti}, Jeff A. and {Fischer}, Debra A.},
  journal  = {\apjs},
  title    = {{Spectroscopic Properties of Cool Stars (SPOCS). I. 1040 F, G, and K Dwarfs from Keck, Lick, and AAT Planet Search Programs}},
  year     = {2005},
  month    = jul,
  number   = {1},
  pages    = {141-166},
  volume   = {159},
  adsurl   = {https://ui.adsabs.harvard.edu/abs/2005ApJS..159..141V},
  doi      = {10.1086/430500},
}

@Article{Muirhead2012,
  author        = {{Muirhead}, Philip S. and {Johnson}, John Asher and {Apps}, Kevin and {Carter}, Joshua A. and {Morton}, Timothy D. and {Fabrycky}, Daniel C. and {Pineda}, John Sebastian and {Bottom}, Michael and {Rojas-Ayala}, B{\'a}rbara and {Schlawin}, Everett and {Hamren}, Katherine and {Covey}, Kevin R. and {Crepp}, Justin R. and {Stassun}, Keivan G. and {Pepper}, Joshua and {Hebb}, Leslie and {Kirby}, Evan N. and {Howard}, Andrew W. and {Isaacson}, Howard T. and {Marcy}, Geoffrey W. and {Levitan}, David and {Diaz-Santos}, Tanio and {Armus}, Lee and {Lloyd}, James P.},
  journal       = {\apj},
  title         = {{Characterizing the Cool KOIs. III. KOI 961: A Small Star with Large Proper Motion and Three Small Planets}},
  year          = {2012},
  month         = mar,
  number        = {2},
  pages         = {144},
  volume        = {747},
  adsurl        = {https://ui.adsabs.harvard.edu/abs/2012ApJ...747..144M},
  archiveprefix = {arXiv},
  doi           = {10.1088/0004-637X/747/2/144},
  eid           = {144},
  eprint        = {1201.2189},
  primaryclass  = {astro-ph.SR},
}

@Article{RojasAyala2012,
  author        = {{Rojas-Ayala}, B{\'a}rbara and {Covey}, Kevin R. and {Muirhead}, Philip S. and {Lloyd}, James P.},
  journal       = {\apj},
  title         = {{Metallicity and Temperature Indicators in M Dwarf K-band Spectra: Testing New and Updated Calibrations with Observations of 133 Solar Neighborhood M Dwarfs}},
  year          = {2012},
  month         = apr,
  number        = {2},
  pages         = {93},
  volume        = {748},
  adsurl        = {https://ui.adsabs.harvard.edu/abs/2012ApJ...748...93R},
  archiveprefix = {arXiv},
  doi           = {10.1088/0004-637X/748/2/93},
  eid           = {93},
  eprint        = {1112.4567},
  primaryclass  = {astro-ph.SR},
}

@Article{Neves2012,
  author        = {{Neves}, V. and {Bonfils}, X. and {Santos}, N.~C. and {Delfosse}, X. and {Forveille}, T. and {Allard}, F. and {Nat{\'a}rio}, C. and {Fernandes}, C.~S. and {Udry}, S.},
  journal       = {\aap},
  title         = {{Metallicity of M dwarfs. II. A comparative study of photometric metallicity scales}},
  year          = {2012},
  month         = feb,
  pages         = {A25},
  volume        = {538},
  adsurl        = {https://ui.adsabs.harvard.edu/abs/2012A&A...538A..25N},
  archiveprefix = {arXiv},
  doi           = {10.1051/0004-6361/201118115},
  eid           = {A25},
  eprint        = {1110.2694},
  primaryclass  = {astro-ph.SR},
}

@Article{Terrien2012,
  author        = {{Terrien}, Ryan C. and {Mahadevan}, Suvrath and {Bender}, Chad F. and {Deshpande}, Rohit and {Ramsey}, Lawrence W. and {Bochanski}, John J.},
  journal       = {\apjl},
  title         = {{An H-band Spectroscopic Metallicity Calibration for M Dwarfs}},
  year          = {2012},
  month         = mar,
  number        = {2},
  pages         = {L38},
  volume        = {747},
  adsurl        = {https://ui.adsabs.harvard.edu/abs/2012ApJ...747L..38T},
  archiveprefix = {arXiv},
  doi           = {10.1088/2041-8205/747/2/L38},
  eid           = {L38},
  eprint        = {1202.1800},
  primaryclass  = {astro-ph.SR},
}

@Article{Mann2013,
  author        = {{Mann}, Andrew W. and {Brewer}, John M. and {Gaidos}, Eric and {L{\'e}pine}, S{\'e}bastien and {Hilton}, Eric J.},
  journal       = {\aj},
  title         = {{Prospecting in Late-type Dwarfs: A Calibration of Infrared and Visible Spectroscopic Metallicities of Late K and M Dwarfs Spanning 1.5 dex}},
  year          = {2013},
  month         = feb,
  number        = {2},
  pages         = {52},
  volume        = {145},
  adsurl        = {https://ui.adsabs.harvard.edu/abs/2013AJ....145...52M},
  archiveprefix = {arXiv},
  doi           = {10.1088/0004-6256/145/2/52},
  eid           = {52},
  eprint        = {1211.4630},
  primaryclass  = {astro-ph.SR},
}

@Article{Mann2014,
  author        = {{Mann}, Andrew W. and {Deacon}, Niall R. and {Gaidos}, Eric and {Ansdell}, Megan and {Brewer}, John M. and {Liu}, Michael C. and {Magnier}, Eugene A. and {Aller}, Kimberly M.},
  journal       = {\aj},
  title         = {{Prospecting in Ultracool Dwarfs: Measuring the Metallicities of Mid- and Late-M Dwarfs}},
  year          = {2014},
  month         = jun,
  number        = {6},
  pages         = {160},
  volume        = {147},
  adsurl        = {https://ui.adsabs.harvard.edu/abs/2014AJ....147..160M},
  archiveprefix = {arXiv},
  doi           = {10.1088/0004-6256/147/6/160},
  eid           = {160},
  eprint        = {1403.5560},
  primaryclass  = {astro-ph.SR},
}

@Article{Newton2014,
  author        = {{Newton}, Elisabeth R. and {Charbonneau}, David and {Irwin}, Jonathan and {Berta-Thompson}, Zachory K. and {Rojas-Ayala}, Barbara and {Covey}, Kevin and {Lloyd}, James P.},
  journal       = {\aj},
  title         = {{Near-infrared Metallicities, Radial Velocities, and Spectral Types for 447 Nearby M Dwarfs}},
  year          = {2014},
  month         = jan,
  number        = {1},
  pages         = {20},
  volume        = {147},
  adsurl        = {https://ui.adsabs.harvard.edu/abs/2014AJ....147...20N},
  archiveprefix = {arXiv},
  doi           = {10.1088/0004-6256/147/1/20},
  eid           = {20},
  eprint        = {1310.1087},
  primaryclass  = {astro-ph.SR},
}

@Article{Bonfils2005,
  author        = {{Bonfils}, X. and {Delfosse}, X. and {Udry}, S. and {Santos}, N.~C. and {Forveille}, T. and {S{\'e}gransan}, D.},
  journal       = {\aap},
  title         = {{Metallicity of M dwarfs. I. A photometric calibration and the impact on the mass-luminosity relation at the bottom of the main sequence}},
  year          = {2005},
  month         = nov,
  number        = {2},
  pages         = {635-642},
  volume        = {442},
  adsurl        = {https://ui.adsabs.harvard.edu/abs/2005A&A...442..635B},
  archiveprefix = {arXiv},
  doi           = {10.1051/0004-6361:20053046},
  eprint        = {astro-ph/0503260},
  primaryclass  = {astro-ph},
}

@Article{Woolf2005,
  author        = {{Woolf}, Vincent M. and {Wallerstein}, George},
  journal       = {\mnras},
  title         = {{Metallicity measurements using atomic lines in M and K dwarf stars}},
  year          = {2005},
  month         = jan,
  number        = {3},
  pages         = {963-968},
  volume        = {356},
  adsurl        = {https://ui.adsabs.harvard.edu/abs/2005MNRAS.356..963W},
  archiveprefix = {arXiv},
  doi           = {10.1111/j.1365-2966.2004.08515.x},
  eprint        = {astro-ph/0410452},
  primaryclass  = {astro-ph},
}

@Article{Onehag2012,
  author        = {{{\"O}nehag}, A. and {Heiter}, U. and {Gustafsson}, B. and {Piskunov}, N. and {Plez}, B. and {Reiners}, A.},
  journal       = {\aap},
  title         = {{M-dwarf metallicities. A high-resolution spectroscopic study in the near infrared}},
  year          = {2012},
  month         = jun,
  pages         = {A33},
  volume        = {542},
  adsurl        = {https://ui.adsabs.harvard.edu/abs/2012A&A...542A..33O},
  archiveprefix = {arXiv},
  doi           = {10.1051/0004-6361/201118101},
  eid           = {A33},
  eprint        = {1112.0141},
  primaryclass  = {astro-ph.SR},
}

@Article{Veyette2017,
  author        = {{Veyette}, Mark J. and {Muirhead}, Philip S. and {Mann}, Andrew W. and {Brewer}, John M. and {Allard}, France and {Homeier}, Derek},
  journal       = {\apj},
  title         = {{A Physically Motivated and Empirically Calibrated Method to Measure the Effective Temperature, Metallicity, and Ti Abundance of M Dwarfs}},
  year          = {2017},
  month         = dec,
  number        = {1},
  pages         = {26},
  volume        = {851},
  adsurl        = {https://ui.adsabs.harvard.edu/abs/2017ApJ...851...26V},
  archiveprefix = {arXiv},
  doi           = {10.3847/1538-4357/aa96aa},
  eid           = {26},
  eprint        = {1710.10259},
  primaryclass  = {astro-ph.SR},
}

@Article{Rajpurohit2018,
  author        = {{Rajpurohit}, A.~S. and {Allard}, F. and {Teixeira}, G.~D.~C. and {Homeier}, D. and {Rajpurohit}, S. and {Mousis}, O.},
  journal       = {\aap},
  title         = {{Photospheric properties and fundamental parameters of M dwarfs}},
  year          = {2018},
  month         = feb,
  pages         = {A19},
  volume        = {610},
  adsurl        = {https://ui.adsabs.harvard.edu/abs/2018A&A...610A..19R},
  archiveprefix = {arXiv},
  doi           = {10.1051/0004-6361/201731507},
  eid           = {A19},
  eprint        = {1708.06211},
  primaryclass  = {astro-ph.SR},
}

@Article{Passegger2018,
  author        = {{Passegger}, V.~M. and {Reiners}, A. and {Jeffers}, S.~V. and {Wende-von Berg}, S. and {Sch{\"o}fer}, P. and {Caballero}, J.~A. and {Schweitzer}, A. and {Amado}, P.~J. and {B{\'e}jar}, V.~J.~S. and {Cort{\'e}s-Contreras}, M. and {Hatzes}, A.~P. and {K{\"u}rster}, M. and {Montes}, D. and {Pedraz}, S. and {Quirrenbach}, A. and {Ribas}, I. and {Seifert}, W.},
  journal       = {\aap},
  title         = {{The CARMENES search for exoplanets around M dwarfs. Photospheric parameters of target stars from high-resolution spectroscopy}},
  year          = {2018},
  month         = jul,
  pages         = {A6},
  volume        = {615},
  adsurl        = {https://ui.adsabs.harvard.edu/abs/2018A&A...615A...6P},
  archiveprefix = {arXiv},
  doi           = {10.1051/0004-6361/201732312},
  eid           = {A6},
  eprint        = {1802.02946},
  primaryclass  = {astro-ph.SR},
}

@Article{Souto2020,
  author        = {{Souto}, Diogo and {Cunha}, Katia and {Smith}, Verne V. and {Allende Prieto}, C. and {Burgasser}, Adam and {Covey}, Kevin and {Garc{\'\i}a-Hern{\'a}ndez}, D.~A. and {Holtzman}, Jon A. and {Johnson}, Jennifer A. and {J{\"o}nsson}, Henrik and {Mahadevan}, Suvrath and {Majewski}, Steven R. and {Masseron}, Thomas and {Shetrone}, Matthew and {Rojas-Ayala}, B{\'a}rbara and {Sobeck}, Jennifer and {Stassun}, Keivan G. and {Terrien}, Ryan and {Teske}, Johanna and {Wanderley}, F{\'a}bio and {Zamora}, Olga},
  journal       = {\apj},
  title         = {{Stellar Characterization of M Dwarfs from the APOGEE Survey: A Calibrator Sample for M-dwarf Metallicities}},
  year          = {2020},
  month         = feb,
  number        = {2},
  pages         = {133},
  volume        = {890},
  adsurl        = {https://ui.adsabs.harvard.edu/abs/2020ApJ...890..133S},
  archiveprefix = {arXiv},
  doi           = {10.3847/1538-4357/ab6d07},
  eid           = {133},
  eprint        = {2001.05597},
  primaryclass  = {astro-ph.SR},
}

@Article{Neves2014,
  author        = {{Neves}, V. and {Bonfils}, X. and {Santos}, N.~C. and {Delfosse}, X. and {Forveille}, T. and {Allard}, F. and {Udry}, S.},
  journal       = {\aap},
  title         = {{Metallicity of M dwarfs. IV. A high-precision [Fe/H] and T$_{eff}$ technique from high-resolution optical spectra for M dwarfs}},
  year          = {2014},
  month         = aug,
  pages         = {A121},
  volume        = {568},
  adsurl        = {https://ui.adsabs.harvard.edu/abs/2014A&A...568A.121N},
  archiveprefix = {arXiv},
  doi           = {10.1051/0004-6361/201424139},
  eid           = {A121},
  eprint        = {1406.6127},
  primaryclass  = {astro-ph.SR},
}

@Article{Mallik1997,
  author   = {{Mallik}, S.~V.},
  journal  = {\aaps},
  title    = {{The CA II triplet lines as diagnostics of luminosity, metallicity and chromospheric activity in cool stars}},
  year     = {1997},
  month    = aug,
  pages    = {359-384},
  volume   = {124},
  adsurl   = {https://ui.adsabs.harvard.edu/abs/1997A&AS..124..359M},
  doi      = {10.1051/aas:1997199},
}

@Article{Mallik1994,
  author  = {{Mallik}, S.~V.},
  journal = {\aaps},
  title   = {{The CA II triplet lines in cool stars}},
  year    = {1994},
  month   = feb,
  pages   = {279-291},
  volume  = {103},
  adsurl  = {https://ui.adsabs.harvard.edu/abs/1994A&AS..103..279M},
}

@Article{Sakari2016,
  author        = {{Sakari}, Charli M. and {Wallerstein}, George},
  journal       = {\mnras},
  title         = {{The integrated calcium II triplet as a metallicity indicator: comparisons with high-resolution [Fe/H] in M31 globular clusters}},
  year          = {2016},
  month         = feb,
  number        = {1},
  pages         = {831-843},
  volume        = {456},
  adsurl        = {https://ui.adsabs.harvard.edu/abs/2016MNRAS.456..831S},
  archiveprefix = {arXiv},
  doi           = {10.1093/mnras/stv2711},
  eprint        = {1511.06766},
  primaryclass  = {astro-ph.GA},
}

@Article{Busa2007,
  author   = {{Bus{\`a}}, I. and {Aznar Cuadrado}, R. and {Terranegra}, L. and {Andretta}, V. and {Gomez}, M.~T.},
  journal  = {\aap},
  title    = {{The Ca II infrared triplet as a stellar activity diagnostic. II. Test and calibration with high resolution observations}},
  year     = {2007},
  month    = may,
  number   = {3},
  pages    = {1089-1098},
  volume   = {466},
  adsurl   = {https://ui.adsabs.harvard.edu/abs/2007A&A...466.1089B},
  doi      = {10.1051/0004-6361:20065588},
}

@Article{Jackson2010,
  author        = {{Jackson}, R.~J. and {Jeffries}, R.~D.},
  journal       = {\mnras},
  title         = {{Chromospheric activity among fast-rotating M dwarfs in the open cluster NGC 2516}},
  year          = {2010},
  month         = sep,
  number        = {1},
  pages         = {465-478},
  volume        = {407},
  adsurl        = {https://ui.adsabs.harvard.edu/abs/2010MNRAS.407..465J},
  archiveprefix = {arXiv},
  doi           = {10.1111/j.1365-2966.2010.16917.x},
  eprint        = {1004.4557},
  primaryclass  = {astro-ph.SR},
}

@Article{Mittag2017,
  author   = {{Mittag}, M. and {Hempelmann}, A. and {Schmitt}, J.~H.~M.~M. and {Fuhrmeister}, B. and {Gonz{\'a}lez-P{\'e}rez}, J.~N. and {Schr{\"o}der}, K. -P.},
  journal  = {\aap},
  title    = {{Stellar rotation periods determined from simultaneously measured Ca II H\&K and Ca II IRT lines}},
  year     = {2017},
  month    = nov,
  pages    = {A87},
  volume   = {607},
  adsurl   = {https://ui.adsabs.harvard.edu/abs/2017A&A...607A..87M},
  doi      = {10.1051/0004-6361/201630262},
  eid      = {A87},
}

@Article{Martin2017,
  author   = {{Martin}, J. and {Fuhrmeister}, B. and {Mittag}, M. and {Schmidt}, T.~O.~B. and {Hempelmann}, A. and {Gonz{\'a}lez-P{\'e}rez}, J.~N. and {Schmitt}, J.~H.~M.~M.},
  journal  = {\aap},
  title    = {{The Ca II infrared triplet's performance as an activity indicator compared to Ca II H and K. Empirical relations to convert Ca II infrared triplet measurements to common activity indices}},
  year     = {2017},
  month    = sep,
  pages    = {A113},
  volume   = {605},
  adsurl   = {https://ui.adsabs.harvard.edu/abs/2017A&A...605A.113M},
  doi      = {10.1051/0004-6361/201630298},
  eid      = {A113},
}

@Article{Terrien2015,
  author        = {{Terrien}, Ryan C. and {Mahadevan}, Suvrath and {Deshpande}, Rohit and {Bender}, Chad F.},
  journal       = {\apjs},
  title         = {{A Near-Infrared Spectroscopic Survey of 886 Nearby M Dwarfs}},
  year          = {2015},
  month         = sep,
  number        = {1},
  pages         = {16},
  volume        = {220},
  adsurl        = {https://ui.adsabs.harvard.edu/abs/2015ApJS..220...16T},
  archiveprefix = {arXiv},
  doi           = {10.1088/0067-0049/220/1/16},
  eid           = {16},
  eprint        = {1509.01565},
  primaryclass  = {astro-ph.SR},
}

@Article{Delfosse2000,
  author        = {{Delfosse}, X. and {Forveille}, T. and {S{\'e}gransan}, D. and {Beuzit}, J. -L. and {Udry}, S. and {Perrier}, C. and {Mayor}, M.},
  journal       = {\aap},
  title         = {{Accurate masses of very low mass stars. IV. Improved mass-luminosity relations}},
  year          = {2000},
  month         = dec,
  pages         = {217-224},
  volume        = {364},
  adsurl        = {https://ui.adsabs.harvard.edu/abs/2000A&A...364..217D},
  archiveprefix = {arXiv},
  doi           = {10.48550/arXiv.astro-ph/0010586},
  eprint        = {astro-ph/0010586},
  primaryclass  = {astro-ph},
}

@InProceedings{Seifahrt2022,
  author        = {{Seifahrt}, Andreas and {Bean}, Jacob L. and {Kasper}, David and {St{\"u}rmer}, Julian and {Brady}, Madison and {Liu}, Robert and {Zechmeister}, Mathias and {Stef{\'a}nsson}, Gudmundur K. and {Montet}, Ben and {White}, John and {Tapia}, Eduardo and {Mocnik}, Teo and {Xu}, Siyi and {Schwab}, Christian},
  booktitle     = {Ground-based and Airborne Instrumentation for Astronomy IX},
  title         = {{MAROON-X: the first two years of EPRVs from Gemini North}},
  year          = {2022},
  editor        = {{Evans}, Christopher J. and {Bryant}, Julia J. and {Motohara}, Kentaro},
  month         = aug,
  pages         = {121841G},
  series        = {Society of Photo-Optical Instrumentation Engineers (SPIE) Conference Series},
  volume        = {12184},
  adsurl        = {https://ui.adsabs.harvard.edu/abs/2022SPIE12184E..1GS},
  archiveprefix = {arXiv},
  doi           = {10.1117/12.2629428},
  eid           = {121841G},
  eprint        = {2210.06563},
  primaryclass  = {astro-ph.IM},
}

@Article{Sanchez2012,
  author        = {{S{\'a}nchez Almeida}, J. and {Terlevich}, R. and {Terlevich}, E. and {Cid Fernandes}, R. and {Morales-Luis}, A.~B.},
  journal       = {\apj},
  title         = {{Qualitative Interpretation of Galaxy Spectra}},
  year          = {2012},
  month         = sep,
  number        = {2},
  pages         = {163},
  volume        = {756},
  adsurl        = {https://ui.adsabs.harvard.edu/abs/2012ApJ...756..163S},
  archiveprefix = {arXiv},
  doi           = {10.1088/0004-637X/756/2/163},
  eid           = {163},
  eprint        = {1207.3928},
  primaryclass  = {astro-ph.CO},
}

@Article{Dittmann2016,
  author        = {{Dittmann}, Jason A. and {Irwin}, Jonathan M. and {Charbonneau}, David and {Newton}, Elisabeth R.},
  journal       = {\apj},
  title         = {{Calibration of the MEarth Photometric System: Optical Magnitudes and Photometric Metallicity Estimates for 1802 Nearby M-Dwarfs}},
  year          = {2016},
  month         = feb,
  number        = {2},
  pages         = {153},
  volume        = {818},
  adsurl        = {https://ui.adsabs.harvard.edu/abs/2016ApJ...818..153D},
  archiveprefix = {arXiv},
  doi           = {10.3847/0004-637X/818/2/153},
  eid           = {153},
  eprint        = {1510.06711},
  primaryclass  = {astro-ph.SR},
}

@Article{Parsons2018,
  author        = {{Parsons}, S.~G. and {G{\"a}nsicke}, B.~T. and {Marsh}, T.~R. and {Ashley}, R.~P. and {Breedt}, E. and {Burleigh}, M.~R. and {Copperwheat}, C.~M. and {Dhillon}, V.~S. and {Green}, M.~J. and {Hermes}, J.~J. and {Irawati}, P. and {Kerry}, P. and {Littlefair}, S.~P. and {Rebassa-Mansergas}, A. and {Sahman}, D.~I. and {Schreiber}, M.~R. and {Zorotovic}, M.},
  journal       = {\mnras},
  title         = {{The scatter of the M dwarf mass-radius relationship}},
  year          = {2018},
  month         = nov,
  number        = {1},
  pages         = {1083-1096},
  volume        = {481},
  adsurl        = {https://ui.adsabs.harvard.edu/abs/2018MNRAS.481.1083P},
  archiveprefix = {arXiv},
  doi           = {10.1093/mnras/sty2345},
  eprint        = {1808.07780},
  primaryclass  = {astro-ph.SR},
}

@Article{Dieterich2021,
  author        = {{Dieterich}, Serge B. and {Simler}, Andrew and {Henry}, Todd J. and {Jao}, Wei-Chun},
  journal       = {\aj},
  title         = {{The Solar Neighborhood. XLVII. Comparing M-dwarf Models with Hubble Space Telescope Dynamical Masses and Spectroscopy}},
  year          = {2021},
  month         = apr,
  number        = {4},
  pages         = {172},
  volume        = {161},
  adsurl        = {https://ui.adsabs.harvard.edu/abs/2021AJ....161..172D},
  archiveprefix = {arXiv},
  doi           = {10.3847/1538-3881/abd2c2},
  eid           = {172},
  eprint        = {2012.00915},
  primaryclass  = {astro-ph.SR},
}

@Article{Olander2021,
  author        = {{Olander}, T. and {Heiter}, U. and {Kochukhov}, O.},
  journal       = {\aap},
  title         = {{Comparative high-resolution spectroscopy of M dwarfs: Exploring non-LTE effects}},
  year          = {2021},
  month         = may,
  pages         = {A103},
  volume        = {649},
  adsurl        = {https://ui.adsabs.harvard.edu/abs/2021A&A...649A.103O},
  archiveprefix = {arXiv},
  doi           = {10.1051/0004-6361/202039747},
  eid           = {A103},
  eprint        = {2102.08836},
  primaryclass  = {astro-ph.SR},
}

@Article{Winters2026,
  author        = {{Winters}, Jennifer G. and {Pass}, Emily K. and {Charbonneau}, David and {Irwin}, Jonathan and {Latham}, David W. and {Mink}, Jessica and {Esquerdo}, Gilbert A. and {Berlind}, Perry and {Calkins}, Michael L. and {Payne}, Matthew J.},
  journal       = {\apj},
  title         = {{Radial and Rotational Velocities of a Volume-complete Sample of M Dwarfs with Masses 0.1─0.3 M$_{{\ensuremath{\odot}}}$ within 15 pc}},
  year          = {2026},
  month         = jun,
  number        = {2},
  pages         = {224},
  volume        = {1004},
  adsurl        = {https://ui.adsabs.harvard.edu/abs/2026ApJ..1004..224W},
  archiveprefix = {arXiv},
  doi           = {10.3847/1538-4357/ae48fb},
  eid           = {224},
  eprint        = {2602.00382},
  primaryclass  = {astro-ph.SR},
}

@Article{Hartman2026,
  author        = {{Hartman}, Zachary D. and {Clark}, Catherine A. and {Lund}, Michael B. and {Lester}, Kathryn V. and {Caballero}, Jos{\'e} A. and {Howell}, Steve B. and {Ciardi}, David and {Deveny}, Sarah and {Everett}, Mark E. and {Furlan}, Elise and {Kalari}, Venu and {Littlefield}, Colin and {Stephens}, Andrew W. and {Burt}, Jennifer A. and {Huber}, Guillaume and {Matson}, Rachel and {Mamajek}, Eric E. and {Tuchow}, Noah},
  journal       = {\aj},
  title         = {{Paving the Road to the Habitable Worlds Observatory with High-resolution Imaging. I. New and Archival Speckle Observations of Potential HWO Target Stars}},
  year          = {2026},
  month         = mar,
  number        = {3},
  pages         = {174},
  volume        = {171},
  adsurl        = {https://ui.adsabs.harvard.edu/abs/2026AJ....171..174H},
  archiveprefix = {arXiv},
  doi           = {10.3847/1538-3881/ae2fe7},
  eid           = {174},
  eprint        = {2601.05387},
  primaryclass  = {astro-ph.SR},
}

@Article{Glidden2026,
  author        = {{Glidden}, Ana and {Witzke}, Veronika and {Shapiro}, Alexander I. and {Seager}, Sara},
  journal       = {\apjl},
  title         = {{Overestimated Pressure Broadening Misleads Model Spectra in Cool M Dwarf Stars}},
  year          = {2026},
  month         = jan,
  number        = {1},
  pages         = {L8},
  volume        = {997},
  adsurl        = {https://ui.adsabs.harvard.edu/abs/2026ApJ...997L...8G},
  archiveprefix = {arXiv},
  doi           = {10.3847/2041-8213/ae3137},
  eid           = {L8},
  eprint        = {2601.05313},
  primaryclass  = {astro-ph.SR},
}

@Article{Marcussen2023,
  author        = {{Marcussen}, Marcus L. and {Albrecht}, Simon H.},
  journal       = {\aj},
  title         = {{Spectroscopic Follow-up of Gaia Exoplanet Candidates: Impostor Binary Stars Invade the Gaia DR3 Astrometric Exoplanet Candidates}},
  year          = {2023},
  month         = jun,
  number        = {6},
  pages         = {266},
  volume        = {165},
  adsurl        = {https://ui.adsabs.harvard.edu/abs/2023AJ....165..266M},
  archiveprefix = {arXiv},
  doi           = {10.3847/1538-3881/acd53d},
  eid           = {266},
  eprint        = {2305.08623},
  primaryclass  = {astro-ph.EP},
}

@Article{Holl2023,
  author        = {{Holl}, B. and {Sozzetti}, A. and {Sahlmann}, J. and {Giacobbe}, P. and {S{\'e}gransan}, D. and {Unger}, N. and {Delisle}, J.-B. and {Barbato}, D. and {Lattanzi}, M.~G. and {Morbidelli}, R. and {Sosnowska}, D.},
  journal       = {\aap},
  title         = {{Gaia Data Release 3. Astrometric orbit determination with Markov chain Monte Carlo and genetic algorithms: Systems with stellar, sub-stellar, and planetary mass companions}},
  year          = {2023},
  month         = jun,
  pages         = {A10},
  volume        = {674},
  adsurl        = {https://ui.adsabs.harvard.edu/abs/2023A&A...674A..10H},
  archiveprefix = {arXiv},
  doi           = {10.1051/0004-6361/202244161},
  eid           = {A10},
  eprint        = {2206.05439},
  primaryclass  = {astro-ph.EP},
}

@Article{Pass2025,
  author        = {{Pass}, Emily K. and {Cargile}, Phillip A. and {DiTomasso}, Victoria and {Rodr{\'\i}guez Mart{\'\i}nez}, Romy and {Charbonneau}, David and {Latham}, David W. and {Vanderburg}, Andrew and {Bieryla}, Allyson and {Quinn}, Samuel N. and {Buchhave}, Lars A.},
  journal       = {\apjs},
  title         = {{Metallicities from High-resolution TRES Spectra with uberMS: Performance Benchmarks and Literature Comparison}},
  year          = {2025},
  month         = dec,
  number        = {2},
  pages         = {27},
  volume        = {281},
  adsurl        = {https://ui.adsabs.harvard.edu/abs/2025ApJS..281...27P},
  archiveprefix = {arXiv},
  doi           = {10.3847/1538-4365/ae0acb},
  eid           = {27},
  eprint        = {2506.18961},
  primaryclass  = {astro-ph.SR},
}

@Book{Kurucz1993,
  author  = {{Kurucz}, Robert L.},
  title   = {{SYNTHE spectrum synthesis programs and line data}},
  year    = {1993},
  adsurl  = {https://ui.adsabs.harvard.edu/abs/1993sssp.book.....K},
}

@Article{Kurucz2005,
  author   = {{Kurucz}, Robert L.},
  journal  = {Memorie della Societa Astronomica Italiana Supplementi},
  title    = {{ATLAS12, SYNTHE, ATLAS9, WIDTH9, et cetera}},
  year     = {2005},
  month    = jan,
  pages    = {14},
  volume   = {8},
  adsurl   = {https://ui.adsabs.harvard.edu/abs/2005MSAIS...8...14K},
}

@Article{Weinberg2019,
  author        = {{Weinberg}, David H. and {Holtzman}, Jon A. and {Hasselquist}, Sten and {Bird}, Jonathan C. and {Johnson}, Jennifer A. and {Shetrone}, Matthew and {Sobeck}, Jennifer and {Allende Prieto}, Carlos and {Bizyaev}, Dmitry and {Carrera}, Ricardo and {Cohen}, Roger E. and {Cunha}, Katia and {Ebelke}, Garrett and {Fernandez-Trincado}, J.~G. and {Garc{\'\i}a-Hern{\'a}ndez}, D.~A. and {Hayes}, Christian R. and {J{\"o}nsson}, Henrik and {Lane}, Richard R. and {Majewski}, Steven R. and {Malanushenko}, Viktor and {M{\'e}sz{\'a}ros}, Szabolcs and {Nidever}, David L. and {Nitschelm}, Christian and {Pan}, Kaike and {Rix}, Hans-Walter and {Rybizki}, Jan and {Schiavon}, Ricardo P. and {Schneider}, Donald P. and {Wilson}, John C. and {Zamora}, Olga},
  journal       = {\apj},
  title         = {{Chemical Cartography with APOGEE: Multi-element Abundance Ratios}},
  year          = {2019},
  month         = mar,
  number        = {1},
  pages         = {102},
  volume        = {874},
  adsurl        = {https://ui.adsabs.harvard.edu/abs/2019ApJ...874..102W},
  archiveprefix = {arXiv},
  doi           = {10.3847/1538-4357/ab07c7},
  eid           = {102},
  eprint        = {1810.12325},
  primaryclass  = {astro-ph.GA},
}

@Article{Majewski2017,
  author        = {{Majewski}, Steven R. and {Schiavon}, Ricardo P. and {Frinchaboy}, Peter M. and {Allende Prieto}, Carlos and {Barkhouser}, Robert and {Bizyaev}, Dmitry and {Blank}, Basil and {Brunner}, Sophia and {Burton}, Adam and {Carrera}, Ricardo and {Chojnowski}, S. Drew and {Cunha}, K{\'a}tia and {Epstein}, Courtney and {Fitzgerald}, Greg and {Garc{\'\i}a P{\'e}rez}, Ana E. and {Hearty}, Fred R. and {Henderson}, Chuck and {Holtzman}, Jon A. and {Johnson}, Jennifer A. and {Lam}, Charles R. and {Lawler}, James E. and {Maseman}, Paul and {M{\'e}sz{\'a}ros}, Szabolcs and {Nelson}, Matthew and {Nguyen}, Duy Coung and {Nidever}, David L. and {Pinsonneault}, Marc and {Shetrone}, Matthew and {Smee}, Stephen and {Smith}, Verne V. and {Stolberg}, Todd and {Skrutskie}, Michael F. and {Walker}, Eric and {Wilson}, John C. and {Zasowski}, Gail and {Anders}, Friedrich and {Basu}, Sarbani and {Beland}, Stephane and {Blanton}, Michael R. and {Bovy}, Jo and {Brownstein}, Joel R. and {Carlberg}, Joleen and {Chaplin}, William and {Chiappini}, Cristina and {Eisenstein}, Daniel J. and {Elsworth}, Yvonne and {Feuillet}, Diane and {Fleming}, Scott W. and {Galbraith-Frew}, Jessica and {Garc{\'\i}a}, Rafael A. and {Garc{\'\i}a-Hern{\'a}ndez}, D. An{\'\i}bal and {Gillespie}, Bruce A. and {Girardi}, L{\'e}o and {Gunn}, James E. and {Hasselquist}, Sten and {Hayden}, Michael R. and {Hekker}, Saskia and {Ivans}, Inese and {Kinemuchi}, Karen and {Klaene}, Mark and {Mahadevan}, Suvrath and {Mathur}, Savita and {Mosser}, Beno{\^\i}t and {Muna}, Demitri and {Munn}, Jeffrey A. and {Nichol}, Robert C. and {O'Connell}, Robert W. and {Parejko}, John K. and {Robin}, A.~C. and {Rocha-Pinto}, Helio and {Schultheis}, Matthias and {Serenelli}, Aldo M. and {Shane}, Neville and {Silva Aguirre}, Victor and {Sobeck}, Jennifer S. and {Thompson}, Benjamin and {Troup}, Nicholas W. and {Weinberg}, David H. and {Zamora}, Olga},
  journal       = {\aj},
  title         = {{The Apache Point Observatory Galactic Evolution Experiment (APOGEE)}},
  year          = {2017},
  month         = sep,
  number        = {3},
  pages         = {94},
  volume        = {154},
  adsurl        = {https://ui.adsabs.harvard.edu/abs/2017AJ....154...94M},
  archiveprefix = {arXiv},
  doi           = {10.3847/1538-3881/aa784d},
  eid           = {94},
  eprint        = {1509.05420},
  primaryclass  = {astro-ph.IM},
}

@Article{Dotter2026,
  author        = {{Dotter}, Aaron and {Bauer}, Evan B. and {Park}, Minjung and {Conroy}, Charlie and {Milone}, Antonino P. and {Joyce}, Meridith and {Cantiello}, Matteo},
  journal       = {\apjs},
  title         = {{MESA Isochrones and Stellar Tracks (MIST). II. Models with {\ensuremath{\alpha}}-enhanced Chemical Composition}},
  year          = {2026},
  month         = apr,
  number        = {2},
  pages         = {64},
  volume        = {283},
  adsurl        = {https://ui.adsabs.harvard.edu/abs/2026ApJS..283...64D},
  archiveprefix = {arXiv},
  doi           = {10.3847/1538-4365/ae48f3},
  eid           = {64},
  eprint        = {2602.22012},
  primaryclass  = {astro-ph.SR},
}

@Article{RodriguezMartinez2026,
  author        = {{Rodr{\'\i}guez Mart{\'\i}nez}, Romy and {Pass}, Emily K. and {Cargile}, Phillip A. and {DiTomasso}, Victoria and {Charbonneau}, David and {Eastman}, Jason D. and {Latham}, David W.},
  journal       = {\apj},
  title         = {{A Uniform Determination of the Bulk Metallicities and {\ensuremath{\alpha}} Enrichments of Confirmed Exoplanet Systems with TRES}},
  year          = {2026},
  month         = jun,
  number        = {1},
  pages         = {91},
  volume        = {1004},
  adsurl        = {https://ui.adsabs.harvard.edu/abs/2026ApJ..1004...91R},
  archiveprefix = {arXiv},
  doi           = {10.3847/1538-4357/ae6ce2},
  eid           = {91},
  eprint        = {2605.11075},
  primaryclass  = {astro-ph.EP},
}

@PhdThesis{Buchhave2010,
  author  = {{Buchhave}, Lars Christian Astrup},
  school  = {University of Copenhagen, Denmark},
  title   = {{Detecting and characterizing transiting extrasolar planets}},
  year    = {2010},
  month   = jan,
  adsurl  = {https://ui.adsabs.harvard.edu/abs/2010PhDT.......450B},
}

@Article{Gullikson2014,
  author        = {{Gullikson}, Kevin and {Dodson-Robinson}, Sarah and {Kraus}, Adam},
  journal       = {\aj},
  title         = {{Correcting for Telluric Absorption: Methods, Case Studies, and Release of the TelFit Code}},
  year          = {2014},
  month         = sep,
  number        = {3},
  pages         = {53},
  volume        = {148},
  adsurl        = {https://ui.adsabs.harvard.edu/abs/2014AJ....148...53G},
  archiveprefix = {arXiv},
  doi           = {10.1088/0004-6256/148/3/53},
  eid           = {53},
  eprint        = {1406.6059},
  primaryclass  = {astro-ph.IM},
}

@Article{Nagel2023,
  author        = {{Nagel}, E. and {Czesla}, S. and {Kaminski}, A. and {Zechmeister}, M. and {Tal-Or}, L. and {Schmitt}, J.~H.~M.~M. and {Reiners}, A. and {Quirrenbach}, A. and {Garc{\'\i}a L{\'o}pez}, A. and {Caballero}, J.~A. and {Ribas}, I. and {Amado}, P.~J. and {B{\'e}jar}, V.~J.~S. and {Cort{\'e}s-Contreras}, M. and {Dreizler}, S. and {Hatzes}, A.~P. and {Henning}, Th. and {Jeffers}, S.~V. and {K{\"u}rster}, M. and {Lafarga}, M. and {L{\'o}pez-Puertas}, M. and {Montes}, D. and {Morales}, J.~C. and {Pedraz}, S. and {Schweitzer}, A.},
  journal       = {\aap},
  title         = {{The CARMENES search for exoplanets around M dwarfs. Telluric absorption corrected high S/N optical and near-infrared template spectra of 382 M dwarf stars}},
  year          = {2023},
  month         = dec,
  pages         = {A73},
  volume        = {680},
  adsurl        = {https://ui.adsabs.harvard.edu/abs/2023A&A...680A..73N},
  archiveprefix = {arXiv},
  doi           = {10.1051/0004-6361/202346524},
  eid           = {A73},
  eprint        = {2310.14715},
  primaryclass  = {astro-ph.SR},
}

@Article{Fuhrmeister2023,
  author        = {{Fuhrmeister}, B. and {Czesla}, S. and {Perdelwitz}, V. and {Nagel}, E. and {Schmitt}, J.~H.~M.~M. and {Jeffers}, S.~V. and {Caballero}, J.~A. and {Zechmeister}, M. and {Montes}, D. and {Reiners}, A. and {L{\'o}pez-Gallifa}, {\'A}. and {Ribas}, I. and {Quirrenbach}, A. and {Amado}, P.~J. and {Galad{\'\i}-Enr{\'\i}quez}, D. and {B{\'e}jar}, V.~J.~S. and {Danielski}, C. and {Hatzes}, A.~P. and {Kaminski}, A. and {K{\"u}rster}, M. and {Morales}, J.~C. and {Zapatero Osorio}, M.~R.},
  journal       = {\aap},
  title         = {{The CARMENES search for exoplanets around M dwarfs. Variability on long timescales as seen in chromospheric indicators}},
  year          = {2023},
  month         = feb,
  pages         = {A71},
  volume        = {670},
  adsurl        = {https://ui.adsabs.harvard.edu/abs/2023A&A...670A..71F},
  archiveprefix = {arXiv},
  doi           = {10.1051/0004-6361/202244829},
  eid           = {A71},
  eprint        = {2212.03514},
  primaryclass  = {astro-ph.SR},
}

@InProceedings{Quirrenbach2014,
  author    = {{Quirrenbach}, A. and {Amado}, P.~J. and {Caballero}, J.~A. and {Mundt}, R. and {Reiners}, A. and {Ribas}, I. and {Seifert}, W. and {Abril}, M. and {Aceituno}, J. and {Alonso-Floriano}, F.~J. and {Ammler-von Eiff}, M. and {Antona Jim{\'e}nez}, R. and {Anwand-Heerwart}, H. and {Azzaro}, M. and {Bauer}, F. and {Barrado}, D. and {Becerril}, S. and {B{\'e}jar}, V.~J.~S. and {Ben{\'\i}tez}, D. and {Berdi{\~n}as}, Z.~M. and {C{\'a}rdenas}, M.~C. and {Casal}, E. and {Claret}, A. and {Colom{\'e}}, J. and {Cort{\'e}s-Contreras}, M. and {Czesla}, S. and {Doellinger}, M. and {Dreizler}, S. and {Feiz}, C. and {Fern{\'a}ndez}, M. and {Galad{\'\i}}, D. and {G{\'a}lvez-Ortiz}, M.~C. and {Garc{\'\i}a-Piquer}, A. and {Garc{\'\i}a-Vargas}, M.~L. and {Garrido}, R. and {Gesa}, L. and {G{\'o}mez Galera}, V. and {Gonz{\'a}lez {\'A}lvarez}, E. and {Gonz{\'a}lez Hern{\'a}ndez}, J.~I. and {Gr{\"o}zinger}, U. and {Gu{\`a}rdia}, J. and {Guenther}, E.~W. and {de Guindos}, E. and {Guti{\'e}rrez-Soto}, J. and {Hagen}, H.-J. and {Hatzes}, A.~P. and {Hauschildt}, P.~H. and {Helmling}, J. and {Henning}, T. and {Hermann}, D. and {Hern{\'a}ndez Casta{\~n}o}, L. and {Herrero}, E. and {Hidalgo}, D. and {Holgado}, G. and {Huber}, A. and {Huber}, K.~F. and {Jeffers}, S. and {Joergens}, V. and {de Juan}, E. and {Kehr}, M. and {Klein}, R. and {K{\"u}rster}, M. and {Lamert}, A. and {Lalitha}, S. and {Laun}, W. and {Lemke}, U. and {Lenzen}, R. and {L{\'o}pez del Fresno}, Mauro and {L{\'o}pez Mart{\'\i}}, B. and {L{\'o}pez-Santiago}, J. and {Mall}, U. and {Mandel}, H. and {Mart{\'\i}n}, E.~L. and {Mart{\'\i}n-Ruiz}, S. and {Mart{\'\i}nez-Rodr{\'\i}guez}, H. and {Marvin}, C.~J. and {Mathar}, R.~J. and {Mirabet}, E. and {Montes}, D. and {Morales Mu{\~n}oz}, R. and {Moya}, A. and {Naranjo}, V. and {Ofir}, A. and {Oreiro}, R. and {Pall{\'e}}, E. and {Panduro}, J. and {Passegger}, V.-M. and {P{\'e}rez-Calpena}, A. and {P{\'e}rez Medialdea}, D. and {Perger}, M. and {Pluto}, M. and {Ram{\'o}n}, A. and {Rebolo}, R. and {Redondo}, P. and {Reffert}, S. and {Reinhardt}, S. and {Rhode}, P. and {Rix}, H.-W. and {Rodler}, F. and {Rodr{\'\i}guez}, E. and {Rodr{\'\i}guez-L{\'o}pez}, C. and {Rodr{\'\i}guez-P{\'e}rez}, E. and {Rohloff}, R.-R. and {Rosich}, A. and {S{\'a}nchez-Blanco}, E. and {S{\'a}nchez Carrasco}, M.~A. and {Sanz-Forcada}, J. and {Sarmiento}, L.~F. and {Sch{\"a}fer}, S. and {Schiller}, J. and {Schmidt}, C. and {Schmitt}, J.~H.~M.~M. and {Solano}, E. and {Stahl}, O. and {Storz}, C. and {St{\"u}rmer}, J. and {Su{\'a}rez}, J.~C. and {Ulbrich}, R.~G. and {Veredas}, G. and {Wagner}, K. and {Winkler}, J. and {Zapatero Osorio}, M.~R. and {Zechmeister}, M. and {Abell{\'a}n de Paco}, F.~J. and {Anglada-Escud{\'e}}, G. and {del Burgo}, C. and {Klutsch}, A. and {Lizon}, J.~L. and {L{\'o}pez-Morales}, M. and {Morales}, J.~C. and {Perryman}, M.~A.~C. and {Tulloch}, S.~M. and {Xu}, W.},
  booktitle = {Ground-based and Airborne Instrumentation for Astronomy V},
  title     = {{CARMENES instrument overview}},
  year      = {2014},
  editor    = {{Ramsay}, Suzanne K. and {McLean}, Ian S. and {Takami}, Hideki},
  month     = jul,
  pages     = {91471F},
  series    = {Society of Photo-Optical Instrumentation Engineers (SPIE) Conference Series},
  volume    = {9147},
  adsurl    = {https://ui.adsabs.harvard.edu/abs/2014SPIE.9147E..1FQ},
  doi       = {10.1117/12.2056453},
  eid       = {91471F},
}

@Article{An2023,
  author        = {{An}, Dong-Sheng and {Xie}, Ji-Wei and {Dai}, Yuan-Zhe and {Zhou}, Ji-Lin},
  journal       = {\aj},
  title         = {{Planetary Orbit Eccentricity Trends (POET). I. The Eccentricity-Metallicity Trend for Small Planets Revealed by the LAMOST-Gaia-Kepler Sample}},
  year          = {2023},
  month         = mar,
  number        = {3},
  pages         = {125},
  volume        = {165},
  adsurl        = {https://ui.adsabs.harvard.edu/abs/2023AJ....165..125A},
  archiveprefix = {arXiv},
  doi           = {10.3847/1538-3881/acb533},
  eid           = {125},
  eprint        = {2303.02941},
  primaryclass  = {astro-ph.EP},
}

@Article{Hua2025,
  author        = {{Hua}, Xinyan and {Wang}, Sharon Xuesong and {An}, Dongsheng and {Wang}, Songhu and {Huang}, Yang and {Chen}, Dichang and {Buchner}, Johannes and {Zhu}, Wei and {Dai}, Fei and {Xie}, Jiwei},
  journal       = {\apjl},
  title         = {{Short-period Small Planets with High Mutual Inclinations are More Common around Metal-rich Stars}},
  year          = {2025},
  month         = feb,
  number        = {2},
  pages         = {L46},
  volume        = {980},
  adsurl        = {https://ui.adsabs.harvard.edu/abs/2025ApJ...980L..46H},
  archiveprefix = {arXiv},
  doi           = {10.3847/2041-8213/adb1dc},
  eid           = {L46},
  eprint        = {2502.00442},
  primaryclass  = {astro-ph.EP},
}

@Article{Spalding2022,
  author        = {{Spalding}, Christopher and {Winn}, Joshua N.},
  journal       = {\apj},
  title         = {{Tidal Erasure of Stellar Obliquities Constrains the Timing of Hot Jupiter Formation}},
  year          = {2022},
  month         = mar,
  number        = {1},
  pages         = {22},
  volume        = {927},
  adsurl        = {https://ui.adsabs.harvard.edu/abs/2022ApJ...927...22S},
  archiveprefix = {arXiv},
  doi           = {10.3847/1538-4357/ac4993},
  eid           = {22},
  eprint        = {2201.03653},
  primaryclass  = {astro-ph.EP},
}

@Article{Zhu2023,
  author        = {{Zhu}, Wei},
  journal       = {RAA},
  title         = {{The Metallicity Dimension of the Super Earth-cold Jupiter Correlation}},
  year          = {2024},
  month         = apr,
  number        = {4},
  pages         = {045013},
  volume        = {24},
  adsurl        = {https://ui.adsabs.harvard.edu/abs/2024RAA....24d5013Z},
  archiveprefix = {arXiv},
  doi           = {10.1088/1674-4527/ad3132},
  eid           = {045013},
  eprint        = {2306.16691},
  primaryclass  = {astro-ph.EP},
}

@Article{Ditomasso2026,
  author   = {{DiTomasso}, Victoria and {Yee}, Samuel W. and {Rodr{\'\i}guez Mart{\'\i}nez}, Romy and {Pass}, Emily K. and {Cargile}, Phillip A. and {Charbonneau}, David and {Vanderburg}, Andrew and {Latham}, David W. and {Chandra}, Vedant},
  journal  = {\aj},
  title    = {{A Sample of Thick-disk Stars Hosting Hot Jupiters}},
  year     = {2026},
  month    = dec,
  number   = {6},
  pages    = {326},
  volume   = {172},
  adsurl   = {https://scixplorer.org/abs/2026AJ....172..326D},
  doi      = {10.3847/1538-3881/aeb6e9},
  eid      = {326},
}

@Article{Rebassa-Mansergas2026,
  author        = {{Rebassa-Mansergas}, Alberto and {Tejero-G{\'o}mez}, Iset and {Raddi}, Roberto},
  journal       = {\aap},
  title         = {{Revisiting the Galactic age─metallicity relation from wide white dwarf─main-sequence binaries}},
  year          = {2026},
  month         = may,
  pages         = {A250},
  volume        = {709},
  adsurl        = {https://ui.adsabs.harvard.edu/abs/2026A&A...709A.250R},
  archiveprefix = {arXiv},
  doi           = {10.1051/0004-6361/202660074},
  eid           = {A250},
  eprint        = {2604.15939},
  primaryclass  = {astro-ph.SR},
}

@Article{Pass2024,
  author        = {{Pass}, Emily K. and {Charbonneau}, David and {Latham}, David W. and {Berlind}, Perry and {Calkins}, Michael L. and {Esquerdo}, Gilbert A. and {Mink}, Jessica},
  journal       = {\apj},
  title         = {{The Mass Dependence of H{\ensuremath{\alpha}} Emission and Stellar Spindown for Fully Convective M Dwarfs}},
  year          = {2024},
  month         = may,
  number        = {2},
  pages         = {231},
  volume        = {966},
  adsurl        = {https://ui.adsabs.harvard.edu/abs/2024ApJ...966..231P},
  archiveprefix = {arXiv},
  doi           = {10.3847/1538-4357/ad3631},
  eid           = {231},
  eprint        = {2401.10167},
  primaryclass  = {astro-ph.SR},
}

@Article{Buder2019,
  author        = {{Buder}, S. and {Lind}, K. and {Ness}, M.~K. and {Asplund}, M. and {Duong}, L. and {Lin}, J. and {Kos}, J. and {Casagrande}, L. and {Casey}, A.~R. and {Bland-Hawthorn}, J. and {de Silva}, G.~M. and {D'Orazi}, V. and {Freeman}, K.~C. and {Martell}, S.~L. and {Schlesinger}, K.~J. and {Sharma}, S. and {Simpson}, J.~D. and {Zucker}, D.~B. and {Zwitter}, T. and {{\v{C}}otar}, K. and {Dotter}, A. and {Hayden}, M.~R. and {Hyde}, E.~A. and {Kafle}, P.~R. and {Lewis}, G.~F. and {Nataf}, D.~M. and {Nordlander}, T. and {Reid}, W. and {Rix}, H.-W. and {Sk{\'u}lad{\'o}ttir}, {\'A}. and {Stello}, D. and {Ting}, Y.-S. and {Traven}, G. and {Wyse}, R.~F.~G. and {GALAH Collaboration}},
  journal       = {\aap},
  title         = {{The GALAH survey: An abundance, age, and kinematic inventory of the solar neighbourhood made with TGAS}},
  year          = {2019},
  month         = apr,
  pages         = {A19},
  volume        = {624},
  adsurl        = {https://ui.adsabs.harvard.edu/abs/2019A&A...624A..19B},
  archiveprefix = {arXiv},
  doi           = {10.1051/0004-6361/201833218},
  eid           = {A19},
  eprint        = {1804.05869},
  primaryclass  = {astro-ph.SR},
}

@Article{DeSilva2015,
  author        = {{De Silva}, G.~M. and {Freeman}, K.~C. and {Bland-Hawthorn}, J. and {Martell}, S. and {de Boer}, E. Wylie and {Asplund}, M. and {Keller}, S. and {Sharma}, S. and {Zucker}, D.~B. and {Zwitter}, T. and {Anguiano}, B. and {Bacigalupo}, C. and {Bayliss}, D. and {Beavis}, M.~A. and {Bergemann}, M. and {Campbell}, S. and {Cannon}, R. and {Carollo}, D. and {Casagrande}, L. and {Casey}, A.~R. and {Da Costa}, G. and {D'Orazi}, V. and {Dotter}, A. and {Duong}, L. and {Heger}, A. and {Ireland}, M.~J. and {Kafle}, P.~R. and {Kos}, J. and {Lattanzio}, J. and {Lewis}, G.~F. and {Lin}, J. and {Lind}, K. and {Munari}, U. and {Nataf}, D.~M. and {O'Toole}, S. and {Parker}, Q. and {Reid}, W. and {Schlesinger}, K.~J. and {Sheinis}, A. and {Simpson}, J.~D. and {Stello}, D. and {Ting}, Y.-S. and {Traven}, G. and {Watson}, F. and {Wittenmyer}, R. and {Yong}, D. and {{\v{Z}}erjal}, M.},
  journal       = {\mnras},
  title         = {{The GALAH survey: scientific motivation}},
  year          = {2015},
  month         = may,
  number        = {3},
  pages         = {2604-2617},
  volume        = {449},
  adsurl        = {https://ui.adsabs.harvard.edu/abs/2015MNRAS.449.2604D},
  archiveprefix = {arXiv},
  doi           = {10.1093/mnras/stv327},
  eprint        = {1502.04767},
  primaryclass  = {astro-ph.GA},
}

@Article{Christiansen2025,
  author        = {{Christiansen}, Jessie L. and {McElroy}, Douglas L. and {Harbut}, Marcy and {Ciardi}, David R. and {Crane}, Megan and {Good}, John and {Hardegree-Ullman}, Kevin K. and {Kesseli}, Aurora Y. and {Lund}, Michael B. and {Lynn}, Meca and {Muthiar}, Ananda and {Nilsson}, Ricky and {Oluyide}, Toba and {Papin}, Michael and {Rivera}, Amalia and {Swain}, Melanie and {Susemiehl}, Nicholas D. and {Tam}, Raymond and {van Eyken}, Julian and {Beichman}, Charles},
  journal       = {\psj},
  title         = {{The NASA Exoplanet Archive and Exoplanet Follow-up Observing Program: Data, Tools, and Usage}},
  year          = {2025},
  month         = aug,
  number        = {8},
  pages         = {186},
  volume        = {6},
  adsurl        = {https://ui.adsabs.harvard.edu/abs/2025PSJ.....6..186C},
  archiveprefix = {arXiv},
  doi           = {10.3847/PSJ/ade3c2},
  eid           = {186},
  eprint        = {2506.03299},
  primaryclass  = {astro-ph.EP},
}

@Article{Dittmann2017,
  author        = {{Dittmann}, Jason A. and {Irwin}, Jonathan M. and {Charbonneau}, David and {Bonfils}, Xavier and {Astudillo-Defru}, Nicola and {Haywood}, Rapha{\"e}lle D. and {Berta-Thompson}, Zachory K. and {Newton}, Elisabeth R. and {Rodriguez}, Joseph E. and {Winters}, Jennifer G. and {Tan}, Thiam-Guan and {Almenara}, Jose-Manuel and {Bouchy}, Fran{\c{c}}ois and {Delfosse}, Xavier and {Forveille}, Thierry and {Lovis}, Christophe and {Murgas}, Felipe and {Pepe}, Francesco and {Santos}, Nuno C. and {Udry}, Stephane and {W{\"u}nsche}, Ana{\"e}l and {Esquerdo}, Gilbert A. and {Latham}, David W. and {Dressing}, Courtney D.},
  journal       = {\nat},
  title         = {{A temperate rocky super-Earth transiting a nearby cool star}},
  year          = {2017},
  month         = apr,
  number        = {7650},
  pages         = {333-336},
  volume        = {544},
  adsurl        = {https://ui.adsabs.harvard.edu/abs/2017Natur.544..333D},
  archiveprefix = {arXiv},
  doi           = {10.1038/nature22055},
  eprint        = {1704.05556},
  primaryclass  = {astro-ph.EP},
}

@Article{Winters2019,
  author        = {{Winters}, Jennifer G. and {Medina}, Amber A. and {Irwin}, Jonathan M. and {Charbonneau}, David and {Astudillo-Defru}, Nicola and {Horch}, Elliott P. and {Eastman}, Jason D. and {Vrijmoet}, Eliot Halley and {Henry}, Todd J. and {Diamond-Lowe}, Hannah and {Winston}, Elaine and {Barclay}, Thomas and {Bonfils}, Xavier and {Ricker}, George R. and {Vanderspek}, Roland and {Latham}, David W. and {Seager}, Sara and {Winn}, Joshua N. and {Jenkins}, Jon M. and {Udry}, St{\'e}phane and {Twicken}, Joseph D. and {Teske}, Johanna K. and {Tenenbaum}, Peter and {Pepe}, Francesco and {Murgas}, Felipe and {Muirhead}, Philip S. and {Mink}, Jessica and {Lovis}, Christophe and {Levine}, Alan M. and {L{\'e}pine}, S{\'e}bastien and {Jao}, Wei-Chun and {Henze}, Christopher E. and {Fur{\'e}sz}, G{\'a}bor and {Forveille}, Thierry and {Figueira}, Pedro and {Esquerdo}, Gilbert A. and {Dressing}, Courtney D. and {D{\'\i}az}, Rodrigo F. and {Delfosse}, Xavier and {Burke}, Christopher J. and {Bouchy}, Fran{\c{c}}ois and {Berlind}, Perry and {Almenara}, Jose-Manuel},
  journal       = {\aj},
  title         = {{Three Red Suns in the Sky: A Transiting, Terrestrial Planet in a Triple M-dwarf System at 6.9 pc}},
  year          = {2019},
  month         = oct,
  number        = {4},
  pages         = {152},
  volume        = {158},
  adsurl        = {https://ui.adsabs.harvard.edu/abs/2019AJ....158..152W},
  archiveprefix = {arXiv},
  doi           = {10.3847/1538-3881/ab364d},
  eid           = {152},
  eprint        = {1906.10147},
  primaryclass  = {astro-ph.EP},
}

@Article{Luque2018,
  author        = {{Luque}, R. and {Nowak}, G. and {Pall{\'e}}, E. and {Kossakowski}, D. and {Trifonov}, T. and {Zechmeister}, M. and {B{\'e}jar}, V.~J.~S. and {Cardona Guill{\'e}n}, C. and {Tal-Or}, L. and {Hidalgo}, D. and {Ribas}, I. and {Reiners}, A. and {Caballero}, J.~A. and {Amado}, P.~J. and {Quirrenbach}, A. and {Aceituno}, J. and {Cort{\'e}s-Contreras}, M. and {D{\'\i}ez-Alonso}, E. and {Dreizler}, S. and {Guenther}, E.~W. and {Henning}, T. and {Jeffers}, S.~V. and {Kaminski}, A. and {K{\"u}rster}, M. and {Lafarga}, M. and {Montes}, D. and {Morales}, J.~C. and {Passegger}, V.~M. and {Schmitt}, J.~H.~M.~M. and {Schweitzer}, A.},
  journal       = {\aap},
  title         = {{The CARMENES search for exoplanets around M dwarfs. The warm super-Earths in twin orbits around the mid-type M dwarfs Ross 1020 (GJ 3779) and LP 819-052 (GJ 1265)}},
  year          = {2018},
  month         = dec,
  pages         = {A171},
  volume        = {620},
  adsurl        = {https://ui.adsabs.harvard.edu/abs/2018A&A...620A.171L},
  archiveprefix = {arXiv},
  doi           = {10.1051/0004-6361/201833423},
  eid           = {A171},
  eprint        = {1810.07572},
  primaryclass  = {astro-ph.EP},
}

@Article{Bauer2020,
  author        = {{Bauer}, F.~F. and {Zechmeister}, M. and {Kaminski}, A. and {Rodr{\'\i}guez L{\'o}pez}, C. and {Caballero}, J.~A. and {Azzaro}, M. and {Stahl}, O. and {Kossakowski}, D. and {Quirrenbach}, A. and {Becerril Jarque}, S. and {Rodr{\'\i}guez}, E. and {Amado}, P.~J. and {Seifert}, W. and {Reiners}, A. and {Sch{\"a}fer}, S. and {Ribas}, I. and {B{\'e}jar}, V.~J.~S. and {Cort{\'e}s-Contreras}, M. and {Dreizler}, S. and {Hatzes}, A. and {Henning}, T. and {Jeffers}, S.~V. and {K{\"u}rster}, M. and {Lafarga}, M. and {Montes}, D. and {Morales}, J.~C. and {Schmitt}, J.~H.~M.~M. and {Schweitzer}, A. and {Solano}, E.},
  journal       = {\aap},
  title         = {{The CARMENES search for exoplanets around M dwarfs. Measuring precise radial velocities in the near infrared: The example of the super-Earth CD Cet b}},
  year          = {2020},
  month         = aug,
  pages         = {A50},
  volume        = {640},
  adsurl        = {https://ui.adsabs.harvard.edu/abs/2020A&A...640A..50B},
  archiveprefix = {arXiv},
  doi           = {10.1051/0004-6361/202038031},
  eid           = {A50},
  eprint        = {2006.01684},
  primaryclass  = {astro-ph.EP},
}

@Article{Harakawa2022,
  author        = {{Harakawa}, Hiroki and {Takarada}, Takuya and {Kasagi}, Yui and {Hirano}, Teruyuki and {Kotani}, Takayuki and {Kuzuhara}, Masayuki and {Omiya}, Masashi and {Kawahara}, Hajime and {Fukui}, Akihiko and {Hori}, Yasunori and {Ishikawa}, Hiroyuki Tako and {Ogihara}, Masahiro and {Livingston}, John and {Brandt}, Timothy D. and {Currie}, Thayne and {Aoki}, Wako and {Beichman}, Charles A. and {Henning}, Thomas and {Hodapp}, Klaus and {Ishizuka}, Masato and {Izumiura}, Hideyuki and {Jacobson}, Shane and {Janson}, Markus and {Kambe}, Eiji and {Kodama}, Takanori and {Kokubo}, Eiichiro and {Konishi}, Mihoko and {Krishnamurthy}, Vigneshwaran and {Kudo}, Tomoyuki and {Kurokawa}, Takashi and {Kusakabe}, Nobuhiko and {Kwon}, Jungmi and {Matsumoto}, Yuji and {McElwain}, Michael W. and {Mitsui}, Koyu and {Nakagawa}, Takao and {Narita}, Norio and {Nishikawa}, Jun and {Nugroho}, Stevanus K. and {Serabyn}, Eugene and {Serizawa}, Takuma and {Takahashi}, Aoi and {Ueda}, Akitoshi and {Uyama}, Taichi and {Vievard}, S{\'e}bastien and {Wang}, Ji and {Wisniewski}, John and {Tamura}, Motohide and {Sato}, Bun'ei},
  journal       = {PASJ},
  title         = {{A super-Earth orbiting near the inner edge of the habitable zone around the M4.5 dwarf Ross 508}},
  year          = {2022},
  month         = aug,
  number        = {4},
  pages         = {904-922},
  volume        = {74},
  adsurl        = {https://ui.adsabs.harvard.edu/abs/2022PASJ...74..904H},
  archiveprefix = {arXiv},
  doi           = {10.1093/pasj/psac044},
  eprint        = {2205.11986},
  primaryclass  = {astro-ph.EP},
}

@Article{Dholakia2024,
  author        = {{Dholakia}, Shishir and {Palethorpe}, Larissa and {Venner}, Alexander and {Mortier}, Annelies and {Wilson}, Thomas G. and {Huang}, Chelsea X. and {Rice}, Ken and {Van Eylen}, Vincent and {Nabbie}, Emma and {Cloutier}, Ryan and {Boschin}, Walter and {Ciardi}, David and {Delrez}, Laetitia and {Dransfield}, Georgina and {Ducrot}, Elsa and {Essack}, Zahra and {Everett}, Mark E. and {Gillon}, Micha{\"e}l and {Hooton}, Matthew J. and {Kunimoto}, Michelle and {Latham}, David W. and {L{\'o}pez-Morales}, Mercedes and {Li}, Bin and {Li}, Fan and {McDermott}, Scott and {Murphy}, Simon J. and {Murray}, Catriona A. and {Seager}, Sara and {Timmermans}, Mathilde and {Triaud}, Amaury and {Turner}, Daisy A. and {Twicken}, Joseph D. and {Vanderburg}, Andrew and {Wang}, Su and {Wittenmyer}, Robert A. and {Wright}, Duncan},
  journal       = {\mnras},
  title         = {{Gliese 12 b, a temperate Earth-sized planet at 12 parsecs discovered with TESS and CHEOPS}},
  year          = {2024},
  month         = jun,
  number        = {1},
  pages         = {1276-1293},
  volume        = {531},
  adsurl        = {https://ui.adsabs.harvard.edu/abs/2024MNRAS.531.1276D},
  archiveprefix = {arXiv},
  doi           = {10.1093/mnras/stae1152},
  eprint        = {2405.13118},
  primaryclass  = {astro-ph.EP},
}

@Article{Moutou2026,
  author        = {{Moutou}, C. and {Petit}, P. and {Charpentier}, P. and {Cristofari}, P. and {Baruteau}, C. and {Th{\'e}bault}, P. and {Arnold}, L. and {Artigau}, E. and {Carmona}, A. and {Cook}, N.~J. and {Debras}, F. and {Delfosse}, X. and {Donati}, J.-F. and {Malo}, L. and {Ould-Elhkim}, M.},
  journal       = {\aap},
  title         = {{Characterising planetary systems with SPIRou: Questions about the magnetic cycle of 55 Cnc A and two new planets around B}},
  year          = {2026},
  month         = jan,
  pages         = {A190},
  volume        = {705},
  adsurl        = {https://ui.adsabs.harvard.edu/abs/2026A&A...705A.190M},
  archiveprefix = {arXiv},
  doi           = {10.1051/0004-6361/202556146},
  eid           = {A190},
  eprint        = {2510.11523},
  primaryclass  = {astro-ph.EP},
}
\bibliographystyle{aa_url}



\end{document}